%% file: main.tex
\documentclass[fleqn,usenatbib]{mnras}

\usepackage{newtxtext,newtxmath}

\usepackage[T1]{fontenc}

\DeclareRobustCommand{\VAN}[3]{#2}
\let\VANthebibliography\thebibliography
\def\thebibliography{\DeclareRobustCommand{\VAN}[3]{##3}\VANthebibliography}

\usepackage{graphicx}	
\usepackage{amsmath}	

\usepackage{ulem}
\usepackage[pdftex]{xcolor} 

\makeatletter
\newcommand{\CASA}{{\small CASA} } 
\newcommand{\kms}{\,km\,s$^{-1}$} 
\newcommand{\WHz}{W Hz$^{-1}$}
\renewcommand{\ion}[2]{#1$\;${\small\rm\@Roman{#2}}\relax}
\newcommand{\phn}{\phantom{0}}

\newcommand{\phs}{\phantom{$-$}}

\newcommand{\myemail}{t.hayashi@nao.ac.jp}

\DeclareRobustCommand{\erase}{\bgroup\markoverwith{\textcolor[gray]{0.6}{\rule[.5ex]{2pt}{1.5pt}}}\ULon}
\makeatother

\title[A GMRT survey of BAL quasars]{
	A Giant Metrewave Radio Telescope Survey of Radio-loud Broad Absorption Line Quasars
}

\author[T.~J.~Hayashi et al.]{
	Takayuki J. Hayashi$^{1,2}$\thanks{E-mail: \myemail}, 
	Akihiro Doi$^{3,4}$, 
	Hiroshi Nagai$^{1,5}$,\\
    $^{1}$National Astronomical Observatory of Japan, 2-21-1 Osawa, Mitaka, Tokyo, 181-8588, Japan\\
	$^{2}$Azabu Junior and Senior High School, 2-3-29 Motoazabu, Minato-ku, Tokyo, 106-0046, Japan\\
	$^{3}$Institute of Space and Astronautical Science, Japan Aerospace Exploration Agency,
		3-1-1 Yoshinodai, Chuo, Sagamihara, Kanagawa, 252-5210, Japan\\
	$^{4}$Department of Space and Astronautical Science, The Graduate University for Advanced Studies, SOKENDAI,\\
        3-1-1 Yoshinodai,  Chuo, Sagamihara, Kanagawa, 229-8510, Japan\\
    $^{5}$Astronomical Science Program, The Graduate University for Advanced Studies, SOKENDAI, 
        2-21-1 Osawa, Mitaka, 181-8588, Japan\\
}
\date{\today}
\pagerange{\pageref{firstpage}--\pageref{lastpage}}
\date{Accepted XXX. Received YYY; in original form ZZZ}

\pubyear{2024}

\begin{document}
\label{firstpage}
\pagerange{\pageref{firstpage}--\pageref{lastpage}}
\maketitle

\begin{abstract}
A substantial fraction of quasars display broad absorption lines (BALs) in their rest-frame ultraviolet spectra.
While the origin of BALs is thought to be related to the accretion disc wind,
it remains unclear whether the observed ratio of BAL to non-BAL quasars is due to orientation.
We conducted observations of 48 BAL quasars and the same number of non-BAL quasars at 322\,MHz using the Giant Metrewave Radio Telescope.
Combined with previous flux measurements ranging from MHz to GHz frequencies, we compared continuum radio spectra between the two quasar groups. 
These data offer insights into low-frequency radio properties that have been difficult to investigate with previous observations only at GHz frequencies.
Our results present that $73\pm13$\,per cent of the BAL quasars exhibit steep or peaked spectra, a higher proportion than $44 \pm 14$\,per cent observed in the non-BAL quasars.
In contrast, there are no discernible differences between the two quasar groups in the radio luminosity, peak frequency, and spectral index distributions of sources with steep or peaked spectra and sources with flat or inverted spectra.
Generally, as the jet axis and line of sight become closer to parallel, quasars exhibit flat or inverted spectra rather than steep or peaked ones. 
Therefore, these results suggest that BAL quasars are more frequently observed farther from the jet axis than non-BAL quasars. 
However, given that a certain proportion of BAL quasars exhibit flat or inverted spectra, more than the simple orientation scenario is required to elucidate the radio properties of BAL quasars.
\end{abstract}

\begin{keywords}
galaxies: active -- galaxies: jets -- galaxies: evolution -- radio continuum: galaxies -- quasars: general --  radiation mechanisms: non-thermal
\end{keywords}


\section{Introduction}\label{sec:intro}
Broad absorption line (BAL) quasars exhibit blue-shifted absorption troughs in their ultraviolet spectra, which result from broad resonance lines \citep{1967ApJ...147..396L,1981ARA&A..19...41W,1991ApJ...373...23W}.
An active galactic nucleus (AGN) wind, accelerated by radiation pressure to a velocity of $\sim$\,0.1$c$, is responsible for the absorption \citep[][]{2002ApJ...569..641L,2007ApJ...665..990G}. 
The winds emanating from the accretion disc are the most potent absorber \citep[e.g.,][]{1995ApJ...451..498M,2000ApJ...545...63E,2000ApJ...543..686P,2013PASJ...65...40N}. 
AGN winds are of significant importance since they provide substantial energy to their host galaxies \citep[e.g.,][]{2013ApJ...762...49B,2019MNRAS.483.1808H,2020ApJ...891...53C,2022ApJ...937...74C}, playing a pivotal role in regulating the coevolution of galaxies and supermassive black holes \citep[][]{2009ApJ...699...89C,2012ApJ...745..178F,2023arXiv230109731B}. 
Therefore, a thorough understanding of BAL quasars is imperative in a cosmological context.

BAL quasars fall into two categories based on the ionisation levels of their absorption troughs. HiBAL quasars display high-ionisation troughs, such as \ion{C}{4} and \ion{Si}{4}, constituting 10--30 per cent of all quasars \citep[e.g.,][]{2006ApJS..165....1T,2008MNRAS.386.1426K,2009ApJ...692..758G,2011MNRAS.410..860A}. 
LoBAL quasars, accounting for $\sim$\,10 per cent of all BAL quasars, exhibit low-ionisation troughs in addition to high-ionisation ones, such as \ion{Mg}{2} and \ion{Al}{3}. 
A small subset of LoBAL quasars, $\sim$\,1 per cent of all BAL quasars, are referred to as FeLoBAL quasars, showing absorption from excited states of \ion{Fe}{2} or \ion{Fe}{3} \citep{2006ApJS..165....1T}. 
The cause of these observed proportions is still an enigmatic puzzle, with two prominent models proposed: an orientation scheme and an evolution scheme. 
The former suggests that all quasars have directional winds and that BAL detection depends on the object's inclination \citep{1991ApJ...373...23W,1995ApJ...448L..73G}.
In this scenario, the ratio of BAL to non-BAL quasars and the proportion of different types of BAL quasars depend on the number of quasars viewed from various angles \citep{2000ApJ...545...63E}. 
In contrast, the evolution scheme argues that the ratio of BAL to non-BAL quasars reflects the period during which quasars exhibit AGN winds. 
LoBAL quasars, in particular, are believed to be young, recently-fueled quasars \citep{1993ApJ...413...95V,2001ApJ...555..719C,2006MNRAS.368.1001L,2022NatAs...6..339C,2023ApJ...949...69L,2024ApJ...963....3P}. 
Ultimately, the question concerns whether BAL and non-BAL quasars share the same central engine.

Previous studies have shown that jet activity is the main factor responsible for the radio emission observed in radio-loud BAL quasars, which is supported by high spatial resolution images obtained through very long baseline interferometry \citep[VLBI; e.g.,][]{2003A&A...397L..13J,2008MNRAS.391..246L,2013A&A...554A..94B,2013ApJ...772....4H,2015A&A...579A.109K,2015MNRAS.450.1123C}.
Since thermal disc winds and nonthermal jets arise from the same central engine, radio observations that detect nonthermal emission can provide valuable insights into the abovementioned question. 
Previous investigations conducted at GHz frequencies have revealed differences between BAL and non-BAL quasars, with BAL quasars found to be less luminous than non-BAL quasars  \citep{1992ApJ...396..487S,2000ApJ...538...72B,2001ApJS..135..227B,2008ApJ...687..859S}. 
Additionally, they exhibit steeper spectra \citep{2011ApJ...743...71D,2012A&A...542A..13B}, 
which supports the orientation scenario where BAL quasars are viewed from an edge-on perspective \citep{2012ApJ...752....6D}. 
However, morphological studies have suggested the presence of a significant proportion of BAL quasars, including polar BAL quasars, at all inclinations \citep{2022MNRAS.511.4946N}.
Moreover, BAL quasars exhibiting significant radio variability, indicative of face-on objects, have been documented \citep{2006ApJ...639..716Z,2007ApJ...661L.139G,2011ApJ...743...71D,2017PASJ...69...77C}.
Therefore, the simple orientation scenario cannot fully account for the radio characteristics of BAL quasars.

Thus far, several observational inquiries have also explored the overall radio spectral shape of BAL quasars, covering both radio-quiet \citep{1997AJ....113..144B} and radio-loud samples \citep{2008MNRAS.388.1853M,2012A&A...542A..13B,2017MNRAS.467.4763T}.
Initially, there was a prevalent belief that most radio-loud BAL quasars displayed compact radio morphology \citep{2000ApJ...538...72B,2001ApJS..135..227B}.
Their spectral characteristics were found to be typical of compact steep-spectrum (CSS) or GHz peaked-spectrum (GPS) sources \citep{2008MNRAS.388.1853M}, which are candidates for lobe-dominated young radio sources \citep{1998PASP..110..493O}.
These findings were part of the basis for the evolution scheme of BAL quasars. 
However, when comparing them with non-BAL quasars, the fractions of sources suggesting a peaked spectrum were similar in the two samples  \citep{2012A&A...542A..13B, 2017MNRAS.467.4763T}.

Despite the efforts mentioned above, observations conducted at GHz frequencies are suboptimal for comprehending the true nature of objects located at high redshifts, where the ultraviolet absorption lines fall within the optical band, owing to their susceptibility to the orientation-dependent relativistic beaming effects of the jets. 
Based solely on high-frequency observations, it is impossible to discern the underlying cause of the radio weakness of BAL quasars, whether attributed to intrinsically weak jet activity in the central engine or weak relativistic beaming effects.
Moreover, due to the limited low-frequency observations in the MHz range in the past, comparisons of the overall spectral shape for  high-redshift sources were restricted to the most extreme cases, where radio spectra peak at high rest frequencies.
Therefore, observations at lower frequencies play a crucial role in addressing these issues. By capturing the low-frequency extended emission, where relativistic effects are weak, we can trace the total amount of intrinsic jet activity from the central engines.

Recently, the all-sky survey conducted at 144\,MHz \citep{2022A&A...659A...1S} has illuminated that the detection of BAL in its optical spectra is uniform across a range of radio luminosities \citep{2019A&A...622A..15M}, differing from the previous findings at GHz frequencies \citep[e.g.,][]{2008ApJ...687..859S}. 
Furthermore, \cite{2019A&A...622A..15M} have reported a potential difference in spectral indices by comparing the 144-MHz data with conventional 1.4-GHz data \citep{1995ApJ...450..559B}. 
Nevertheless, prior estimates of spectral indices based on the limited number of data points in the MHz regime are inaccurate, and the underlying causes of these discrepancies have yet to be discussed. 
Therefore, multifrequency data obtained at lower frequencies are imperative for capturing the overall spectral shape and determining the physical parameters at the same rest frequencies across all sources.

This paper presents the findings of a survey on BAL quasars and a comparison sample at 322\,MHz using the Giant Metrewave Radio Telescope (GMRT). 
The study compares the low-frequency radio spectra of BAL quasars with those of non-BAL quasars. 
The samples are described in Section~\ref{sec:sample}, while Section~\ref{sec:obs} explains the observations and data reduction. The results are presented in Section~\ref{sec:res}, along with discussions of the origin of the radio properties in Section~\ref{sec:discuss}. 
The paper defines spectral index, $\alpha$, as $f_\nu \propto \nu^\alpha$, where $f_\nu$ represents flux density at frequency, $\nu$. 
We utilised the standard cosmological model with cold dark matter and a cosmological constant having $H_0 = 70$\,km\,s$^{-1}$\,Mpc$^{-1}$, $\Omega_{\rm M} = 0.3$, and $\Omega_\Lambda = 0.7$, which is supported by observational studies from the past decades \cite[e.g.,][]{2020A&A...641A...6P}.

\section{Sample}\label{sec:sample}
\subsection{BAL sample}
We selected our targets from the quasar catalogue \citep{2007AJ....134..102S}  obtained from the fifth data release \citep[DR5;][]{2007ApJS..172..634A} of the Sloan Digital Sky Survey \citep[SDSS;][]{2000AJ....120.1579Y}. 
Initially, we extracted our BAL sample from a subset of quasars, specifically those that have had their \ion{C}{4} BAL reported by \cite{2009ApJ...692..758G}.  
The definition of BAL is a subject of debate, and various studies have addressed this issue \citep{2002ApJS..141..267H,2007ApJ...665..990G,2008MNRAS.386.1426K}.  
Here, the traditional balnicity index \citep[BI;][]{1991ApJ...373...23W} is a strict criterion, and some studies have suggested that it may underestimate the presence of BAL \citep{2003AJ....125.1711R,2008MNRAS.386.1426K}. 
Therefore, in this study, we adopted the modified balnicity index, BI$_0$, proposed by \cite{2009ApJ...692..758G}, which is more lenient than BI.

\begin{figure*}
	\centering
	\includegraphics[height=50mm]{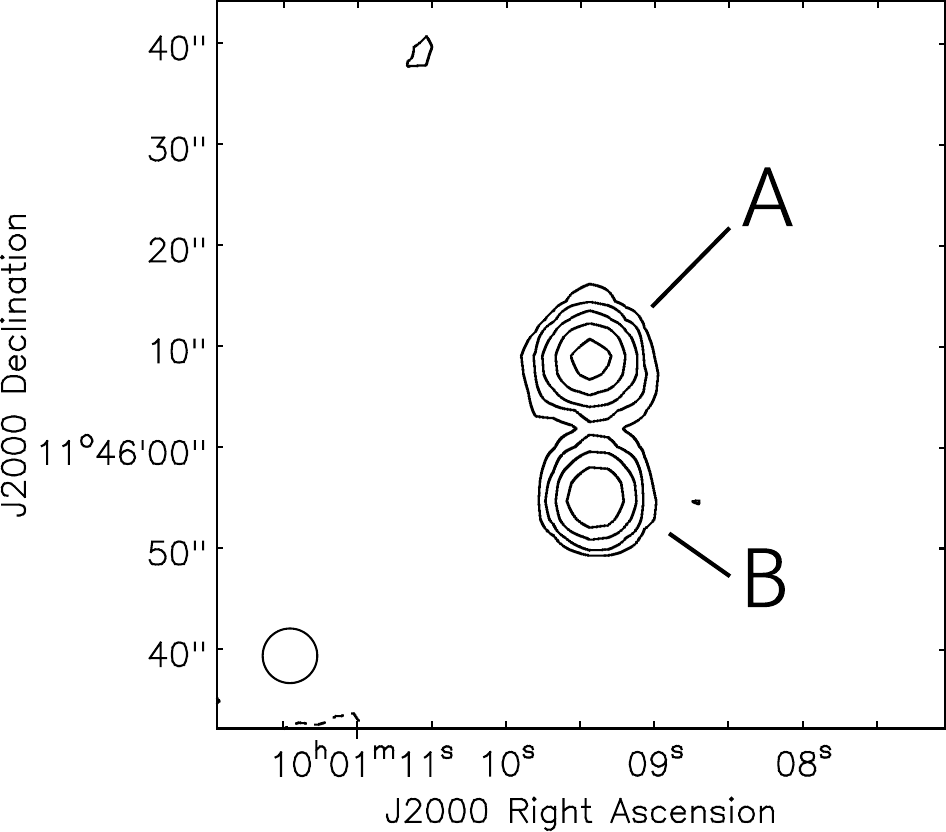} 
    \hspace{3mm}
	\includegraphics[height=50mm]{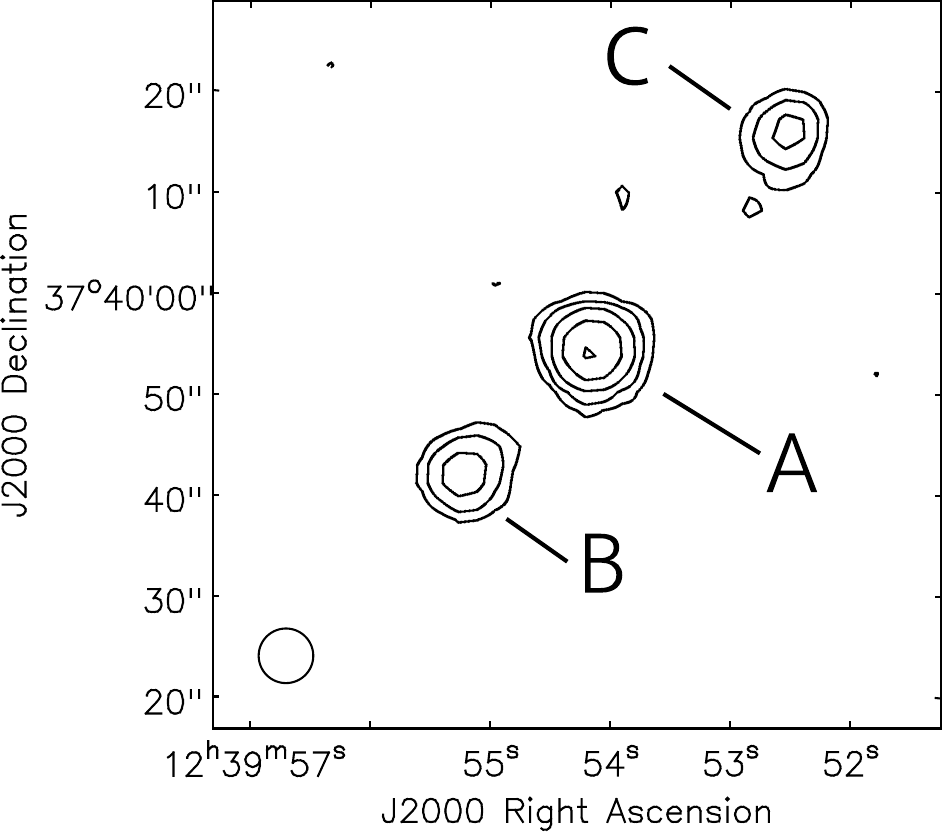} 
    \hspace{3mm}
	\includegraphics[height=50mm]{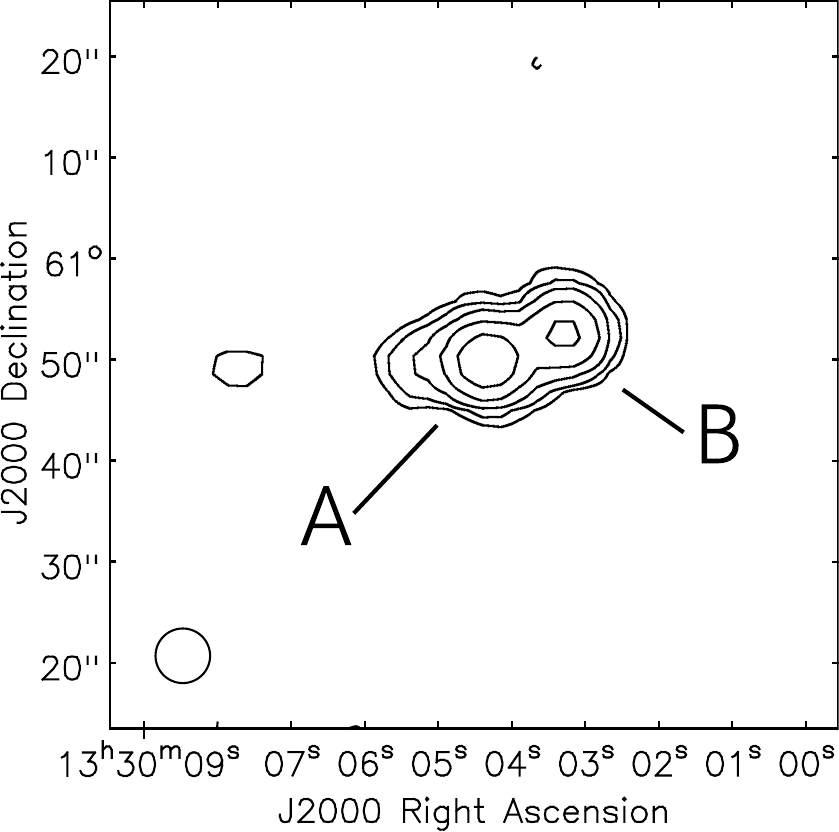}\\
    \vspace{3mm}
    \hspace{3mm}
	\includegraphics[height=50mm]{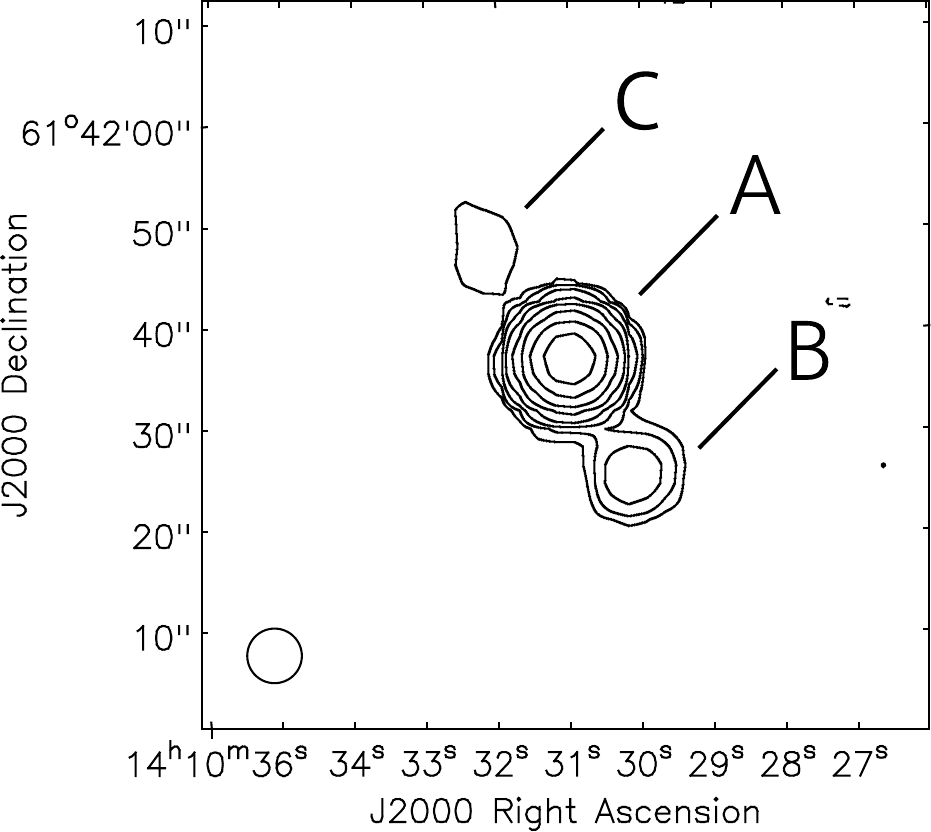} 
    \hspace{3mm}
	\includegraphics[height=50mm]{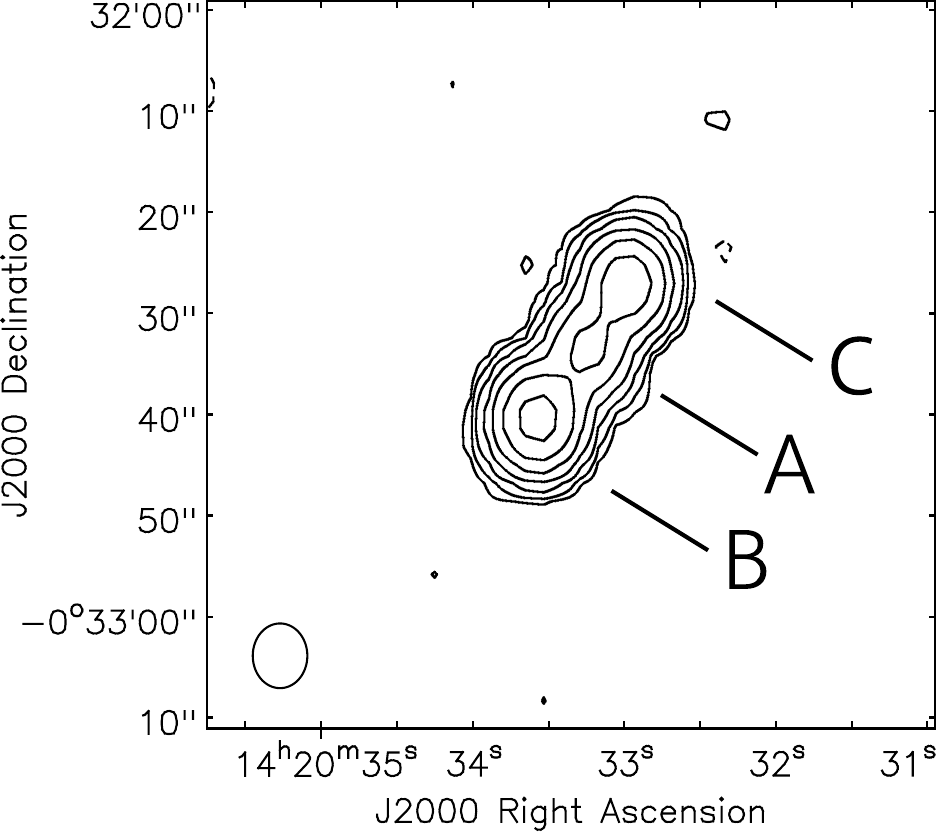} 
     \caption{
        Radio images at 1.4\,GHz for b01 (J100109$+$114608), b20 (J123954$+$373954), b25 (J133004$+$605949), n24 (J141031$+$614136), and n27 (J142033$-$003233) from top left to bottom right, all of which are resolved BAL quasars, provided by the FIRST survey. 
        The optical position of the quasar corresponds to the component A in each image.
        From top left to bottom right, the contour levels commence at 520, 546, 479, 494, and 394\,$\mu$Jy\,beam$^{-1}$ (equivalent to the 3$\sigma$ noise level of the image) and are then multiplied by a factor of 2.
	}
	\label{fig:FIRST_resolved}
\end{figure*}

To ensure that the blue-shifted and high-velocity  \ion{C}{4} absorption line was within the SDSS spectral bandpass, we limited our target range to sources with redshifts of $1.68\leq z\leq 4.93$.
Our identification of radio-loud quasars involved cross-referencing the optical positions from the SDSS with the Faint Images of the Radio Sky at Twenty centimeters survey \citep[FIRST survey;][]{1995ApJ...450..559B}, restricting our sample to sources with an integrated flux density greater than 10\,mJy and located within 3\,arcsecs of the FIRST radio positions. 
We focused on objects within the right ascension range of 10 to 17\,h to accommodate the allocated observing time.
Our sample, detailed in Table~\ref{tbl:BALsample}, comprises 48 sources, 35 of which intersect with the sample in \cite{2011ApJ...743...71D}. 
The FIRST survey spatially resolved three sources in our sample; 
Fig.~\ref{fig:FIRST_resolved} presents their images. 
It is worth noting that three sources in our sample, 
b20 (J123954.15$+$373954.5), 
b34 (J141437.99$+$045537.4), 
and b36 (J143340.35$+$512019.3), 
are also classified as LoBAL quasars by \cite{2009ApJ...692..758G}, showing BAL in their \ion{Al}{3} line.
Also, b34 shows BAL in its \ion{Mg}{2} line.

\input{tab_sample_BAL.tex}

Subsequently, we identified a discrepancy in the redshift of b02 (100424.88$+$122922.2; $z=2.64$) reported by \cite{2007AJ....134..102S} and \cite{2009ApJ...692..758G}, with the latter presenting a value of $z=4.66$. 
Further analysis, based on a broader SDSS spectrum obtained during a different epoch \citep{2020ApJS..250....8L}, suggested a redshift closer to that of \citeauthor{2007AJ....134..102S}, which we ultimately adopted. 
Assuming the redshift of $z=2.64$, we detected no \ion{C}{4} absorption from this source; hence, the BI and BI$_0$ presented in Table~\ref{tbl:BALsample} should be 0. 
However, the source displayed potential BAL in \ion{Al}{3} and \ion{Mg}{2}, indicating characteristics of a LoBAL quasar. 
Despite this issue, we could not acquire GMRT observational data for this source due to radio frequency interference (RFI).
Even in statistical analyses that do not require the GMRT data, we provide both examinations, considering and excluding this specific object.

\subsection{Non-BAL sample}
To establish a comparison group, we created a non-BAL sample where each object was paired with objects in the BAL sample based on each physical quantity. 
The criteria for selection comprises a redshift of 10 per cent of the corresponding BAL quasar within the range of $1.68\leq z\leq 4.93$, a flux density at 1.4\,GHz of 20 per cent of the BAL quasar, and an $i$-band magnitude within $\pm$1.0 of the BAL quasar. 
In addition, we tried to select objects that could be observed using the same phase calibrators as the BAL sample to minimise the total observation time. 
Table~\ref{tbl:nBALsample} lists the resulting non-BAL sample, which contains two sources that the FIRST survey resolved; 
Fig.~\ref{fig:FIRST_resolved} also shows their images. 
Additionally, we visually inspected the SDSS spectra of the non-BAL sample in the rest-frame wavelength range of 1400--1600\,\AA. 
We confirmed they were free from significant broad absorption of \ion{Si}{4} or \ion{C}{4}.
We note, however, that we inadvertently incorporated a source, 
n24 (J141031.00$+$614136.9), 
that displays a BAL in \ion{Al}{3} at 1857.4\,\AA~in our sample.
We will present both analyses, incorporating and excluding this object.

\begin{figure*}
	\centering
	\includegraphics[width=0.66\columnwidth]{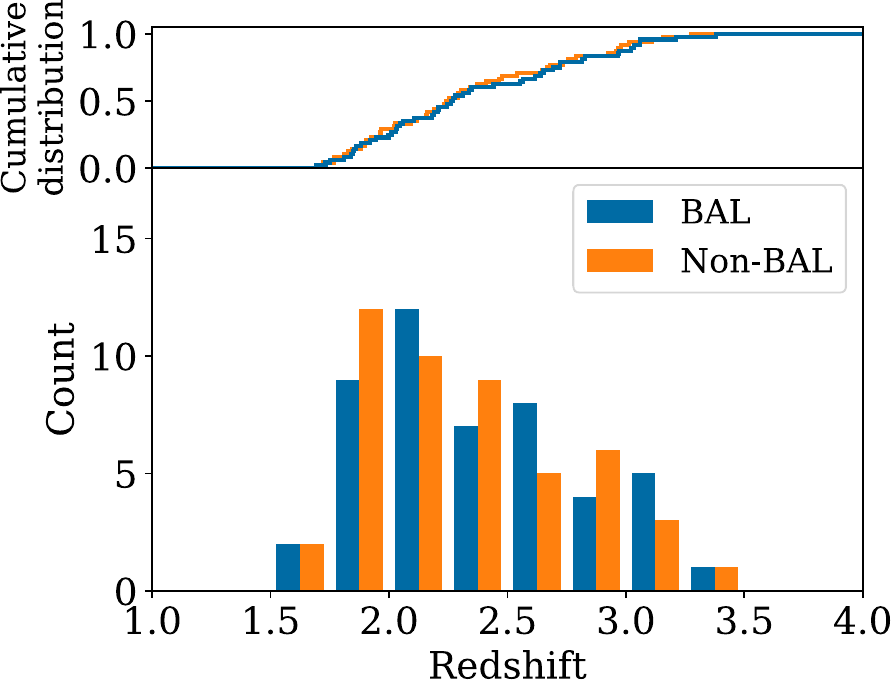}
	\includegraphics[width=0.66\columnwidth]{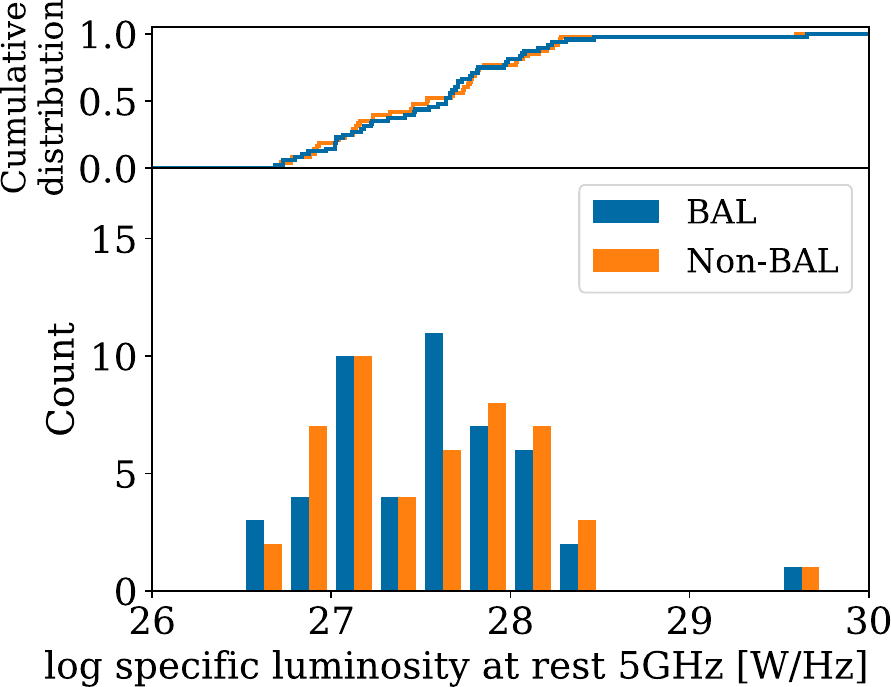}
	\includegraphics[width=0.66\columnwidth]{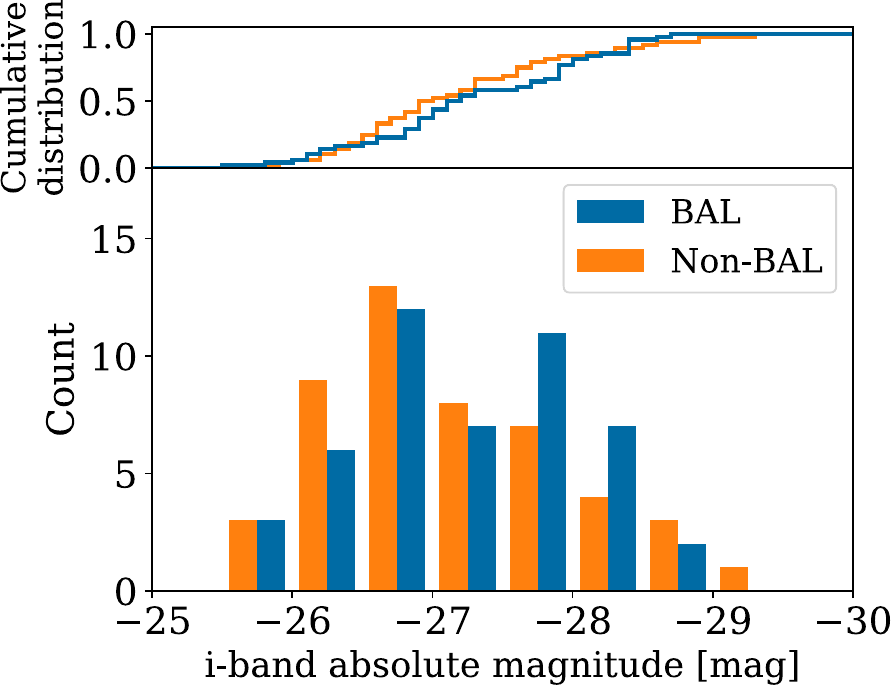}
	\caption{
        Comparison of the characteristics of the BAL and non-BAL samples: 
        redshift $z$, specific radio luminosity at rest 5.0\,GHz, $L_{5.0}$, and $i$-band absolute magnitude, $M_i$, from left to right. 
        Each panel shows the distribution of the two samples, represented respectively in blue and orange, both in terms of number and cumulative frequency. 
        We have applied two statistical tests, the KS and RS tests, to evaluate the likelihood that the two distributions do not originate from different parent populations.
        The KS test yields a test statistic of $D_{\rm KS}$ = 0.0833, 0.0833, and 0.1875 for $z$, $L_{5.0}$, and $M_i$, respectively, with corresponding probabilities of  $p = 0.996$, 0.996 and 0.368. 
        The RS test give a test statistic of $Z_{\rm RS}$ = 1194, 1162, and 1018 for $z$, $L_{5.0}$, and $M_i$, respectively, with corresponding probabilities of $p = 0.764$, 0.945 and 0.326.
        These $p$-values indicate the probability of obtaining a result equal to or more extreme than the observed result, assuming the null hypothesis that the two samples are drawn from the same parent population.
        }
	\label{fig:sample}
\end{figure*}

Fig.~\ref{fig:sample} depicts the distributions of redshift, specific luminosity at 5\,GHz with a spectral index of $\alpha = -0.7$, and $i$-band absolute magnitude for both the BAL and non-BAL samples. 
The Kolmogorov-Smirnov (KS) and Wilcoxon rank sum (RS) tests indicate no notable distinctions between the distributions of the two samples. 
Consequently, we can deduce that the two samples exhibit homogeneity, except for BALs.

\input{tab_sample_nBAL.tex}

\subsection{Flux variability}
\begin{figure}
	\centering
	\includegraphics[width=0.9\columnwidth]{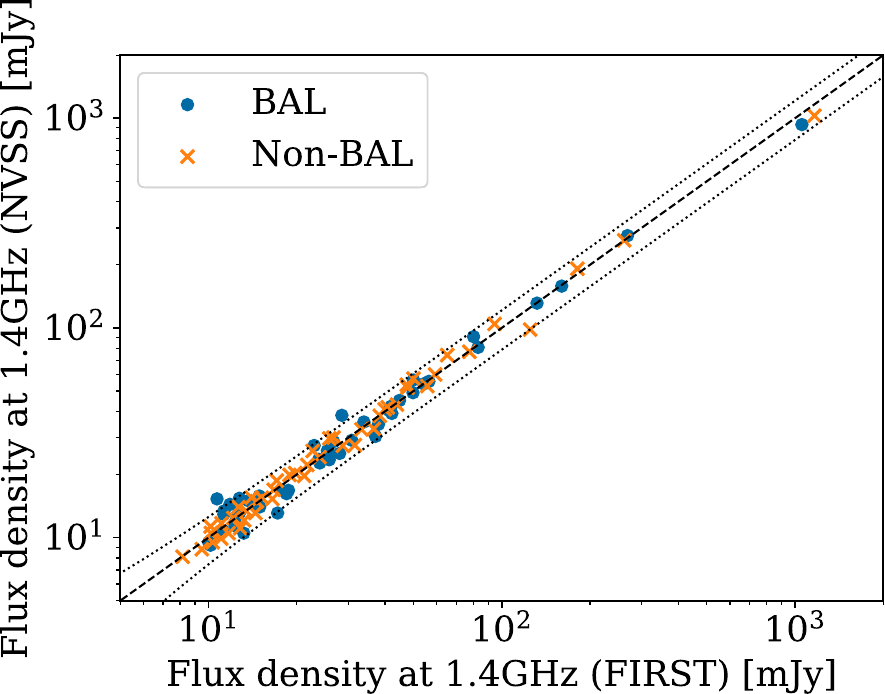}\\
	\caption{
        Flux density distribution at 1.4\,GHz obtained from the FIRST survey and the NVSS. 
        Blue-filled circles and orange crosses indicate BAL and non-BAL quasars, respectively.
        The dashed line indicates where the two survey fluxes are equal, and the dotted lines show the 3$\sigma$ variation of the sources around the line of equal flux. 
        Two BAL quasars that appear to lie well above the line may have a second object contaminating the NVSS flux density and are likely not variable.
	}
	\label{fig:variability}
\end{figure}

The primary objective of this investigation is to examine the radio spectra of quasars using new GMRT observations and previous flux measurements.
Generally, quasars display variations in flux density at radio wavelengths 
\citep{1992ApJ...386..473H,2005ApJ...618..108B,2007A&A...469..899H}, which may impact the data collected at different epochs.
To assess this impact on the samples, we compared the 1.4-GHz flux density measurements acquired from the FIRST survey with those obtained from the NRAO VLA Sky Survey \citep[NVSS;][]{1998AJ....115.1693C}.
Figure~\ref{fig:variability} depicts the comparison of the integrated flux densities.
To evaluate the extent of flux-density variability, we utilised the index introduced by \citet{2006ApJ...639..716Z}, given by
\begin{eqnarray}
V = \frac{|f_1-f_2|}{\sqrt{\sigma_1^2+\sigma_2^2}},
\end{eqnarray}
where $f_i$ and $\sigma_i$ denote the flux density and measurement error for the $i$-th epoch data, respectively.
The errors in flux density are the root mean square of the thermal noise and calibration uncertainty, which were assumed to be 5 per cent.
When comparing data acquired from different array configurations, objects with a high-resolution peak flux density greater than the integrated flux density at a lower resolution are considered potential variable objects.

We have identified two candidates, b04 (J104452.41$+$104005.9) and n24 (J141031.00$+$614136.9), representing significant variability of $V > 3$, with $V = 3.57$ and 3.52, respectively. 
While we acknowledge that only two epochs of observations may not be sufficient to provide conclusive evidence regarding the stability or variability of the targets, we will present both analyses, encompassing and omitting these candidates for variable sources.

\section{Observations and Data Reduction}\label{sec:obs}
\subsection{New 322-MHz observation}\label{sec:myobs}
We conducted the new GMRT observations between 16--19 May 2013 under project code 24\_006. 
To reduce the impact of ionospheric variations, we initiated the observations in the early evening and continued throughout the night as a contiguous block lasting ten hours. 
During the observations, we utilised the default parameters of the 306-MHz band of the legacy GMRT system, with a 32-MHz wide intermediate frequency centred at 322\,MHz for both LL and RR polarisations. 
We collected the data using 256 spectral channels, each with a width of 12.5\,kHz. 
We monitored the flux and bandpass calibrator, 3C\,286, every 3--4\,hours, 
and we observed each target for 6--8\,minute scans at two different hour angles. 
Additionally, we observed phase calibrators approximately every 30 minutes, with a typical angular separation of 6\,degrees and a maximum separation of 13.5 degrees between targets and each calibrator. 
Please refer to Table~\ref{tbl:obs} for an overview of the observations.

\begin{table*}
\begin{minipage}{112mm}
\caption{Observation summary.}
\label{tbl:obs}
\begin{tabular}{cccccccccccccc}
\hline		
Date	&	Phase calibrator	&	Observed targets												\\\hline
2013 May 16	&	1131$+$4514	&	b05, 	b09, 	b11, 	n01, 	n03, 	n08							\\
	&	1338$+$3851	&	b20, 	b29, 	b32, 	b33, 	b35, 	n18, 	n19, 	n26, 	n28, 	n31			\\
	&	1531$+$3533	&	b40, 	n35, 	n36, 	n37, 	n38								\\
	&	1653$+$3945	&	b45, 	b48, 	n45, 	n46									\\
2013 May 17	&	0753$+$4231	&	n47												\\
	&	1123$+$0530	&	b03, 	b04, 	b06, 	b07, 	b12, 	b13, 	b14, 	b15, 	n02, 	n04, 	n05, 	n06	\\
	&	1308$-$0950	&	b26												\\
	&	1419$+$0628	&	b31, 	b34, 	b37, 	n23, 	n27, 	n29, 	n32, 	n33					\\
	&	1510$-$0543	&	b41												\\
2013 May 18 	&	1252$+$5634	&	b19, 	n09, 	n11, 	n14, 	n16								\\
	&	1313$+$6735	&	b10, 	b17, 	b25, 	b28									\\
	&	3C\,286	&	b27, 	n21											\\
	&	1411$+$5212	&	b30, 	b36, 	n20, 	n25, 	n30								\\
	&	1449$+$6316	&	b42, 	n24, 	n34										\\
	&	1545$+$4751	&	b38, 	b39, 	b46, 	n41									\\
2013 May 19 	&	0958$+$3224	&	n48												\\
	&	1021$+$2159	&	\textbf{b01}, 	\textbf{b02}											\\
	&	1156$+$3128	&	b08, 	n07											\\
	&	1242$-$0446	&	b16, 	\textbf{b18}, 	b21, 	b22, 	\textbf{n10}, 	n12, 	n13, 	n15					\\
	&	1347$+$1217	&	b23, 	b24, 	n17, 	n22									\\
	&	1602$+$3326	&	b43, 	b44, 	b47, 	n39, 	n40, 	n42, 	n43, 	n44					\\
\hline
\end{tabular}

\textit{Note.} IDs in observed targets column are object IDs in Table~\ref{tbl:BALsample} and \ref{tbl:nBALsample}. 
Targets whose flux measurements were not made due to severe RFI are indicated in bold.
\end{minipage}
\end{table*}

We processed the data using the Common Astronomy Software Applications \citep[{\small CASA};][]{2007ASPC..376..127M} package, with reference to the calibration and imaging pipeline developed by \cite{2020MNRAS.497.5383I}.
We started with an initial delay correction using the \CASA task \texttt{gaincal} with a bright calibrator, 3C\,286. 
We then performed amplitude and phase bandpass calibration using the same calibrator.
Next, the flux density scale was determined using the formula from \cite{2017ApJS..230....7P}, with 3C\,286 as the calibrator. 
Finally, we calibrated the antenna-based amplitude and phase time variations using phase calibrators. 
We identified and flagged radio frequency interference in the time and frequency domains at each stage using the \texttt{rflag} and \texttt{tfcrop} modes of the \CASA task \texttt{flagdata}.
Unfortunately, due to significant scintillation of the ionosphere, we could not achieve proper phase calibration for three BAL quasars (b01, b02, and b18) and one non-BAL quasar (n10). 
Due to the absence of low-frequency flux measurements potentially impacting the statistical outcomes, we present both analyses, inclusive and exclusive of these objects.
These sources requiring careful consideration in subsequent statistical analyses are listed in Table~\ref{tbl:object-w-note}.

\begin{table*}
\begin{minipage}{145mm}
\caption{Annotation of objects requiring consideration in statistical analyses} \label{tbl:object-w-note}
\begin{tabular}{cl}
\hline							
ID    & comments \\\hline
b01   & RFI in the 322-MHz data, source with multi components\\
b02   & RFI in the 322-MHz data, no \ion{C}{4} BAL but potential presence of \ion{Al}{3} BAL\\
b04   & candidate exhibiting variability\\
b11   & no detection at both 150 and 322\,MHz\\
b18   & RFI in the 322-MHz data\\
b20   & source with multi components\\
b25   & source with multi components\\
n10   & RFI in the 322-MHz data\\
n24   & candidate exhibiting variability, source with multi components, no \ion{C}{4} BAL but potential presence of \ion{Al}{3} BAL\\
n27   & source with multi components\\
n32   & no detection at both 150 and 322\,MHz\\
\hline
\end{tabular}

\end{minipage}
\end{table*}

We utilised the \CASA task \texttt{tclean} equipped with widefield imaging mode and multiterm multifrequency synthesis capabilities to carry out imaging processes. 
First, we reduced the size of the data set by averaging the data to 32\,channels, each 2\,MHz wide. 
In order to resolve the antenna-based amplitude and phase corrections for Stokes $I$, self-calibration was executed. 
Initially, we performed phase-only calibration to the data with $u$-$v$ length greater than $2\,\textrm{k}\lambda$ against a basic point source model of $\sim$\,$1 \times 1$ square degrees, obtained from the FIRST survey \citep{1995ApJ...450..559B} utilising the Montage Image Mosaic Engine \citep{2010arXiv1005.4454J}.
Subsequently, we gradually integrated short baselines and resolved radio sources, followed by several rounds of widefield imaging and phase and amplitude calibration. 
Ultimately, we generated final images using a Briggs weighting of $\texttt{robust} = 0.5$ \citep{1995PhDT.......238B}, encompassing a field of view as large as $2.5\times 2.5$ square degrees to include probable strong sources positioned beyond the primary lobe. 
The typical synthesised beam size for the resulting images was $\sim$\,11 and $\sim$\,8\,arcsecs for the major and minor axes, respectively.

To determine the flux densities of our targets, we used the \CASA task \texttt{IMFIT}, which fits an elliptical Gaussian component within a small box enclosing the radio source. 
We estimated the flux-density errors by taking the root mean square of a calibration uncertainty and the error provided by \texttt{IMFIT}, which contains thermal noise and fitting errors. 
To be conservative, we assumed a calibration uncertainty of 10 per cent \citep[e.g.,][]{2004ApJ...612..974C}.
The typical rms noise level obtained from \texttt{IMFIT} was approximately 0.5\,mJy\,beam$^{-1}$, consistent with the previous snapshot survey observations made using the legacy GMRT system \citep[e.g.,][]{2016MNRAS.457..730P}.

\subsection{Archival data}\label{sec:survey}
We additionally used data points from previous publications to determine the overall spectral shape.

For 144-MHz data, we initially endeavoured to utilise data obtained from the second data release of the LOFAR Two-metre Sky Survey \citep[LoTSS-DR2;][]{2022A&A...659A...1S}.
This survey observes at a frequency centred on 144\,MHz and has a spatial resolution of 6\,arcsecs. 
Although the survey's catalogue tables provide error estimates, these estimations only reflect thermal noise and do not account for the uncertainty associated with amplitude calibration. 
Therefore, we calculated the flux density error by taking the root mean square of thermal noise and amplitude calibration error. 
The latter was assumed to be 10\,per cent of the flux density measurement \citep{2022A&A...659A...1S}. 
Since the survey observations are still ongoing, flux density measurements are not accessible for some sources. For such sources, we relied on the flux density at 147.5\,MHz provided by the TIFR GMRT Sky Survey \citep[TGSS;][]{2017A&A...598A..78I}, whose spatial resolution is 25\,arcsecs. 
We directly utilised the flux measurement errors listed in the survey's catalogue, which account for thermal noise and amplitude calibration errors. 
The TGSS did not detect some sources that LoTSS has yet to observe. For these sources, we estimated the thermal noise from the image cutouts provided by the TGSS and adopted 5$\sigma$ as an upper limit.

For GHz-frequency data, we acquired flux density measurements at 1.4 and 3.0\,GHz from the FIRST survey and the VLA Sky Survey \citep[VLASS;][]{2020PASP..132c5001L}, respectively. 
These surveys have an angular resolution of 5.0 and 2.5\,arcsecs, respectively. 
We calculated the flux density error by taking the square root of the sum of the squares of thermal noise from the catalogues and a 5 per cent amplitude calibration error \citep{VLASSmemo...13}. 
As \cite{2021ApJS..255...30G} recommend, we corrected the catalogue values by multiplying them by 1/0.87 in our analysis due to a systematic underestimation of the flux density measurements. 

In addition to the data mentioned above, where flux measurements at the same frequency are provided for all objects in the samples, the following archive data were also utilised to evaluate the overall spectral shape of targets:
    the LOFAR LBA Sky Survey (LoLSS) at 54\,MHz with a resolution of 15\,arcsecs \citep{2023A&A...673A.165D},
    the VLA Low-Frequency Sky Survey (VLSS) at 74\,MHz with a resolution of 80\,arcsecs \citep{2007AJ....134.1245C},
    the GaLactic and Extragalactic All-sky Murchison Widefield Array (GLEAM) survey at 200\,MHz with a resolution of 120\,arcsecs \citep{2017MNRAS.464.1146H},
    the Texas Survey at 365\,MHz with a resolution of 20\,arcsecs  \citep{1996AJ....111.1945D},
    the Rapid ASKAP Continuum Survey (RACS) at 887.5\,MHz with a resolution of 25\,arcsecs \citep{2021PASA...38...58H}, and
    the Cosmic Lens All-Sky Survey (CLASS) at 8.4\,GHz,  with a resolution of 0.2\,arcsecs \citep{2003MNRAS.341....1M}.
As these data are not available for all objects in the samples, a statistical comparison of spectral indices including them was not performed.
If the flux error in the catalogue does not consider calibration uncertainty, we estimated the error by calculating the root mean square of the thermal noise and amplitude calibration error as described in the source paper.
Furthermore, we also utilised prior flux measurements carried out by VLA specifically for BAL quasars, as documented in studies by \cite{2011ApJ...743...71D}, \cite{2008MNRAS.388.1853M}, and \cite{2012A&A...542A..13B}.

\section{Results}\label{sec:res}
\subsection{Flux measurement and spectral indices}
\begin{figure*}
	\centering
	\includegraphics[height=50mm]{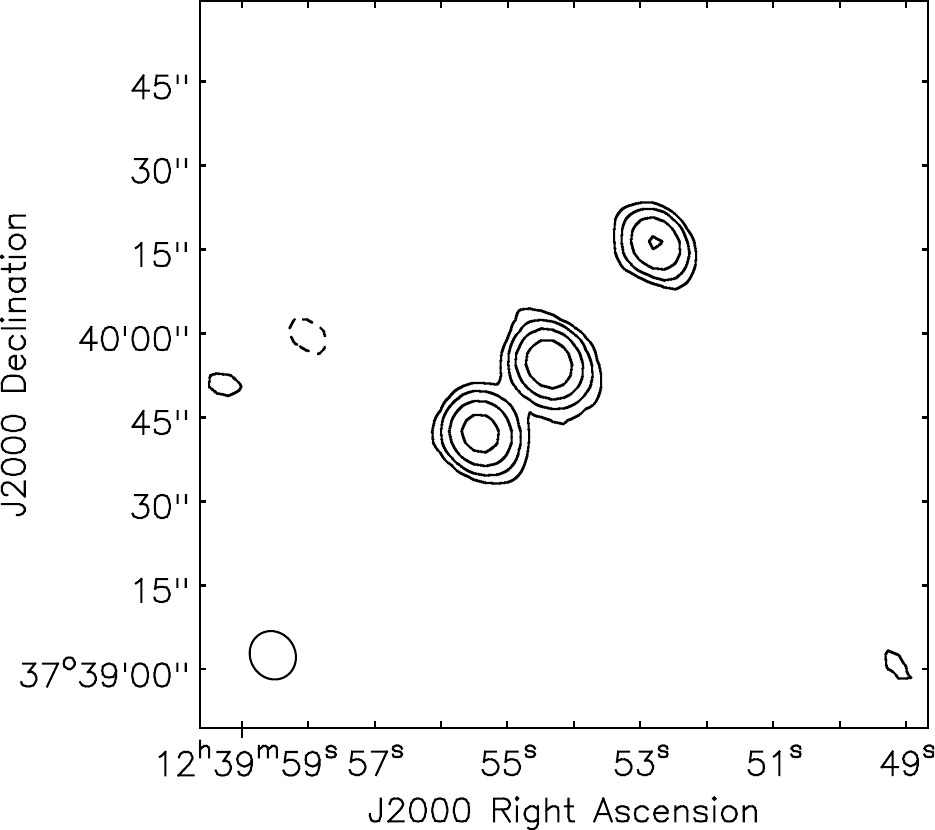}
    \hspace{3mm}
	\includegraphics[height=50mm]{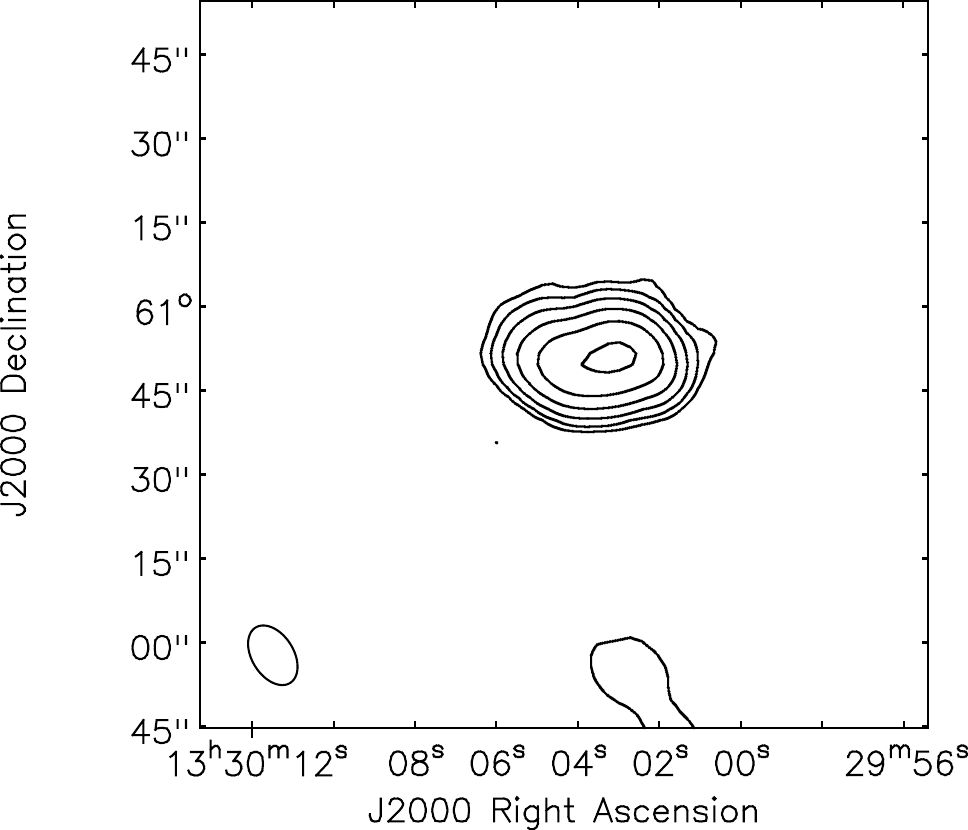}\\
    \vspace{3mm}
	\includegraphics[height=50mm]{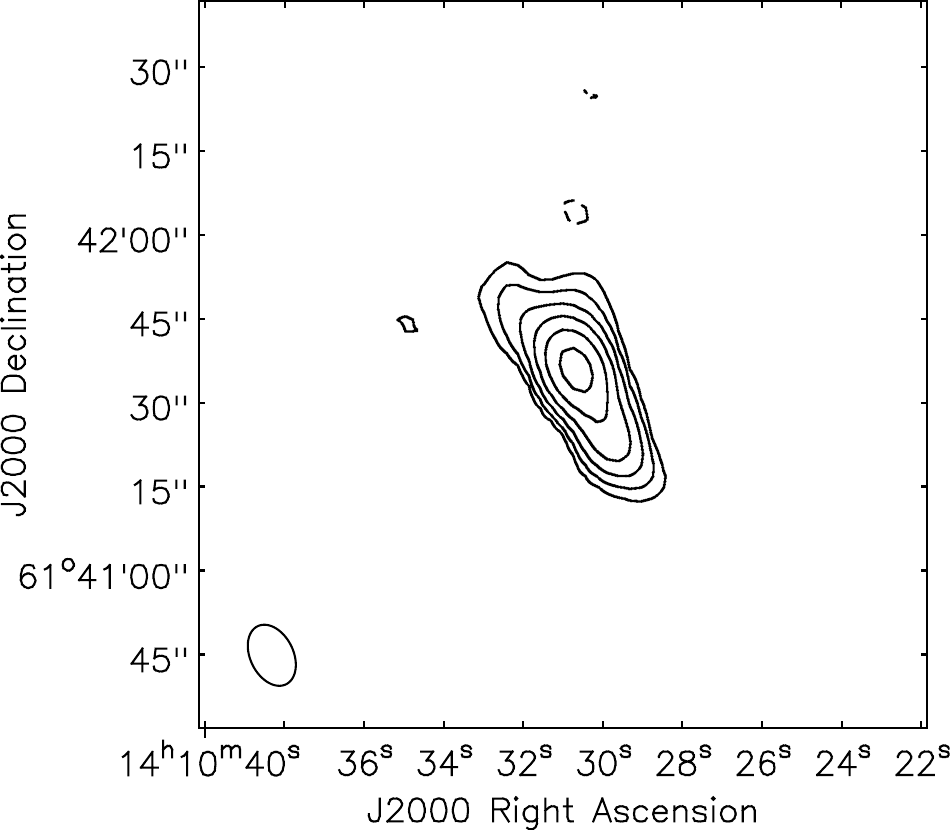} 
    \hspace{3mm}
	\includegraphics[height=50mm]{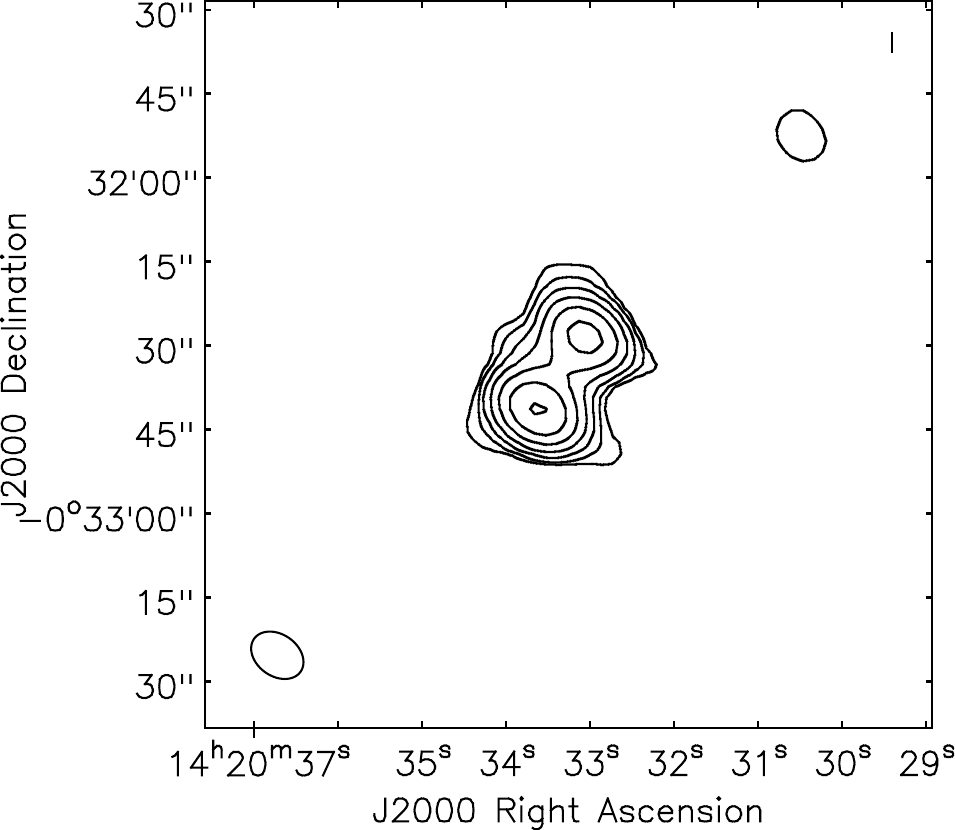} 
	\caption{
        Radio images at 322\,MHz for b20 (J123954$+$373954), b25 (J133004$+$605949), n24 (J141031$+$614136), and n27 (J142033$-$003233) from top left to bottom right, whose contour levels commence at 1.7, 1.0, 1.4, and 2.2\,mJy\,beam$^{-1}$ (equivalent to the 3$\sigma$ noise level of the image), respectively, and are then multiplied by a factor of 2.
 }
	\label{fig:GMRT_resolved}
\end{figure*}

\begin{figure*}
	\centering
	\includegraphics[width=0.9\textwidth]{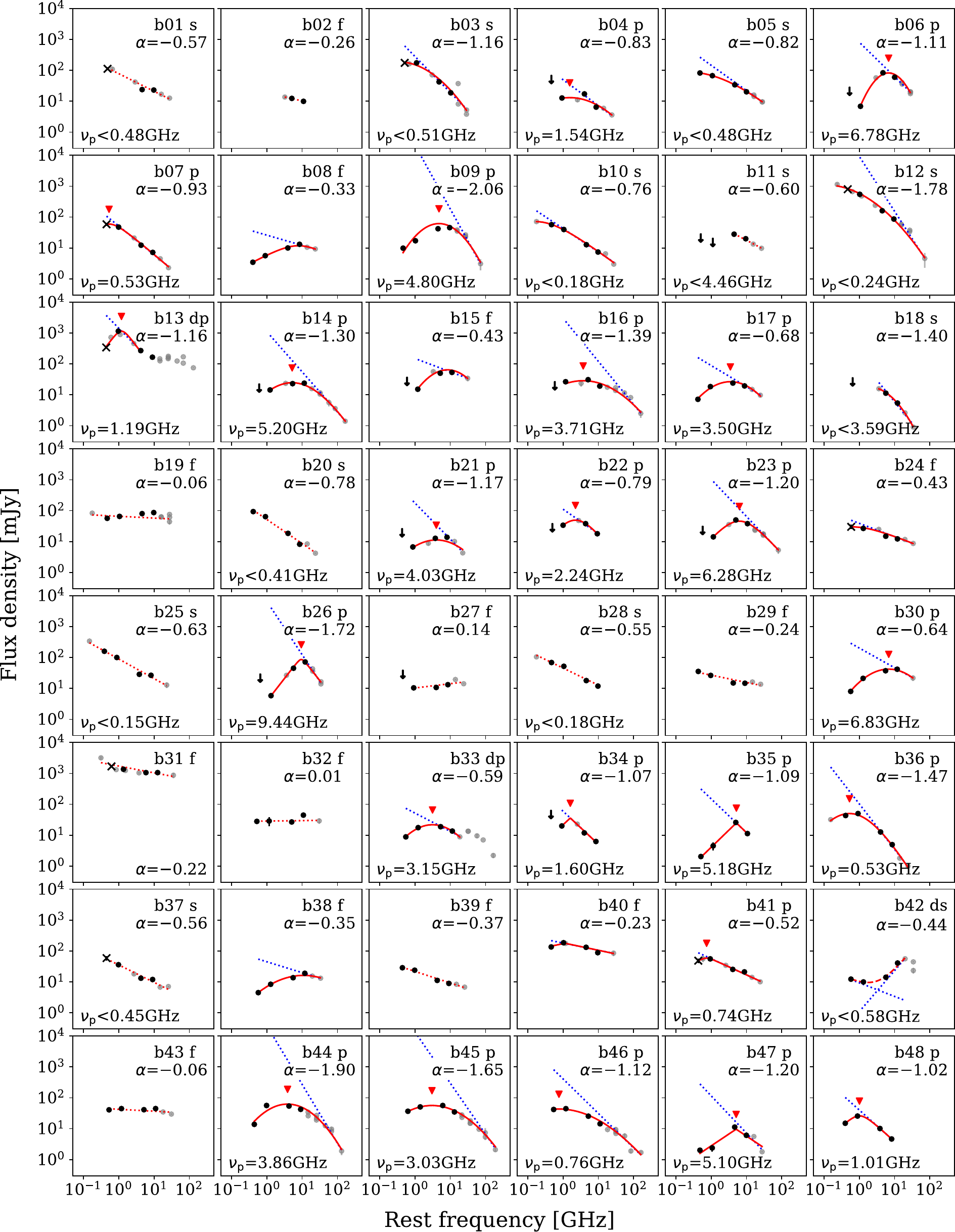}
	\caption{
        Radio spectra of the BAL sample. 
        Although error bars are illustrated for each measurement result, in most cases, they are substantially smaller than the data points and concealed by the markers. 
        Black circles represent data obtained by LoTSS \citep{2022A&A...659A...1S}, the FIRST survey \citep{1995ApJ...450..559B}, VLASS \citep{2020ApJS..250....8L}, and our GMRT observations, employed in the computation of spectral indices.
        Crosses and down arrows, respectively, indicate the flux measurement and its upper limit obtained from the TGSS \citep{2017A&A...598A..78I}.
        Gray circles show additional data only  utilised for estimating overall spectral shapes (see Section~\ref{sec:survey} for details).
        The results of fitting with a hyperbola, power law, and double power law are illustrated as red solid, dotted, and dashed lines, respectively.
        The blue dotted lines represent the power-law fitting for the data within the optically thin regime.
        The spectral indices, $\alpha$, given at each panel was obtained from the fit.
        The spectral classification is denoted to the right of the object name in each panel (see Section\,\ref{sec:shape} for more details).
        As for sources with a steep/peaked spectrum, the estimated peak frequency, $\nu_\mathrm{p}$, is denoted by a red triangle, and its value is provided in each panel.
    }
	\label{fig:spec_BAL}
\end{figure*}

\begin{figure*}
	\centering
	\includegraphics[width=0.9\textwidth]{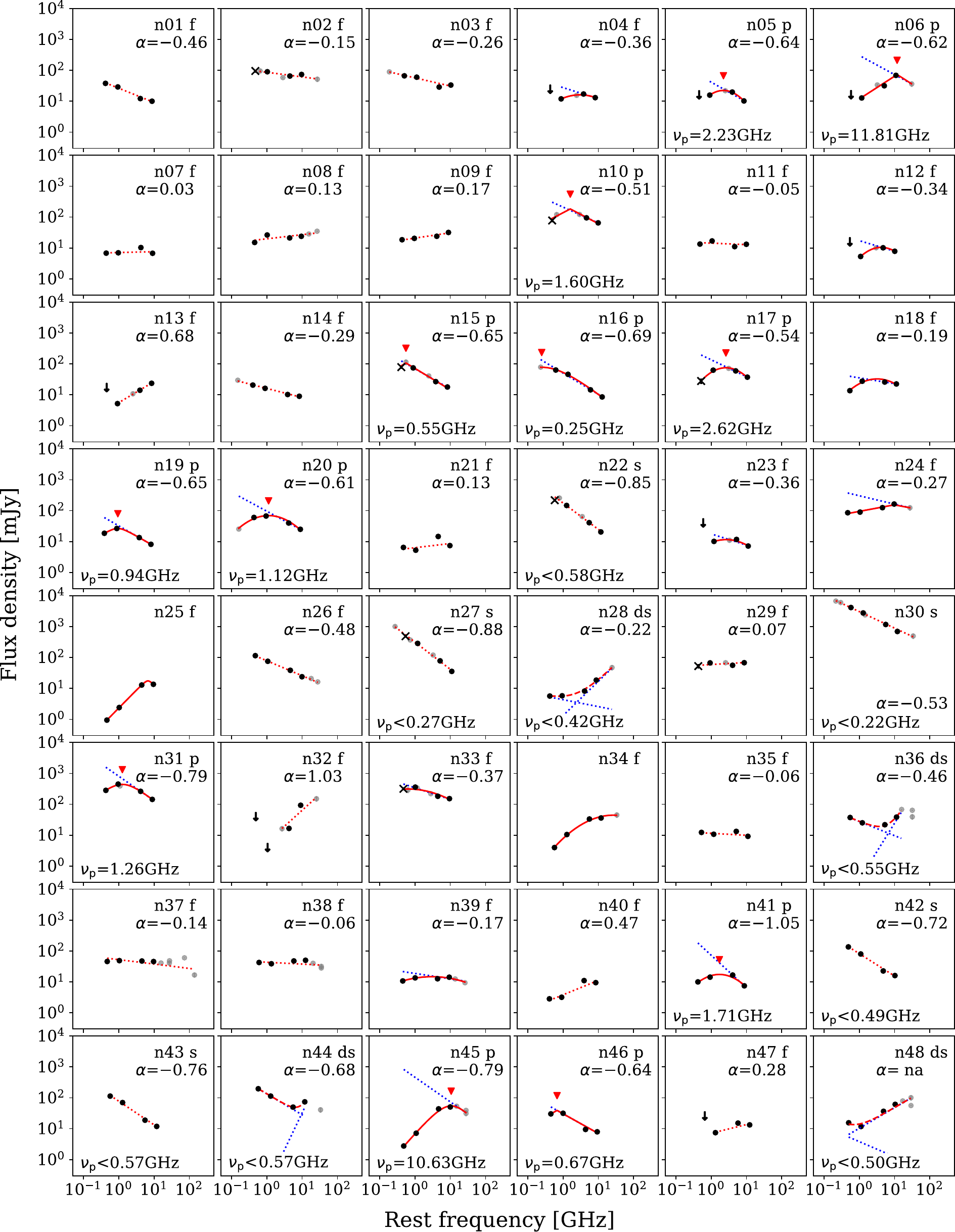}
	\caption{
        Radio spectra of the non-BAL sample. 
        Same as Fig.~\ref{fig:spec_BAL}.
        In the case of n48, because the fitting with the double power law did not converge, $\alpha=-0.8$ was set as the spectral index for its low-frequency component.
 }
	\label{fig:spec_nBAL}
\end{figure*}

We found that the images of all targets at all bands are consistent with those from the FIRST survey (Fig.~\ref{fig:FIRST_resolved}), and we did not detect any new components. 
Our GMRT observations have revealed that all targets, except those shown in Fig.~\ref{fig:GMRT_resolved}, are point sources. 
Tables~\ref{tbl:res-BAL} and \ref{tbl:res-nBAL} show the results of the GMRT observations and prior all-sky surveys, including the corresponding spectral indices and radio luminosities. 
Please see Figs.~\ref{fig:spec_BAL} and \ref{fig:spec_nBAL} for the radio spectra of the BAL and non-BAL samples, respectively.
As described in Section~\ref{sec:myobs}, RFI prevented us from measuring the flux density at 322\,MHz for three BAL quasars (b01, b02, and b18) and one non-BAL quasar (n10). 
In addition, we did not detect one source from each group at this frequency (b11 and n32), which was also undetected at 144\,MHz; 
thus, we could not establish constraints on their low-frequency spectral index between 144 and 322\,MHz. 
Table~\ref{tbl:object-w-note} summarises the objects requiring careful consideration in subsequent statistical analyses.

\input{tab_res_BAL.tex}

\input{tab_res_nBAL.tex}

\subsection{Spectral classification}\label{sec:shape}
To classify the overall spectral shape, 
we fitted the observed spectra with the following function:
\begin{eqnarray}
\log f_\nu = \log f_\mathrm{p} + b - \sqrt{b^2+c^2(\log\nu-\log\nu_\mathrm{p})^2},
\label{eq:analytic}
\end{eqnarray}
where 
$f_\mathrm{p}$ and $\nu_\mathrm{p}$ are the peak flux density and peak frequency, while $b$ and $c$ represent purely numerical values without conveying direct physical information \citep{2000A&A...363..887D,2005A&A...432...31T}.
For objects with spectra lacking significant curvature, we employed a simple power law model, whereas a double power law model was used for objects with convex downward spectra.
Figs.~\ref{fig:spec_BAL} and \ref{fig:spec_nBAL} also show fitting results.
If the estimated peak frequency was below the observed frequency range, the upper limit of $\nu_\mathrm{p}$ was set by the minimum observed frequency.

Based on the fitting results, we classified the radio spectra of the objects into steep/peaked or flat/inverted spectra. 
Following \cite{2000A&A...363..887D}, for sources fitted with a power law or hyperbolic function, we identified spectra with a spectral index, $\alpha$, smaller than $-0.5$ in the optically thin region above the peak frequency as steep or peaked spectra (denoted as 's' or 'p' in Figs.~\ref{fig:spec_BAL} and \ref{fig:spec_nBAL}, respectively).
In contrast, spectra with $\alpha>-0.5$ were classified as flat/inverted spectra (denoted as 'f' in Figs.~\ref{fig:spec_BAL} and \ref{fig:spec_nBAL}). 
If the number of data points in the optically thin regime was insufficient to confirm $\alpha<-0.5$, we classified them as flat/inverted spectra. 
Our classification was based on the low-frequency component for sources with two components in their spectra. 
This focus aligns with the main objective of this study, aiming to scrutinise low-frequency features that are less susceptible to relativistic effects and have not been explored in preceding studies conducted in the GHz regime.
These sources with two spectral components were categorised as sources with steep/peaked spectra (denoted as 'ds' or 'dp' in Figs.~\ref{fig:spec_BAL} and \ref{fig:spec_nBAL}, respectively), regardless of the value of the spectral index. 
Table~\ref{tbl:res-fit} presents the results of the spectral classification and estimated physical values.
Note that because the distinction between a steep and peaked spectrum is based on whether the peak frequency lies within the observed frequency range, the dissimilarity between these two classifications is a matter of degree.

\input{tab_fit_BAL}

\subsection{Comparison of the spectral class composition between the BAL and non-BAL samples}
Table~\ref{tbl:spec-class} presents the number of sources with steep/peaked spectra and sources with flat/inverted spectra in the BAL and non-BAL samples. 
The proportion of steep/peaked spectra is higher in the BAL sample compared to the non-BAL sample; 
the former contains $73\pm13$\,per cent of sources with steep/peaked spectra, whereas 
the latter contains $44\pm14$\,per cent, where errors are given by 95\,per cent confidential level. 
We conducted a chi-square test to examine the disparities in radio spectra between the BAL and non-BAL samples, resulting in a $p$-value of 0.0038, which represents the likelihood of obtaining a result equal to or more extreme than the observed result, assuming the null hypothesis that the two samples are derived from the same parent population.
The distinction between the two groups is considered significant at the 5\,per cent significance level.
Nevertheless, the obtained $p$-value still corresponds to $\sim2.9\sigma$. 
This trend remains even after excluding objects with spatially resolved source structures (b01, b20, b25, n24, and n27), candidates showing flux variations (b04 and n24), and objects with no absorption lines in \ion{C}{4} but absorption lines in \ion{Al}{3} (b02, n24), yielding the $p$-value of 0.0064.
Therefore, we have obtained marginal evidence indicating that BAL quasars harbour more sources with steep/peaked spectra than non-BAL quasars.

\begin{table}
\begin{minipage}{\linewidth}
\caption{Results of spectral classification} \label{tbl:spec-class}
\begin{tabular}{lccccccccccccc}
\hline							
                &	steep/peaked   &	flat/inverted      & total    \\\hline
BAL sample      &	      35 (31) 	&	13 (12)	&	48 (43) \\
non-BAL sample  &	      21 \phn(7)	&	27 \phn(9)  &	48 (16) \\[0.5em]
both samples    &         56 (38)   &   40 (21)  &   96 (59) \\
\hline									
\end{tabular}

\textit{Note.} The numbers within brackets denote the count of sources for which flux measurements at 8.4\,GHz are available.
\end{minipage}
\end{table}

Our samples include 43 BAL quasars and 16 non-BAL quasars with data above 8.4\,GHz.
During the spectral classification, the key criterion for identifying steep/peaked spectra is to detect high-frequency spectra with $\alpha<-0.5$.
Objects with intrinsically optically thin spectra at high frequencies but no data available at those frequencies can be classified as sources with flat/inverted spectra.
Hence, the fewer sources with high-frequency data in the non-BAL sample could introduce a bias towards a reduced count of sources with steep/peaked spectra compared to the BAL sample.
In order to evaluate this effect, Table~\ref{tbl:spec-class} also shows statistics confined to sources with flux density measurements at 8.4\,GHz.
As a result, 31 out of 43 BAL quasars with 8.4-GHz flux measurements were classified as sources with steep/peaked spectra ($72\pm13$\,per cent).
Of 16 non-BAL quasars with flux measurements at 8.4\,GHz, eight were classified as a sources with steep/peaked spectra ($44\pm24$\,per cent). 
The statistics for both sources with and without flux measurements at 8.4\,GHz do not differ significantly.
Therefore, the absence of high-frequency data does not affect our result.

\subsection{Comparison of the spectral properties between the BAL and non-BAL samples}
In the previous subsection, we described the difference in the compositional ratios of spectral types within the BAL and non-BAL samples. 
To investigate the origins of this difference, we analysed the spectral characteristics of both samples for each spectral type.
This study focused on the peak frequencies, spectral indices and radio luminosities at the rest frame.
In order to address the challenge posed by sources with only upper or lower limits on these physical values and to compare them between the BAL and non-BAL samples, we employed the Kaplan-Meier estimator, which is a non-parametric, maximum likelihood statistical estimator that estimates the cumulative distribution of the parent population \citep{kaplan58}. 
We applied the \texttt{KaplanMeierFitter} module from the Python package, \texttt{lifelines}.

\subsubsection{Sources with steep/peaked spectra}
\begin{figure}
	\centering
	\includegraphics[width=0.8\columnwidth]{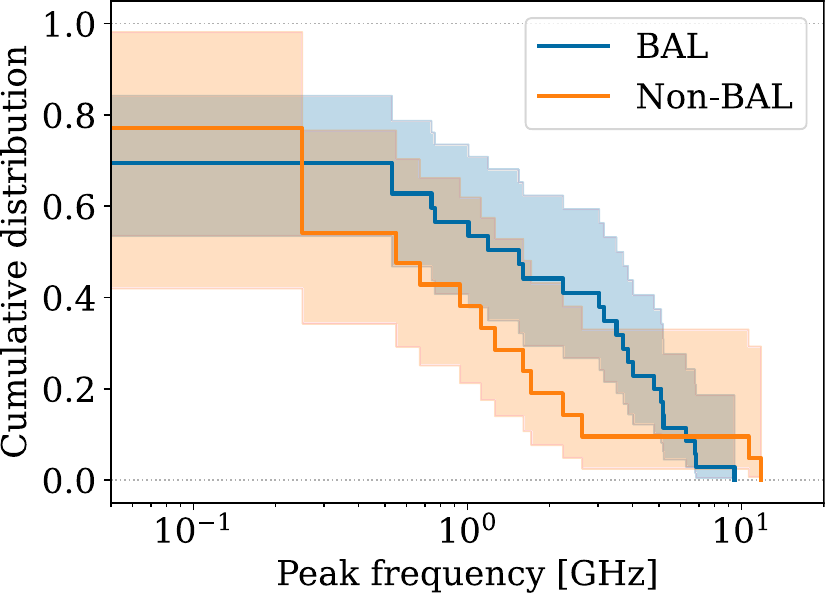}\\[3mm]
	\includegraphics[width=0.8\columnwidth]{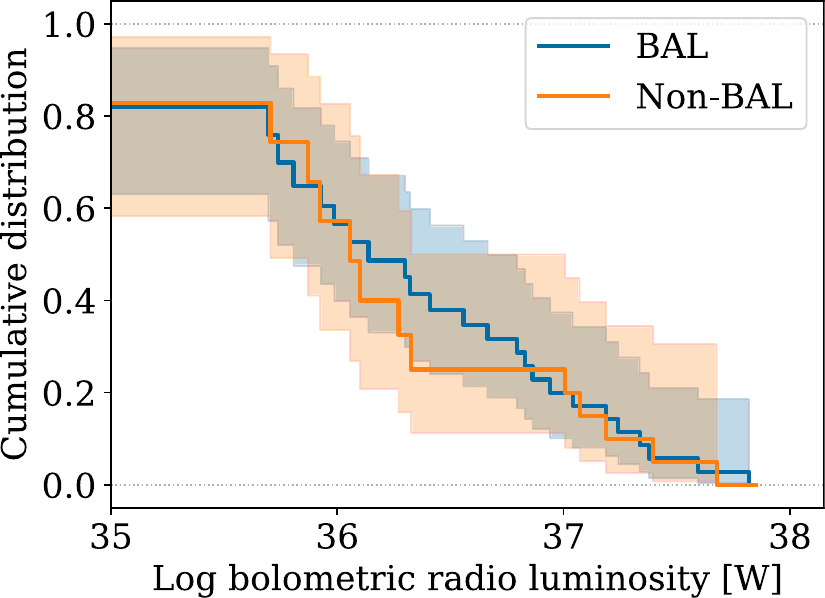}
	\caption{
    Cumulative distributions of peak frequency and bolometric luminosity of sources with steep/peaked spectra for the BAL (blue) and non-BAL (orange) samples using the Kaplan-Meier estimator with a 95\,per cent confidence interval. 
    (Top) peak frequency, $\nu_\mathrm{p}$. 
    (Bottom) bolometric luminosity, $L_\mathrm{bol}$.
    A log-rank test yields a $p$-value of 0.28 and 0.87 for $\nu_\mathrm{p}$ and $L_\mathrm{bol}$, respectively.
    These $p$-values indicate the probability of obtaining a result equal to or more extreme than the observed result, assuming the null hypothesis that the two samples are drawn from the same parent population.
    }
	\label{fig:KM-peak}
\end{figure}

We examined whether there were any distinctions in the spectral features between the BAL and non-BAL samples for sources with steep/peaked spectra.
Fig.~\ref{fig:KM-peak} represents comparisons of the peak frequencies, $\nu_\mathrm{p}$, and bolometric radio luminosities, $L_\mathrm{bol}$, derived from the model fitting to the spectra.
We approximated $L_\mathrm{bol} \simeq 4\pi D_{\mathrm L}^2 f_{\mathrm p}\nu_{\mathrm p}$ for the bolometric radio luminosity \citep[e.g.,][]{2006A&A...456...97F}, where $D_{\mathrm L}$ is luminosity distance.
For all steep spectral sources where only an upper limit of $\nu_\mathrm{p}$ was constrained, the spectrum around the lowest observed frequency was observed as flatter than the spectral index of $-1$ (see Tables~\ref{tbl:res-BAL} and \ref{tbl:res-nBAL}).
Assuming this trend continues even at lower frequencies, we can impose an upper limit on $L_\mathrm{bol}$.
In Fig.~\ref{fig:KM-peak}, we find that the BAL sample indicates a higher median peak frequency than the non-BAL sample, but a log-rank test on the distributions of the BAL and non-BAL samples for $\nu_\mathrm{p}$ and $L_\mathrm{bol}$ generates $p$-values of 0.28 and 0.87, respectively. 
These $p$-values indicate the probability of achieving a result equal to or more extreme than observed, assuming a null hypothesis that the two distributions originate from the same parent populations.
The statistical tests show no evidence of a distinction between the two groups at the 5\,per cent significance level.
Even after excluding objects with spatially resolved source structures (b01, b20, b25, and n27) and candidates showing flux variations (b04), these trends do not change, yielding $p$-values of 0.17 and 0.38 for the distributions of $\nu_\mathrm{p}$ and $L_\mathrm{bol}$, respectively.

\begin{figure*}
	\centering
	\includegraphics[width=0.32\textwidth]{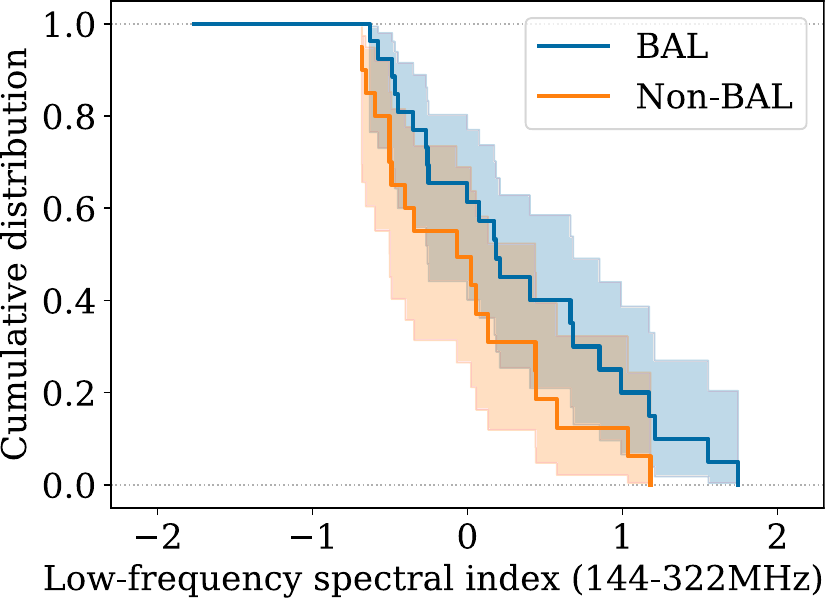}
    \hspace{0.01\textwidth}
	\includegraphics[width=0.32\textwidth]{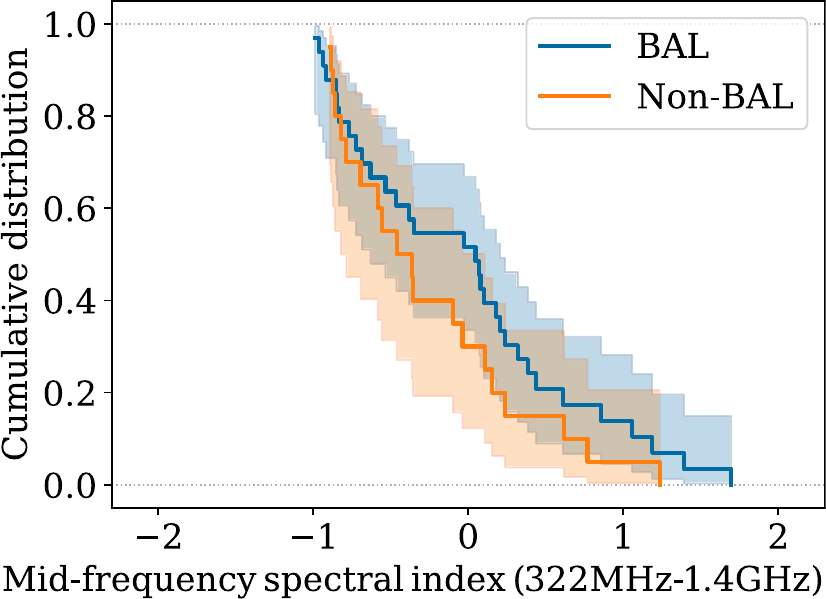}
    \hspace{0.01\textwidth}
	\includegraphics[width=0.32\textwidth]{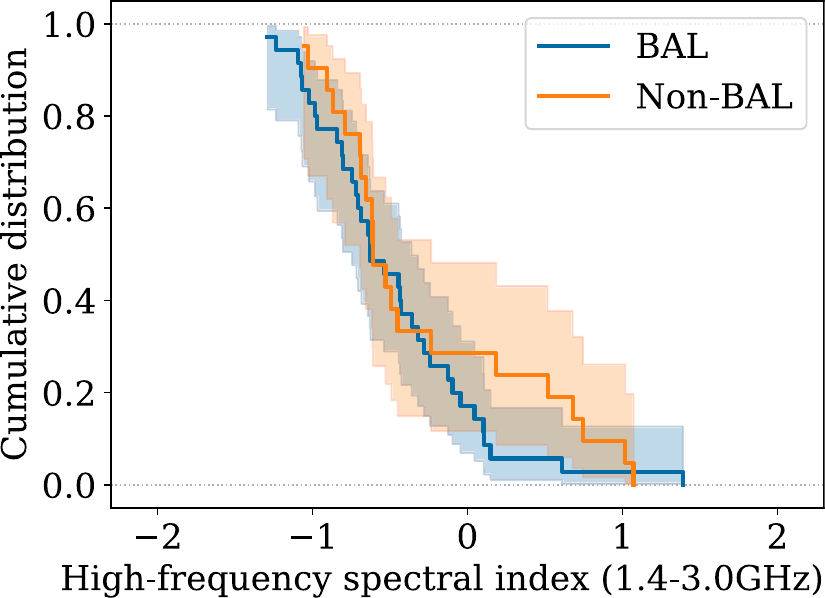}\\[1em]
	\includegraphics[width=0.32\textwidth]{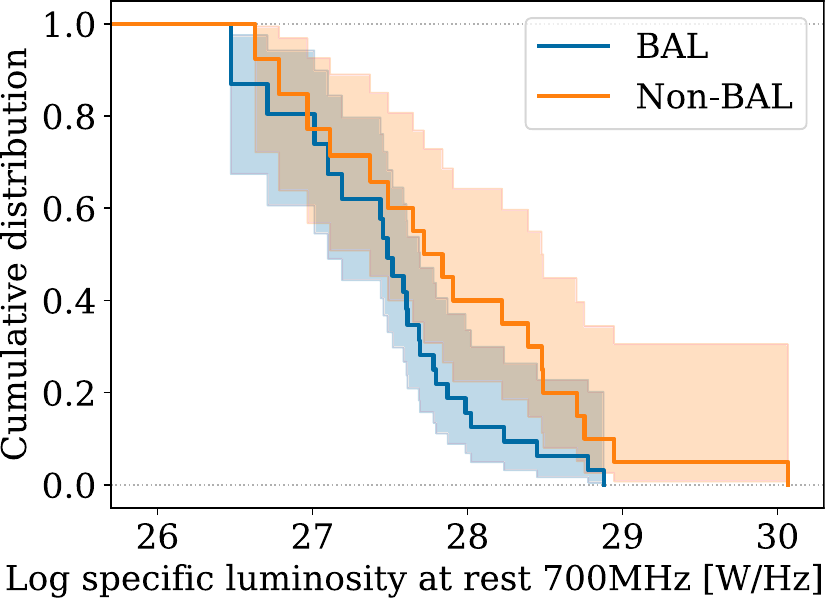}
    \hspace{0.01\textwidth}
	\includegraphics[width=0.32\textwidth]{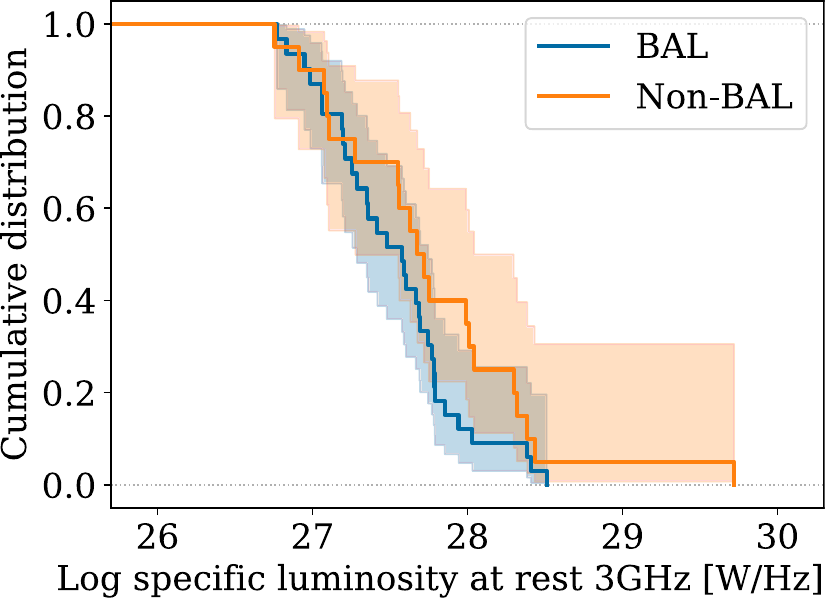}
    \hspace{0.01\textwidth}
	\includegraphics[width=0.32\textwidth]{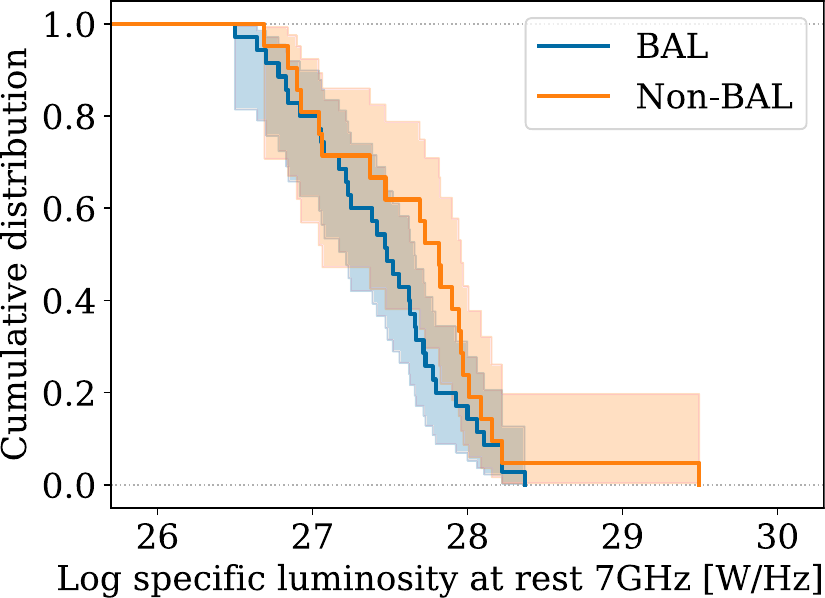}
	\caption{
    Cumulative distributions of spectral indices and specific luminosities for sources with steep/peaked spectra in the BAL (blue) and non-BAL (orange) samples using the Kaplan-Meier estimator with a 95\,per cent confidence interval. 
    (Top) 
    The distributions of 
    low-frequency spectral index, $\alpha_\textrm{low}$, 
    between 144 and 322\,MHz, 
    mid-frequency spectral index, $\alpha_\textrm{mid}$, 
    between 322\,MHz and 1.4\,GHz, and 
    high-frequency spectral index, $\alpha_\textrm{high}$, 
    between 1.4 and 3.0\,GHz 
    are displayed from left to right.
    A log-rank test produces a $p$-value of 0.066, 0.23, and 0.33 for $\alpha_\textrm{low}$, $\alpha_\textrm{mid}$, and $\alpha_\textrm{high}$, respectively.
    (bottom)
    The distributions of 
    rest-frame luminosity at 700\,MHz, $L_{0.7}$, interpolated by $\alpha_\textrm{low}$, 
    rest-frame specific luminosity at 3.0\,GHz, $L_{3}$, interpolated by $\alpha_\textrm{mid}$, and 
    rest-frame specific luminosity at 7.0\,GHz, $L_{7}$, interpolated by $\alpha_\textrm{high}$ 
    are displayed from left to right.
    A log-rank test produces a $p$-value of 0.19, 0.65, and 0.22 for $L_{0.7}$, $L_{3}$, and $L_{7}$, respectively.
    }
	\label{fig:KM-sp}
\end{figure*}

We also compared spectral indices and luminosities derived from observed flux densities between the two samples. 
Fig.~\ref{fig:KM-sp} displays the cumulative distributions for spectral indices at low, mid, and high frequencies (labelled as $\alpha_\textrm{low}$, $\alpha_\textrm{mid}$, and $\alpha_\textrm{high}$, respectively) that were obtained over the frequency ranges of 144–322\,MHz, 322\,MHz–1.4\,GHz, and 1.4–3.0\,GHz, respectively. 
These values are independent of the model applied to the flux measurement data.
Our analysis indicates that the BAL sample exhibits a higher proportion of objects with optically thick spectral indices at low frequencies than the non-BAL sample, which aligns with the observation of a higher median peak frequency, $\nu_\mathrm{p}$, in the BAL sample (Fig.~\ref{fig:KM-peak}).
Nevertheless, as in $\nu_\mathrm{p}$, a log-rank test on the distributions of the BAL and non-BAL samples for $\alpha_\textrm{low}$, $\alpha_\textrm{mid}$, and $\alpha_\textrm{high}$ results in $p$-values of 0.066, 0.23, and 0.33, respectively, providing limited evidence for the distinction between the two groups at the 5\,per cent significance level.
Even after excluding objects suffering from significant RFI in the 322-MHz data, spatially resolved sources, and candidates showing flux variations (b01, b04, b18, b20, b25, n10, and n27; Table~\ref{tbl:object-w-note}), these trends remain, yielding $p$-values of 0.052, 0.23, and 0.32 for the distributions of $\alpha_\textrm{low}$, $\alpha_\textrm{mid}$, and $\alpha_\textrm{high}$, respectively.

We can now compute the specific luminosity at low frequency without making assumptions based on the acquired spectral index with the newly obtained 322-MHz data. 
We selected our targets based on the redshift criterion of $1.68 \leq z \leq 4.73$, and the observed redshift range in this study spans $1.69 \leq z \leq 3.38$, resulting in observed rest frequencies of 387--631\,MHz, 866--1410\,MHz, 3.7--6.2\,GHz, and 8.0--13.2\,GHz for observations at 144\,MHz, 322\,MHz, 1.4\,GHz, and 3.0\,GHz, respectively. 
Therefore, by using the spectral indices obtained from the data, we calculated the rest-frame specific radio luminosities at 700\,MHz, 3.0\,GHz, and 7.0\,GHz through interpolation (labelled as $L_{0.7}$, $L_{3}$, and $L_{7}$, respectively).
Tables~\ref{tbl:res-BAL} and \ref{tbl:res-nBAL} present the computed values of $L_{0.7}$, $L_{3}$, and $L_{7}$, while the cumulative distributions of them obtained for both the BAL and non-BAL samples are depicted in Fig.~\ref{fig:KM-sp}.
Our analysis has not revealed any apparent differences between the plots of the two groups. 
In order to further evaluate this, we conducted a log-rank test on the distributions of the two samples for $L_{0.7}$, $L_{3}$, and $L_{7}$, which produces $p$-values of 0.19, 0.65, and 0.22, respectively, indicating a lack of evidence for a difference between the two samples.
These results persist even when considering only the compact, non-variable objects without RFI in the 322-MHz data, yielding $p$-values for $L_{0.7}$, $L_{3}$, and $L_{7}$ as 0.27, 0.89, and 0.51, respectively.

\begin{figure*}
	\centering
 	\includegraphics[width=0.32\textwidth]{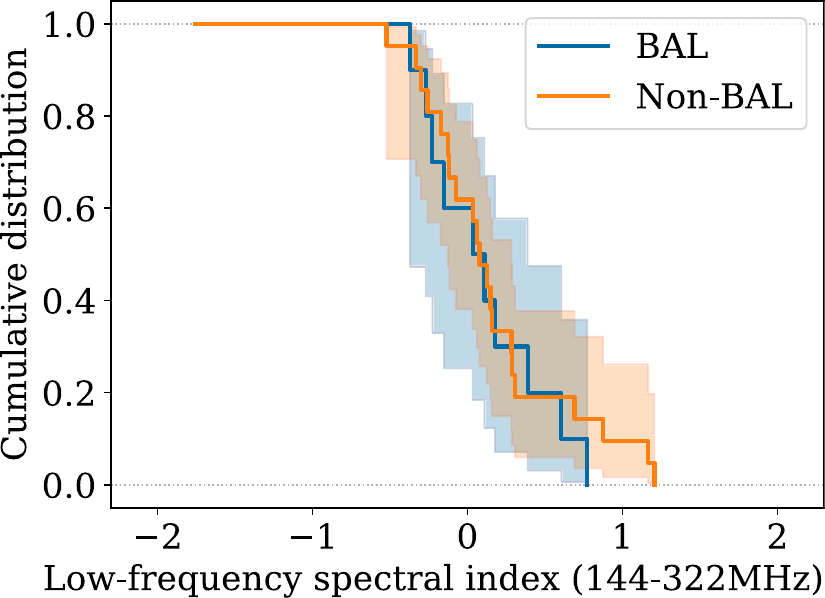}
    \hspace{0.01\textwidth}
	\includegraphics[width=0.32\textwidth]{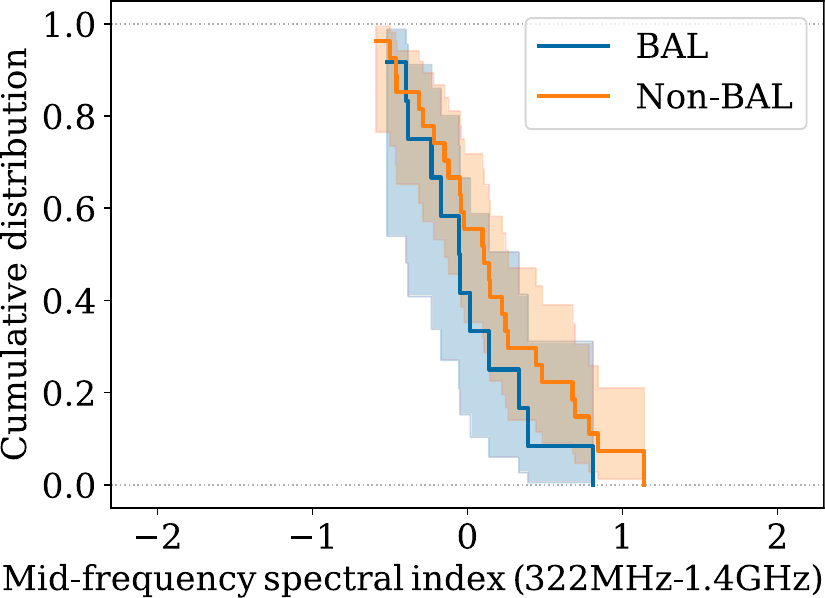}
    \hspace{0.01\textwidth}
	\includegraphics[width=0.32\textwidth]{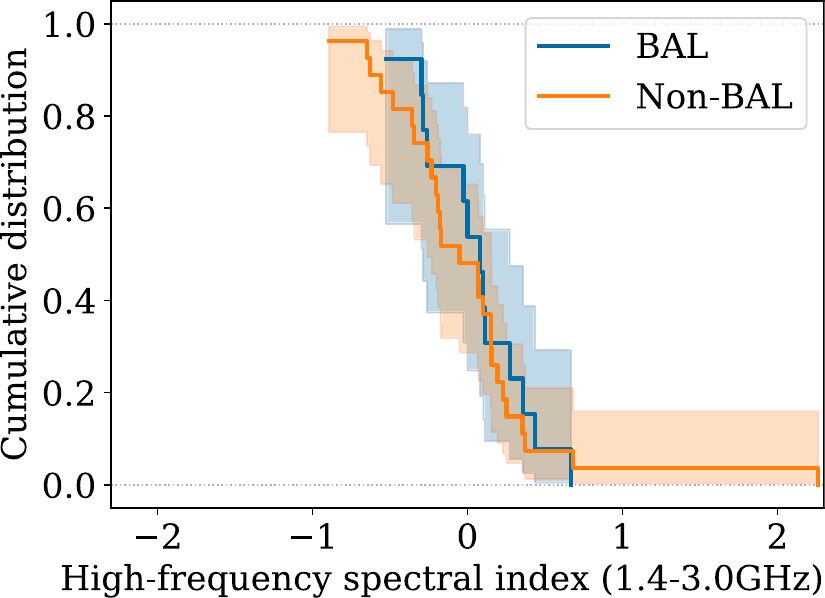}\\[1em]
	\includegraphics[width=0.32\textwidth]{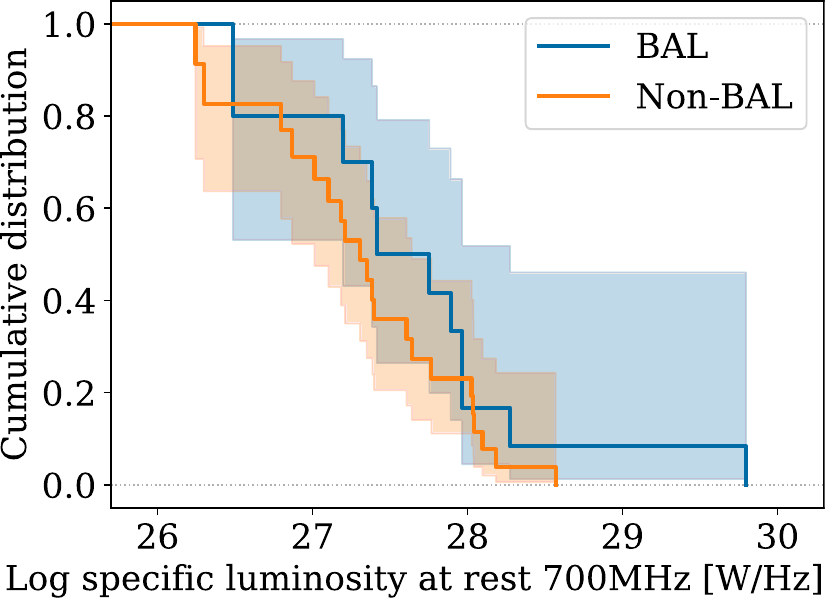}
    \hspace{0.01\textwidth}
	\includegraphics[width=0.32\textwidth]{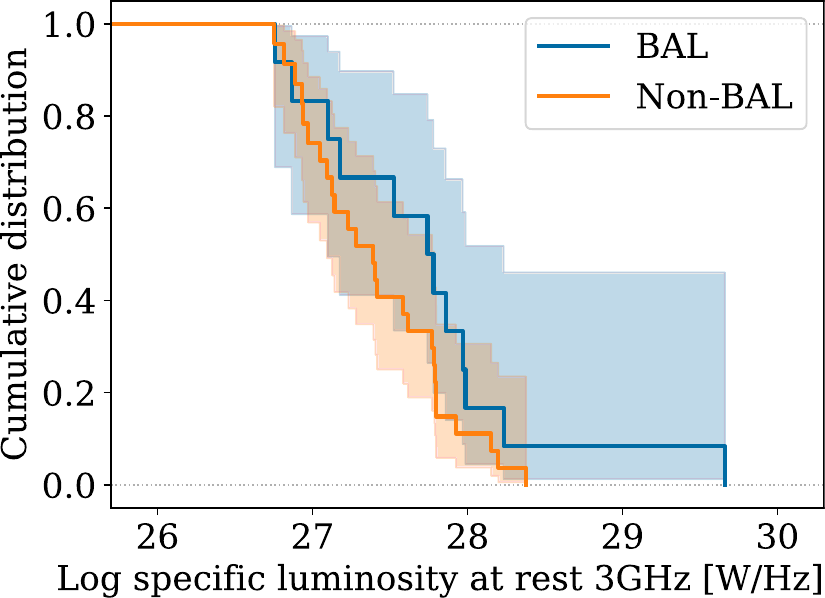}
    \hspace{0.01\textwidth}
	\includegraphics[width=0.32\textwidth]{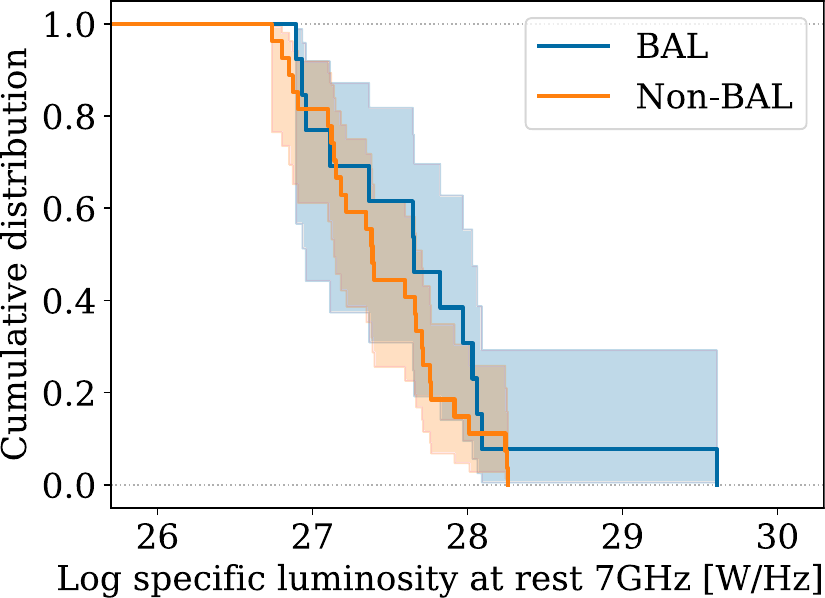}
    \caption{
    Cumulative distributions of spectral indices and specific luminosities for sources with flat/inverted spectra in the BAL (blue) and non-BAL (orange) samples using the Kaplan-Meier estimator with a 95\,per cent confidence interval. 
    (Top) 
    The distributions of 
    low-frequency spectral index, $\alpha_\textrm{low}$, 
    between 144 and 322\,MHz, 
    mid-frequency spectral index, $\alpha_\textrm{mid}$, 
    between 322\,MHz and 1.4\,GHz, and 
    high-frequency spectral index, $\alpha_\textrm{high}$, 
    between 1.4 and 3.0\,GHz 
    are displayed from left to right.
    A log-rank test produces a $p$-value of 0.62, 0.22, and 0.73 for $\alpha_\textrm{low}$, $\alpha_\textrm{mid}$, and $\alpha_\textrm{high}$, respectively.
    (bottom)
    The distributions of 
    rest-frame luminosity at 700\,MHz, $L_{0.7}$, interpolated by $\alpha_\textrm{low}$, 
    rest-frame specific luminosity at 3.0\,GHz, $L_{3}$, interpolated by $\alpha_\textrm{mid}$, and 
    rest-frame specific luminosity at 7.0\,GHz, $L_{7}$, interpolated by $\alpha_\textrm{high}$ 
    are displayed from left to right.
    A log-rank test produces a $p$-value of 0.27, 0.46, and 0.29 for $L_{0.7}$, $L_{3}$, and $L_{7}$, respectively.
	}
	\label{fig:KM-fi}
\end{figure*}

\subsubsection{Sources with flat/inverted spectra}
We also investigated the difference in spectral properties between the BAL and non-BAL samples for sources with flat/inverted spectra. 
Fig.~\ref{fig:KM-fi} displays the estimated cumulative distribution of the parent population for spectral indices and specific radio luminosities, respectively, employing the Kaplan-Meier estimator.

We conducted a log-rank test on the distributions of the BAL and non-BAL samples for $\alpha_\textrm{low}$, $\alpha_\textrm{mid}$, and $\alpha_\textrm{high}$, resulting in $p$-values of 0.62, 0.22, and 0.73, respectively.
We do not recognise a significant difference in spectral indices at the 5\,per cent significance level.
Even after excluding objects with no absorption lines in \ion{C}{4} but absorption lines in \ion{Al}{3} (b02 and n24), these trends remain, yielding $p$-values of 0.59, 0.23, and 0.55 for the distributions of $\alpha_\textrm{low}$, $\alpha_\textrm{mid}$, and $\alpha_\textrm{high}$, respectively.
Additionally, a log-rank test on the distributions for $L_{0.7}$, $L_{3}$, and $L_{7}$ results in $p$-values of 0.27, 0.46, and 0.29, respectively, indicating a lack of evidence for a difference between the two samples.
These results remain unchanged even when excluding the objects with no absorption lines in \ion{C}{4} but absorption lines in \ion{Al}{3}, yielding $p$-values for $L_{0.7}$, $L_{3}$, and $L_{7}$ as 0.22, 0.40, and 0.15, respectively.
Consequently, our analyses show no significant differences in the spectral indices and luminosities of sources with flat/inverted spectra between the two groups. 


\section{Discussion}\label{sec:discuss}

Our investigation, involving data down to the 150-MHz band, corresponding to 387–631\,MHz in the rest frame, has revealed a higher prevalence of a steep/peaked spectrum in BAL quasars than their non-BAL counterparts within flux-limited samples.
Because all objects in our samples, except those shown in Figs.~\ref{fig:FIRST_resolved} and \ref{fig:GMRT_resolved}, are point sources at the resolution of the FIRST survey ($\sim$5\,arcsecs), most of the sources with steep/peaked spectra identified in this study can be categorised into GPS/CSS sources \citep{1998PASP..110..493O}, which are candidates for lobe-dominated young radio sources observed from a large viewing angle with respect to the jet axis.
On the other hand, point sources displaying a flat/inverted spectrum are interpreted as core-dominated radio sources observed from a pole-on perspective \citep{1982MNRAS.198..673B,1982MNRAS.200.1067O}. 
Because no significant differences between the two groups have been observed in the spectral index or luminosity of radio sources with each spectral class, even at low frequencies, we do not expect any difference in their intrinsic jet activity between the two samples.
These outcomes favour the orientation scheme for BAL quasars \citep[e.g.,][]{2020MNRAS.492.4553R}, which is suggested by a steeper spectral index between 4.9 and 8.4\,GHz for BAL quasars than non-BAL quasars \citep{2011ApJ...743...71D,2012ApJ...752....6D}.

Note that our BAL and non-BAL samples are constructed to have similar luminosity distributions based on flux density at 1.4\,GHz (Fig.~\ref{fig:sample}).
On the contrary, whole non-BAL quasars are known to be brighter than our current samples \citep[e.g.,][]{2008ApJ...687..859S}.
These bright quasars in the radio may experience more intense relativistic beaming effects in their jets \citep[cf.][]{1982MNRAS.200.1067O}.
It would be crucial to investigate the fraction of flat/inverted spectra in those bright non-BAL quasars to test the orientation scheme further.


While we have found a difference in the composition of the spectral classes between the two samples, the BAL and non-BAL samples also include a considerable number of sources with flat/inverted spectra and sources with steep/peaked spectra, respectively. 
Therefore, a simple orientation scheme may not comprehensively account for the distinction between BAL and non-BAL quasars \citep[cf.][]{2017MNRAS.467.2571M,2018MNRAS.479.4153Y}. 
Similar issues have been raised regarding radio morphology \citep{2015A&A...579A.109K,2015MNRAS.450.1123C,2022MNRAS.511.4946N} and radio flux variations \citep{2006ApJ...639..716Z,2007ApJ...661L.139G,2011ApJ...743...71D,2017PASJ...69...77C}, pointing towards the possibility of polar BAL quasars \citep{2005ApJ...623L.101P,2006MNRAS.372L..58B,2008ApJ...676L..97W,2013ApJ...773L..10R,2013MNRAS.436.3321B}.

As a complement to the orientation scheme, an intermittent BAL phase associated with periods of restarting jet activity was proposed to explain multiple spectral components at low frequencies \citep{2015A&A...582A...9B}.
However, we have found that the number of sources
with multiple spectral components is comparable between the two samples; three and four sources are identified as having a spectral classification of ’dp’ or ’ds’ in our BAL and non-BAL samples, respectively.
Therefore, it is challenging to infer factors beyond the orientation scheme from the results of low-frequency observations in this study.

Previous VLBI observations have revealed that a subset of GPS/CSS sources associated with radio-loud BAL quasars demonstrates characteristics of lobe-dominated young radio sources in terms of little flux variability, a two-sided structure on a VLBI scale, and unpolarised radio emission in the core region \citep[e.g.,][]{2013A&A...554A..94B,2013PASJ...65...57D,2013ApJ...772....4H}. 
Nevertheless, flat spectral radio quasars can exhibit a peaked spectrum during a flare  \citep{2001AJ....121.1306T,2005A&A...435..839T,2007A&A...469..451T}, and GPS sources may involve core-jet objects resembling blazars \citep{1997A&A...325..943S,2001A&A...377..377S}.
Therefore, VLBI observations are crucial to ascertain whether the identified sources with steep/peaked spectra genuinely represent lobe-dominated compact radio sources.
We are currently conducting a VLBI program on the samples in this study, the results of which will be presented in a forthcoming paper.

\section{Conclusions}\label{sec:conclusion}
We have performed new GMRT observations on 48 radio-loud BAL quasars, chosen from \cite{2009ApJ...692..758G}, and a corresponding number of radio-loud non-BAL quasars. 
Using these observational data in conjunction with previous flux measurements, we have studied low-frequency spectra of the two samples. 
Our key findings are as follows:

\begin{enumerate}
\item 
    In our BAL sample, $73 \pm 13$\,per cent of objects exhibit a steep/peaked spectrum, in contrast to $44 \pm 14$\,per cent for the non-BAL sample, with errors reported at the 95\,per cent confidence level.
    While the statistical evidence is limited to a $\sim2.9\sigma$ level, the BAL sample reveals a higher incidence of steep/peaked spectra than the comparison sample.
\item 
    No significant difference between the two quasar groups is found in peak frequencies ($\nu_\mathrm{p}$) and bolometric radio luminosities ($\nu_\mathrm{bol}$) of the sources with steep/peaked spectra.
    Additionally, no significant distinctions between the two samples are observed in spectral indices ($\alpha_\mathrm{low}$, $\alpha_\mathrm{mid}$, and $\alpha_\mathrm{high}$) and specific radio luminosities ($L_{0.7}$, $L_{3}$, and $L_{7}$) of sources with steep/peaked spectra and sources with flat/inverted spectra.
\end{enumerate}

The orientation scenario of BAL quasars can account for these spectral features, attributing the higher prevalence of steep/peaked spectra in the BAL sample to a relatively edge-on perspective.
Nevertheless, more than the orientation scheme is also needed to explain all the observational results, including the presence of flat/inverted spectra in the BAL sample. 
Additionally, our analyses do not rule out the possibility that peaked spectral sources in both samples are contaminated by blazars in a flaring state.
In this regard, acquiring more densely packed data along the time direction and implementing high-resolution observations with VLBI would provide substantial advantages.

\section*{Acknowledgements}
We express our gratitude to the anonymous reviewer for their invaluable feedback, which has  contributed to enhancing the quality of the paper.
We thank the staff of the GMRT that made these observations possible. 
The GMRT is run by the National Centre for Radio Astrophysics of the Tata Institute of Fundamental Research.
This work was partially supported by a Grant-in-Aid for Scientific Research (C; 21540250, A.D.),  Global COE Program "the Physical Sciences Frontier" from the Japanese Ministry of Education, Culture, Sports, Science and Technology.
We used the NASA/IPAC Extragalactic Database (NED), which the Jet Propulsion Laboratory, California Institute of Technology operate. 
This research used the VizieR catalogue access tool, CDS, Strasbourg, France. 
This research also made use of Montage, which is funded by the National Science Foundation under Grant Number ACI-1440620. 
It was previously funded by the National Aeronautics and Space Administration's Earth Science Technology Office, Computation Technologies Project, under Cooperative Agreement Number NCC5-626 between NASA and the California Institute of Technology.

\section*{Data Availability}
The raw data underlying this article are available via the GMRT online archive facility:
\href{https://naps.ncra.tifr.res.in/goa/data/search}{https://naps.ncra.tifr.res.in/goa/data/search}.
All data analysis packages used in this work are publicly available.



\bibliographystyle{mnras}
\bibliography{BAL} 






\bsp	
\label{lastpage}
\end{document}

%% file: tab_sample_BAL.tex
\begin{table*}
\begin{minipage}{\textwidth}
\renewcommand{\tabcolsep}{2mm}
\caption{BAL sample for GMRT Observation.}
\label{tbl:BALsample}
\begin{tabular}{cccccccccccccc}
\hline																											
ID	&	Source Name	&	$z$	&	$M_{\mathrm i}$	&		BI	&		BI$_0$	&	Compo-	&	R.A.	&	Dec.	&		$f_{1.4}^{\textrm{FIRST}}$	&		$f_{1.4}^{\textrm{NVSS}}$	&	$\log L_{\rm 5}$	\\
	&	(SDSS~J)	&		&	(mag)	&		(\kms)	&		(\kms)	&	nent	&	(J2000)	&	(J2000)	&		(mJy)	&		(mJy)	&	(\WHz)	\\
(1)	&	(2)	&	(3)	&	(4)	&		(5)	&		(6)	&	(7)	&	(8)	&	(9)	&		(10)	&		(11)	&	(12)	\\\hline
b01	&	100109.42$+$114608.8	&	2.278 	&	$-$27.1 	&	\phn	716.1 	&		3030.5 	&	total	&	…	&	…	&	\phn\phn	23.8 	&	\phn\phn	23.6 	&	27.5 	\\
	&	…	&	…	&	…	&		…	&		…	&	A	&	10 01 09.431	&	$+$11 46 08.71	&	\phn\phn	13.7 	&		…	&	27.2 	\\
	&	…	&	…	&	…	&		…	&		…	&	B	&	10 01 09.392	&	$+$11 45 55.13	&	\phn\phn	10.1 	&		…	&	27.1 	\\
b02	&	100424.88$+$122922.2	&	2.640 	&	$-$25.8 	&		4534.0$^{\dag}$	 	&		4534.0$^{\dag}$	 	&	…	&	10 04 24.872	&	$+$12 29 22.39	&	\phn\phn	12.3 	&	\phn\phn	11.8 	&	27.4 	\\
b03	&	104059.79$+$055524.4	&	2.443 	&	$-$26.6 	&		4120.3 	&		4120.3 	&	…	&	10 40 59.802	&	$+$05 55 24.78	&	\phn\phn	42.2 	&	\phn\phn	39.1 	&	27.8 	\\
b04	&	104452.41$+$104005.9	&	1.882 	&	$-$27.9 	&		2463.3 	&		3283.9 	&	…	&	10 44 52.417	&	$+$10 40 05.91	&	\phn\phn	17.2 	&	\phn\phn	13.1 	&	27.0 	\\
b05	&	105416.51$+$512326.0	&	2.341 	&	$-$27.2 	&	\phn	337.5 	&		1056.2 	&	…	&	10 54 16.528	&	$+$51 23 26.21	&	\phn\phn	33.9 	&	\phn\phn	35.6 	&	27.7 	\\
b06	&	110206.66$+$112104.9	&	2.351 	&	$-$27.7 	&	\phn\phn\phn	0.0 	&	\phn\phn	75.3 	&	…	&	11 02 06.657	&	$+$11 21 04.74	&	\phn\phn	83.1 	&	\phn\phn	80.8 	&	28.1 	\\
b07	&	110531.42$+$151215.9	&	2.063 	&	$-$27.0 	&		4556.4 	&		4556.4 	&	…	&	11 05 31.490	&	$+$15 12 17.55	&	\phn\phn	12.3 	&	\phn\phn	11.4 	&	27.0 	\\
b08	&	112241.46$+$303534.9	&	1.810 	&	$-$28.6 	&	\phn\phn\phn	0.1 	&	\phn\phn\phn	0.1 	&	…	&	11 22 41.467	&	$+$30 35 34.88	&	\phn\phn	10.0 	&	\phn\phn\phn	9.5 	&	26.7 	\\
b09	&	112938.47$+$440325.0	&	2.212 	&	$-$27.8 	&	\phn\phn\phn	0.0 	&	\phn	806.5 	&	…	&	11 29 38.475	&	$+$44 03 25.05	&	\phn\phn	42.0 	&	\phn\phn	42.3 	&	27.7 	\\
b10	&	113152.56$+$584510.2	&	2.262 	&	$-$26.8 	&	\phn\phn\phn	2.5 	&	\phn	925.3 	&	…	&	11 31 52.546	&	$+$58 45 10.26	&	\phn\phn	12.8 	&	\phn\phn	13.2 	&	27.2 	\\
b11	&	113445.83$+$431858.0	&	2.184 	&	$-$26.9 	&		4443.4 	&		4443.4 	&	…	&	11 34 45.834	&	$+$43 18 57.87	&	\phn\phn	28.0 	&	\phn\phn	25.2 	&	27.5 	\\
b12	&	115901.75$+$065619.0	&	2.191 	&	$-$26.6 	&		1675.4 	&		1675.4 	&	…	&	11 59 01.713	&	$+$06 56 18.91	&	\phn	160.1 	&	\phn	158.3 	&	28.2 	\\
b13	&	115944.82$+$011206.9	&	2.000 	&	$-$28.4 	&	\phn\phn\phn	0.0 	&	\phn	937.9 	&	…	&	11 59 44.832	&	$+$01 12 06.87	&	\phn	268.5 	&	\phn	275.6 	&	28.3 	\\
b14	&	121323.94$+$010414.7	&	2.829 	&	$-$26.0 	&		1551.0 	&		1551.0 	&	…	&	12 13 23.934	&	$+$01 04 14.82	&	\phn\phn	22.9 	&	\phn\phn	27.5 	&	27.8 	\\
b15	&	121539.66$+$090607.4	&	2.723 	&	$-$28.0 	&	\phn	116.3 	&	\phn	116.3 	&	…	&	12 15 39.670	&	$+$09 06 07.48	&	\phn\phn	49.8 	&	\phn\phn	49.1 	&	28.1 	\\
b16	&	122848.21$-$010414.5	&	2.655 	&	$-$28.1 	&	\phn\phn	17.1 	&	\phn\phn	17.1 	&	…	&	12 28 48.190	&	$-$01 04 14.27	&	\phn\phn	30.8 	&	\phn\phn	29.1 	&	27.8 	\\
b17	&	123411.73$+$615832.6	&	1.946 	&	$-$26.9 	&		4907.9 	&		4907.9 	&	…	&	12 34 11.747	&	$+$61 58 32.40	&	\phn\phn	24.0 	&	\phn\phn	22.7 	&	27.2 	\\
b18	&	123511.59$+$073330.7	&	3.035 	&	$-$27.9 	&	\phn\phn\phn	0.0 	&	\phn\phn	93.5 	&	…	&	12 35 11.609	&	$+$07 33 30.80	&	\phn\phn	11.3 	&	\phn\phn	12.9 	&	27.6 	\\
b19	&	123717.44$+$470807.0	&	2.271 	&	$-$27.3 	&	\phn	868.7 	&	\phn	868.7 	&	…	&	12 37 17.438	&	$+$47 08 07.05	&	\phn\phn	80.2 	&	\phn\phn	90.9 	&	28.0 	\\
b20	&	123954.15$+$373954.5	&	1.841 	&	$-$25.5 	&	\phn	959.3 	&	\phn	959.3 	&	total	&	…	&	…	&	\phn\phn	18.5 	&	\phn\phn	16.2 	&	27.0 	\\
	&	…	&	…	&	…	&		…	&		…	&	A	&	12 39 54.149	&	$+$37 39 54.45	&	\phn\phn	10.7 	&		…	&	26.8 	\\
	&	…	&	…	&	…	&		…	&		…	&	B	&	12 39 55.205	&	$+$37 39 42.20	&	\phn\phn\phn	4.3 	&		…	&	26.4 	\\
	&	…	&	…	&	…	&		…	&		…	&	C	&	12 39 52.533	&	$+$37 40 15.77	&	\phn\phn\phn	3.5 	&	   …	&	26.3 	\\
b21	&	125243.85$+$005320.1	&	1.693 	&	$-$27.1 	&	\phn\phn	94.6 	&	\phn\phn	94.6 	&	…	&	12 52 43.860	&	$+$00 53 20.17	&	\phn\phn	12.8 	&	\phn\phn	15.4 	&	26.7 	\\
b22	&	130332.05$+$014407.4	&	2.109 	&	$-$26.2 	&	\phn\phn\phn	0.0 	&	\phn	136.0 	&	…	&	13 03 32.061	&	$+$01 44 07.41	&	\phn\phn	38.0 	&	\phn\phn	34.7 	&	27.5 	\\
b23	&	130448.06$+$130416.5	&	2.568 	&	$-$27.3 	&	\phn\phn\phn	4.8 	&	\phn\phn\phn	4.8 	&	…	&	13 04 48.050	&	$+$13 04 16.67	&	\phn\phn	50.0 	&	\phn\phn	56.5 	&	28.0 	\\
b24	&	130756.73$+$042215.5	&	3.022 	&	$-$28.7 	&	\phn	879.0 	&	\phn	879.0 	&	…	&	13 07 56.723	&	$+$04 22 15.50	&	\phn\phn	14.9 	&	\phn\phn	15.8 	&	27.7 	\\
b25	&	133004.34$+$605949.7	&	1.734 	&	$-$26.2 	&	\phn\phn	72.1 	&	\phn\phn	72.1 	&	total	&	…	&	…	&	\phn\phn	28.5 	&	\phn\phn	38.3 	&	27.1 	\\
	&	…	&	…	&	…	&		…	&		…	&	A	&	13 30 04.335	&	$+$60 59 49.93	&	\phn\phn	18.1 	&		…	&	26.9 	\\
	&	…	&	…	&	…	&		…	&		…	&	B	&	13 30 03.250	&	$+$60 59 52.71	&	\phn\phn	10.4 	&		…	&	26.7 	\\
b26	&	133701.39$-$024630.3	&	3.064 	&	$-$27.9 	&	\phn\phn\phn	0.0 	&	\phn\phn\phn	2.3 	&	…	&	13 37 01.399	&	$-$02 46 29.89	&	\phn\phn	44.8 	&	\phn\phn	45.1 	&	28.2 	\\
b27	&	135550.30$+$361627.6	&	1.855 	&	$-$26.3 	&	\phn\phn\phn	0.0 	&	\phn	604.4 	&	…	&	13 55 50.294	&	$+$36 16 27.55	&	\phn\phn	10.7 	&	\phn\phn	15.3 	&	26.8 	\\
b28	&	135910.45$+$563617.4	&	2.248 	&	$-$27.9 	&	\phn\phn	56.2 	&	\phn	156.9 	&	…	&	13 59 10.410	&	$+$56 36 17.01	&	\phn\phn	17.9 	&	\phn\phn	16.8 	&	27.3 	\\
b29	&	135910.77$+$400218.6	&	2.013 	&	$-$26.8 	&	\phn	432.9 	&	\phn	432.9 	&	…	&	13 59 10.774	&	$+$40 02 18.66	&	\phn\phn	15.0 	&	\phn\phn	14.0 	&	27.1 	\\
b30	&	140126.15$+$520834.6	&	2.972 	&	$-$28.0 	&	\phn\phn	80.3 	&	\phn\phn	80.3 	&	…	&	14 01 26.163	&	$+$52 08 34.63	&	\phn\phn	37.1 	&	\phn\phn	30.4 	&	28.1 	\\
b31	&	140501.12$+$041535.7	&	3.215 	&	$-$27.1 	&	\phn\phn	12.5 	&	\phn\phn	12.5 	&	…	&	14 05 01.114	&	$+$04 15 35.87	&		1055.9 	&	\phn	933.0 	&	29.7 	\\
b32	&	141313.63$+$411637.8	&	2.616 	&	$-$28.4 	&		1053.3 	&		1053.3 	&	…	&	14 13 13.622	&	$+$41 16 37.91	&	\phn\phn	26.7 	&	\phn\phn	28.7 	&	27.7 	\\
b33	&	141334.38$+$421201.7	&	2.817 	&	$-$27.9 	&	\phn\phn\phn	0.0 	&		1446.9 	&	…	&	14 13 34.404	&	$+$42 12 01.76	&	\phn\phn	18.7 	&	\phn\phn	16.8 	&	27.7 	\\
b34	&	141437.99$+$045537.4	&	1.851 	&	$-$26.1 	&		2015.1 	&		2201.2 	&	…	&	14 14 37.977	&	$+$04 55 37.22	&	\phn\phn	11.8 	&	\phn\phn	14.4 	&	26.8 	\\
b35	&	141736.05$+$372825.9	&	2.554 	&	$-$26.1 	&	\phn\phn\phn	0.0 	&	\phn	177.3 	&	…	&	14 17 36.045	&	$+$37 28 26.04	&	\phn\phn	25.8 	&	\phn\phn	23.5 	&	27.7 	\\
b36	&	143340.35$+$512019.3	&	1.863 	&	$-$26.9 	&		1085.7 	&		1085.7 	&	…	&	14 33 40.387	&	$+$51 20 19.70	&	\phn\phn	12.6 	&	\phn\phn	13.0 	&	26.9 	\\
b37	&	144434.80$+$003305.3	&	2.036 	&	$-$26.5 	&	\phn\phn\phn	0.0 	&	\phn	163.0 	&	…	&	14 44 34.816	&	$+$00 33 05.49	&	\phn\phn	13.2 	&	\phn\phn	10.5 	&	27.0 	\\
b38	&	145910.13$+$425213.2	&	2.967 	&	$-$28.2 	&	\phn\phn	87.1 	&	\phn\phn	87.1 	&	…	&	14 59 10.134	&	$+$42 52 13.20	&	\phn\phn	13.7 	&	\phn\phn	15.0 	&	27.6 	\\
b39	&	150332.93$+$440120.6	&	2.050 	&	$-$27.0 	&		1405.1 	&		1405.1 	&	…	&	15 03 32.948	&	$+$44 01 20.63	&	\phn\phn	11.2 	&	\phn\phn	10.8 	&	27.0 	\\
b40	&	150823.71$+$334700.7	&	2.208 	&	$-$27.6 	&	\phn	663.5 	&	\phn	663.5 	&	…	&	15 08 23.730	&	$+$33 47 00.63	&	\phn	132.0 	&	\phn	131.3 	&	28.2 	\\
b41	&	151630.30$-$005625.5	&	1.921 	&	$-$27.0 	&	\phn\phn\phn	0.0 	&	\phn	517.0 	&	…	&	15 16 30.350	&	$-$00 56 24.67	&	\phn\phn	25.5 	&	\phn\phn	25.8 	&	27.2 	\\
b42	&	153729.54$+$583224.6	&	3.059 	&	$-$26.8 	&	\phn\phn\phn	0.0 	&		1432.0 	&	…	&	15 37 29.553	&	$+$58 32 24.79	&	\phn\phn	14.2 	&	\phn\phn	14.5 	&	27.7 	\\
b43	&	155429.40$+$300118.9	&	2.693 	&	$-$28.4 	&	\phn	574.8 	&	\phn	574.8 	&	…	&	15 54 29.403	&	$+$30 01 19.04	&	\phn\phn	41.2 	&	\phn\phn	40.5 	&	28.0 	\\
b44	&	160354.15$+$300208.6	&	2.030 	&	$-$27.7 	&	\phn\phn\phn	0.0 	&	\phn	480.0 	&	…	&	16 03 54.159	&	$+$30 02 08.88	&	\phn\phn	54.2 	&	\phn\phn	54.1 	&	27.6 	\\
b45	&	162453.47$+$375806.6	&	3.381 	&	$-$28.4 	&	\phn	900.5 	&	\phn	900.5 	&	…	&	16 24 53.470	&	$+$37 58 06.65	&	\phn\phn	56.4 	&	\phn\phn	55.6 	&	28.5 	\\
b46	&	162559.90$+$485817.5	&	2.724 	&	$-$28.4 	&		3447.9 	&		3447.9 	&	…	&	16 25 59.905	&	$+$48 58 17.50	&	\phn\phn	25.5 	&	\phn\phn	26.0 	&	27.8 	\\
b47	&	162656.74$+$295328.0	&	2.312 	&	$-$26.9 	&	\phn	283.4 	&	\phn	283.4 	&	…	&	16 26 56.741	&	$+$29 53 28.02	&	\phn\phn	11.3 	&	\phn\phn	13.4 	&	27.2 	\\
b48	&	165543.24$+$394519.9	&	1.753 	&	$-$27.2 	&		3635.3 	&		3635.3 	&	…	&	16 55 43.235	&	$+$39 45 19.91	&	\phn\phn	10.2 	&	\phn\phn\phn	9.2 	&	26.7 	\\
\hline																
\end{tabular}

\textit{Note.} Columns are as follows:
(1) Object ID;
(2) SDSS source name; 
(3),\,(4) Redshift and $i$-band absolute magnitude from the SDSS quasars catalog DR5 \citep{2007AJ....134..102S}; 
(5),\,(6) balnicity index, BI, and modified balnicity index, BI$_0$, of \ion{C}{4} absorption line, respectively \citep{2009ApJ...692..758G}.
BI and BI$_0$ of b02 indicated by $^{\dag}$ were measured based on the estimated redshift of $z=4.66$. See text for details;
(7) Name of each component if multiple radio source is associated;
(8),\,(9) Radio positions;
(10) Flux density at 1.4\,GHz from the FIRST survey.
Reference for the columns (7)--(9) is \citet{1995ApJ...450..559B};
(11) Flux density at 1.4\,GHz from the NVSS \citep{1998AJ....115.1693C};
(12) Log specific luminosity at rest 5\,GHz calculated by flux density at 1.4\,GHz assuming spectral index of $\alpha=-0.7$.
For quasars associated with multiple radio sources, we provide information on each source.
\end{minipage}
\end{table*}

%% file: tab_sample_nBAL.tex
\begin{table*}
\begin{minipage}{150mm}
\caption{Non-BAL sample for GMRT Observation.}
\label{tbl:nBALsample}
\begin{tabular}{cccccccccccccc}
\hline																					
ID	&	Source Name	&	$z$	&	$M_{\mathrm i}$	&	Compo-	&	R.A.	&	Dec.	&		$f_{1.4}^{\textrm{FIRST}}$	&		$f_{1.4}^{\textrm{NVSS}}$	&	$\log L_{\rm 5}$	\\
	&	(SDSS~J)	&		&	(mag)	&	nent	&	(J2000)	&	(J2000)	&		(mJy)	&		(mJy)	&	(\WHz)	\\
(1)	&	(2)	&	(3)	&	(4)	&	(7)	&	(5)	&	(6)	&		(7)	&		(8)	&	(9)	\\\hline
n01	&	103808.94$+$464249.1	&	1.924 	&	$-$26.9 	&	…	&	10 38 08.950	&	$+$46 42 49.46	&	\phn\phn	12.2 	&	\phn\phn	12.4 	&	26.9 	\\
n02	&	111048.93$+$045608.0	&	2.208 	&	$-$26.8 	&	…	&	11 10 48.963	&	$+$04 56 07.13	&	\phn\phn	65.2 	&	\phn\phn	74.2 	&	27.9 	\\
n03	&	111336.10$+$494034.7	&	2.466 	&	$-$27.6 	&	…	&	11 13 36.087	&	$+$49 40 34.63	&	\phn\phn	28.7 	&	\phn\phn	27.4 	&	27.7 	\\
n04	&	111434.01$+$041434.0	&	1.719 	&	$-$26.0 	&	…	&	11 14 34.018	&	$+$04 14 33.97	&	\phn\phn	17.1 	&	\phn\phn	18.7 	&	26.9 	\\
n05	&	112854.24$+$035341.4	&	1.829 	&	$-$26.9 	&	…	&	11 28 54.258	&	$+$03 53 41.05	&	\phn\phn	19.8 	&	\phn\phn	20.3 	&	27.0 	\\
n06	&	113017.37$+$073212.9	&	2.647 	&	$-$28.9 	&	…	&	11 30 17.371	&	$+$07 32 13.09	&	\phn\phn	31.6 	&	\phn\phn	27.6 	&	27.8 	\\
n07	&	113716.36$+$371046.4	&	2.027 	&	$-$27.3 	&	…	&	11 37 16.365	&	$+$37 10 46.67	&	\phn\phn	10.3 	&	\phn\phn\phn	9.5 	&	26.9 	\\
n08	&	113854.52$+$394553.6	&	2.159 	&	$-$27.2 	&	…	&	11 38 54.533	&	$+$39 45 53.62	&	\phn\phn	21.2 	&	\phn\phn	19.7 	&	27.3 	\\
n09	&	115534.50$+$575156.4	&	1.967 	&	$-$26.4 	&	…	&	11 55 34.509	&	$+$57 51 56.49	&	\phn\phn	23.9 	&	\phn\phn	24.2 	&	27.2 	\\
n10	&	121911.23$-$004345.5	&	2.293 	&	$-$27.7 	&	…	&	12 19 11.247	&	$-$00 43 45.42	&	\phn\phn	94.5 	&	\phn	104.5 	&	28.1 	\\
n11	&	123215.09$+$554049.4	&	2.307 	&	$-$26.3 	&	…	&	12 32 15.115	&	$+$55 40 49.53	&	\phn\phn	11.1 	&	\phn\phn	32.2 	&	27.2 	\\
n12	&	123545.38$-$033610.9	&	2.375 	&	$-$26.6 	&	…	&	12 35 45.391	&	$-$03 36 10.79	&	\phn\phn	10.2 	&	\phn\phn	11.3 	&	27.2 	\\
n13	&	123856.09$-$005930.8	&	1.844 	&	$-$27.3 	&	…	&	12 38 56.101	&	$-$00 59 30.89	&	\phn\phn	14.0 	&	\phn\phn	15.5 	&	26.9 	\\
n14	&	124409.64$+$554823.4	&	1.768 	&	$-$25.9 	&	…	&	12 44 09.639	&	$+$55 48 23.49	&	\phn\phn	10.2 	&	\phn\phn	10.5 	&	26.7 	\\
n15	&	125321.59$-$032315.7	&	1.771 	&	$-$26.6 	&	…	&	12 53 21.602	&	$-$03 23 15.96	&	\phn\phn	26.7 	&	\phn\phn	30.0 	&	27.1 	\\
n16	&	131003.35$+$535348.2	&	3.278 	&	$-$27.2 	&	…	&	13 10 03.352	&	$+$53 53 48.16	&	\phn\phn	14.5 	&	\phn\phn	13.1 	&	27.8 	\\
n17	&	131926.27$+$143439.9	&	2.541 	&	$-$27.6 	&	…	&	13 19 26.285	&	$+$14 34 39.88	&	\phn\phn	59.3 	&	\phn\phn	60.0 	&	28.0 	\\
n18	&	133754.41$+$451239.1	&	2.758 	&	$-$28.3 	&	…	&	13 37 54.428	&	$+$45 12 39.32	&	\phn\phn	25.8 	&	\phn\phn	29.7 	&	27.8 	\\
n19	&	134253.64$+$390223.6	&	1.723 	&	$-$26.2 	&	…	&	13 42 53.653	&	$+$39 02 23.68	&	\phn\phn	13.4 	&	\phn\phn	13.4 	&	26.8 	\\
n20	&	134303.25$+$502832.0	&	1.962 	&	$-$26.6 	&	…	&	13 43 03.252	&	$+$50 28 32.16	&	\phn\phn	39.9 	&	\phn\phn	40.9 	&	27.5 	\\
n21	&	134520.40$+$324112.5	&	2.255 	&	$-$27.7 	&	…	&	13 45 20.421	&	$+$32 41 12.59	&	\phn\phn	14.6 	&	\phn\phn	14.9 	&	27.2 	\\
n22	&	140637.60$+$141530.0	&	2.926 	&	$-$27.3 	&	…	&	14 06 37.619	&	$+$14 15 30.19	&	\phn\phn	40.9 	&	\phn\phn	41.8 	&	28.1 	\\
n23	&	140909.74$+$071226.1	&	2.734 	&	$-$27.1 	&	…	&	14 09 09.748	&	$+$07 12 26.26	&	\phn\phn	11.7 	&	\phn\phn	10.5 	&	27.4 	\\
n24	&	141031.00$+$614136.9	&	2.246 	&	$-$27.0 	&	total	&	…	&	…	&	\phn	125.0 	&	\phn\phn	98.2 	&	28.2 	\\
	&	…	&	…	&	…	&	A	&	14 10 30.992	&	$+$61 41 36.89	&	\phn	118.4 	&		…	&	28.1 	\\
	&	…	&	…	&	…	&	B	&	14 10 30.113	&	$+$61 41 25.51	&	\phn\phn\phn	4.9 	&		…	&	26.8 	\\
	&	…	&	…	&	…	&	C	&	14 10 32.179	&	$+$61 41 48.03	&	\phn\phn\phn	1.8 	&		…	&	26.3 	\\
n25	&	141846.20$+$482308.2	&	2.193 	&	$-$26.2 	&	…	&	14 18 46.211	&	$+$48 23 08.23	&	\phn\phn	12.8 	&	\phn\phn	11.1 	&	27.1 	\\
n26	&	142009.33$+$392738.5	&	2.295 	&	$-$26.9 	&	…	&	14 20 09.377	&	$+$39 27 38.77	&	\phn\phn	38.4 	&	\phn\phn	38.1 	&	27.7 	\\
n27	&	142033.25$-$003233.3	&	2.682 	&	$-$27.3 	&	total	&	…	&	…	&	\phn\phn	77.6 	&	\phn\phn	77.0 	&	28.2 	\\
	&	…	&	…	&	…	&	A	&	14 20 33.238	&	$-$00 32 33.54	&	\phn\phn	16.4 	&		…	&	27.6 	\\
	&	…	&	…	&	…	&	B	&	14 20 33.569	&	$-$00 32 40.42	&	\phn\phn	38.5 	&		…	&	27.9 	\\
	&	…	&	…	&	…	&	C	&	14 20 32.982	&	$-$00 32 27.06	&	\phn\phn	22.7 	&		…	&	27.7 	\\
n28	&	142326.05$+$325220.3	&	1.905 	&	$-$29.3 	&	…	&	14 23 26.073	&	$+$32 52 20.34	&	\phn\phn\phn	8.2 	&	\phn\phn\phn	8.1 	&	26.7 	\\
n29	&	142352.38$+$031125.8	&	1.884 	&	$-$26.9 	&	…	&	14 23 52.396	&	$+$03 11 26.08	&	\phn\phn	55.8 	&	\phn\phn	52.8 	&	27.5 	\\
n30	&	142921.87$+$540611.2	&	3.013 	&	$-$26.5 	&	…	&	14 29 21.891	&	$+$54 06 10.95	&		1165.4 	&		1028.3 	&	29.6 	\\
n31	&	143243.29$+$410327.9	&	1.970 	&	$-$27.8 	&	…	&	14 32 43.322	&	$+$41 03 28.04	&	\phn	261.7 	&	\phn	261.9 	&	28.3 	\\
n32	&	143708.18$+$040534.3	&	2.025 	&	$-$27.5 	&	…	&	14 37 08.201	&	$+$04 05 34.65	&	\phn\phn	16.5 	&	\phn\phn	15.3 	&	27.1 	\\
n33	&	143737.06$+$094847.7	&	2.162 	&	$-$26.8 	&	…	&	14 37 37.042	&	$+$09 48 47.81	&	\phn	181.2 	&	\phn	191.5 	&	28.3 	\\
n34	&	144752.46$+$582420.3	&	2.983 	&	$-$28.1 	&	…	&	14 47 52.441	&	$+$58 24 20.39	&	\phn\phn	33.2 	&	\phn\phn	32.8 	&	28.0 	\\
n35	&	145627.72$+$414944.2	&	2.668 	&	$-$26.5 	&	…	&	14 56 27.731	&	$+$41 49 44.43	&	\phn\phn	13.2 	&	\phn\phn	12.1 	&	27.5 	\\
n36	&	145924.24$+$340113.1	&	2.790 	&	$-$28.3 	&	…	&	14 59 24.251	&	$+$34 01 13.11	&	\phn\phn	21.7 	&	\phn\phn	22.2 	&	27.7 	\\
n37	&	151258.36$+$352533.2	&	2.236 	&	$-$26.7 	&	…	&	15 12 58.373	&	$+$35 25 33.32	&	\phn\phn	47.5 	&	\phn\phn	53.6 	&	27.7 	\\
n38	&	152314.87$+$381402.0	&	3.159 	&	$-$27.9 	&	…	&	15 23 14.870	&	$+$38 14 02.13	&	\phn\phn	47.7 	&	\phn\phn	51.9 	&	28.3 	\\
n39	&	154644.24$+$311711.3	&	2.122 	&	$-$26.3 	&	…	&	15 46 44.250	&	$+$31 17 11.47	&	\phn\phn	12.7 	&	\phn\phn	14.0 	&	27.1 	\\
n40	&	154935.74$+$314338.2	&	1.815 	&	$-$25.8 	&	…	&	15 49 35.725	&	$+$31 43 38.38	&	\phn\phn	11.0 	&	\phn\phn\phn	9.9 	&	26.8 	\\
n41	&	155816.63$+$502953.7	&	1.900 	&	$-$26.5 	&	…	&	15 58 16.632	&	$+$50 29 53.77	&	\phn\phn	16.7 	&	\phn\phn	16.9 	&	27.0 	\\
n42	&	160911.26$+$374635.7	&	2.412 	&	$-$26.4 	&	…	&	16 09 11.287	&	$+$37 46 36.42	&	\phn\phn	22.7 	&	\phn\phn	25.9 	&	27.5 	\\
n43	&	161920.20$+$375502.7	&	2.966 	&	$-$26.7 	&	…	&	16 19 19.962	&	$+$37 55 02.77	&	\phn\phn	18.9 	&	\phn\phn	19.9 	&	27.8 	\\
n44	&	162004.73$+$351554.6	&	2.960 	&	$-$27.6 	&	…	&	16 20 04.690	&	$+$35 15 54.93	&	\phn\phn	50.0 	&	\phn\phn	57.3 	&	28.2 	\\
n45	&	165137.53$+$400219.0	&	2.343 	&	$-$28.5 	&	…	&	16 51 37.565	&	$+$40 02 18.71	&	\phn\phn	43.9 	&	\phn\phn	43.1 	&	27.8 	\\
n46	&	165508.72$+$373244.6	&	2.092 	&	$-$26.6 	&	…	&	16 55 08.724	&	$+$37 32 44.77	&	\phn\phn\phn	9.5 	&	\phn\phn\phn	8.8 	&	26.9 	\\
n47	&	074927.90$+$415242.3	&	3.111 	&	$-$28.9 	&	…	&	07 49 27.911	&	$+$41 52 42.20	&	\phn\phn	15.2 	&	\phn\phn	15.5 	&	27.8 	\\
n48	&	095537.94$+$333503.9	&	2.477 	&	$-$28.6 	&	…	&	09 55 37.936	&	$+$33 35 03.97	&	\phn\phn	36.6 	&	\phn\phn	33.0 	&	27.8 	\\
\hline																					

\end{tabular}
\textit{Note.} Columns are as follows:
(1) Object ID;
(2) SDSS source name; 
(3),\,(4) Redshift and $i$-band absolute magnitude from SDSS DR5 \citep{2007ApJS..172..634A}; 
(5) Name of each component if multiple radio source is associated;
(6),\,(7) Radio positions;
(8) Flux density at 1.4\,GHz from the FIRST survey.
Reference for the columns (5)--(7) is \citet{1995ApJ...450..559B};
(9) Flux density at 1.4\,GHz from the NVSS \citep{1998AJ....115.1693C}.
(10) Log specific luminosity at rest 5\,GHz calculated by flux density at 1.4\,GHz assuming spectral index of $\alpha=-0.7$.
For quasars associated with multiple radio sources, we provide information on each source.
\end{minipage}
\end{table*}

%% file: tab_res_BAL.tex
\begin{table*}
\begin{minipage}{\textwidth}
\renewcommand{\tabcolsep}{0.8mm}
\caption{Flux densities, spectral indices, and radio luminosities of the BAL sample.}
\label{tbl:res-BAL}
\begin{tabular}{cccccccccccccc}
\hline							
ID	&	Compo-	&		$f_{0.144}$						&		$f_{0.322}$				&		$f_{1.4}$				&		$f_{3.0}$				&		$\alpha_{\rm low}$					&		$\alpha_{\rm mid}$			&		$\alpha_{\rm high}$			&		$\log L_{0.7}$		&		$\log L_{3}$		&		$\log L_{7}$		\\
	&	nent	&		(mJy)						&		(mJy)				&		(mJy)				&		(mJy)				&							&					&					&		(\WHz)		&		(\WHz)		&		(\WHz)		\\
(1)	&	(2)	&		(3)						&		(4)				&		(5)				&		(6)				&		(7)					&		(8)			&		(9)			&		(10)		&		(11)		&		(12)		\\\hline
b01	&	total	&	\phn	112.1 	$\pm$	\phn	12.6$^{\dag}$ 			&			…			&	\phn\phn	23.8 	$\pm$	\phn	1.2 	&	\phn\phn	23.0 	$\pm$	\phn	1.1 	&			…				&			…		&		$-$0.04 	$\pm$	0.09 	&		…		&		…		&	27.5 			\\
	&	A	&			…					&			…			&	\phn\phn	13.7 	$\pm$	\phn	0.7 	&	\phn\phn	17.9 	$\pm$	\phn	0.6	&			…				&			…		&	\phs	0.35 	$\pm$	0.08 	&		…		&		…		&	27.3 			\\
	&	B	&			…					&			…			&	\phn\phn	10.1 	$\pm$	\phn	0.5 	&	\phn\phn\phn	5.1 	$\pm$	\phn	0.4	&			…				&			…		&		$-$0.90 	$\pm$	0.12 	&		…		&		…		&	27.0 			\\
b02	&	…	&			…					&			…			&	\phn\phn	12.3 	$\pm$	\phn	0.6 	&	\phn\phn\phn	9.9 	$\pm$	\phn	0.4	&			…				&			…		&		$-$0.29 	$\pm$	0.09 	&		…		&		…		&	27.4 			\\
b03	&	…	&	\phn	174.2 	$\pm$	\phn	17.9$^{\dag}$	&	\phn	174.1 	$\pm$	\phn	17.5 	&	\phn\phn	42.2 	$\pm$	\phn	2.1 	&	\phn\phn	18.7 	$\pm$	\phn	0.7	&	\phs	0.00 	$\pm$	0.18 			&		$-$0.96 	$\pm$	0.15 	&		$-$1.07 	$\pm$	0.08 	&	28.4 			&	28.0 			&	27.7 			\\
b04	&	…	&			$<$			50.3$^{\dag}$	&	\phn\phn	12.7 	$\pm$	\phn\phn	1.8 	&	\phn\phn	17.2 	$\pm$	\phn	0.9 	&	\phn\phn\phn	6.4 	$\pm$	\phn	0.3	&			$>$			$-$1.76 	&	\phs	0.21 	$\pm$	0.20 	&		$-$1.29 	$\pm$	0.09 	&		$<$	27.2 	&	27.1 			&	26.8 			\\
b05	&	…	&	\phn\phn	81.9 	$\pm$	\phn\phn	8.4 			&	\phn\phn	66.9 	$\pm$	\phn\phn	6.7 	&	\phn\phn	33.9 	$\pm$	\phn	1.7 	&	\phn\phn	20.1 	$\pm$	\phn	0.7	&		$-$0.25 	$\pm$	0.18 			&		$-$0.46 	$\pm$	0.15 	&		$-$0.68 	$\pm$	0.08 	&	28.0 			&	27.8 			&	27.6 			\\
b06	&	…	&			$<$			20.9$^{\dag}$	&	\phn\phn\phn	6.9 	$\pm$	\phn\phn	1.2 	&	\phn\phn	83.1 	$\pm$	\phn	4.2 	&	\phn\phn	59.6 	$\pm$	\phn	1.8	&			$>$			$-$1.43 	&	\phs	1.70 	$\pm$	0.24 	&		$-$0.44 	$\pm$	0.08 	&		$<$	27.3 	&	27.7 			&	28.0 			\\
b07	&	…	&	\phn\phn	58.4 	$\pm$	\phn\phn	7.1$^{\dag}$	&	\phn\phn	47.4 	$\pm$	\phn\phn	4.8 	&	\phn\phn	12.3 	$\pm$	\phn	0.6 	&	\phn\phn\phn	7.2 	$\pm$	\phn	0.3	&		$-$0.27 	$\pm$	0.20 			&		$-$0.92 	$\pm$	0.15 	&		$-$0.70 	$\pm$	0.09 	&	27.7 			&	27.2 			&	26.9 			\\
b08	&	…	&	\phn\phn\phn	3.5 	$\pm$	\phn\phn	0.4 			&	\phn\phn\phn	5.6 	$\pm$	\phn\phn	0.6 	&	\phn\phn	10.0 	$\pm$	\phn	0.5 	&	\phn\phn	13.2 	$\pm$	\phn	0.5	&	\phs	0.61 	$\pm$	0.19 			&	\phs	0.39 	$\pm$	0.16 	&	\phs	0.36 	$\pm$	0.08 	&	26.5 			&	26.8 			&	26.9 			\\
b09	&	…	&	\phn\phn\phn	9.9 	$\pm$	\phn\phn	1.0 			&	\phn\phn	17.1 	$\pm$	\phn\phn	2.2 	&	\phn\phn	42.0 	$\pm$	\phn	2.1 	&	\phn\phn	45.6 	$\pm$	\phn	1.4	&	\phs	0.68 	$\pm$	0.20 			&	\phs	0.61 	$\pm$	0.18 	&	\phs	0.11 	$\pm$	0.08 	&	27.2 			&	27.6 			&	27.7 			\\
b10	&	…	&	\phn\phn	56.9 	$\pm$	\phn\phn	5.9 			&	\phn\phn	39.7 	$\pm$	\phn\phn	4.1 	&	\phn\phn	12.8 	$\pm$	\phn	0.7 	&	\phn\phn\phn	7.4 	$\pm$	\phn	0.3	&		$-$0.45 	$\pm$	0.18 			&		$-$0.77 	$\pm$	0.15 	&		$-$0.72 	$\pm$	0.09 	&	27.8 			&	27.4 			&	27.1 			\\
b11	&	…	&			$<$			21.0$^{\dag}$	&			$<$		15.1 	&	\phn\phn	28.0 	$\pm$	\phn	1.4 	&	\phn\phn	20.0 	$\pm$	\phn	0.7	&	\phs		…				&			$>$	0.42 	&		$-$0.44 	$\pm$	0.08 	&		$<$	27.4		&		$<$	27.4 	&	27.4 			\\
b12	&	…	&	\phn	797.3 	$\pm$	\phn	79.9$^{\dag}$	&	\phn	546.1 	$\pm$	\phn	54.7 	&	\phn	160.1 	$\pm$	\phn	8.0 	&	\phn\phn	86.4 	$\pm$	\phn	2.7	&		$-$0.48 	$\pm$	0.18 			&		$-$0.83 	$\pm$	0.15 	&		$-$0.81 	$\pm$	0.08 	&	28.9 			&	28.4 			&	28.1 			\\
b13	&	…	&	\phn	340.2 	$\pm$	\phn	34.3$^{\dag}$	&		1144.9 	$\pm$		114.7 	&	\phn	268.5 	$\pm$		13.4 	&	\phn	164.6 	$\pm$	\phn	5.0	&	\phs	1.55 	$\pm$	0.18 			&		$-$0.99 	$\pm$	0.15 	&		$-$0.64 	$\pm$	0.08 	&	28.8 			&	28.5 			&	28.2 			\\
b14	&	…	&			$<$			16.4$^{\dag}$	&	\phn\phn	14.3 	$\pm$	\phn\phn	1.6 	&	\phn\phn	22.9 	$\pm$	\phn	1.2 	&	\phn\phn	23.7 	$\pm$	\phn	0.8	&			$>$			$-$0.17 	&	\phs	0.32 	$\pm$	0.16 	&	\phs	0.05 	$\pm$	0.08 	&		$<$	27.6 	&	27.7 			&	27.8 			\\
b15	&	…	&			$<$			27.0$^{\dag}$	&	\phn\phn	15.2 	$\pm$	\phn\phn	1.7 	&	\phn\phn	49.8 	$\pm$	\phn	2.5 	&	\phn\phn	53.0 	$\pm$	\phn	1.6	&			$>$			$-$0.74 	&	\phs	0.81 	$\pm$	0.16 	&	\phs	0.08 	$\pm$	0.08 	&		$<$	27.7 	&	27.9 			&	28.1 			\\
b16	&	…	&			$<$			19.5$^{\dag}$	&	\phn\phn	26.5 	$\pm$	\phn\phn	4.6 	&	\phn\phn	30.8 	$\pm$	\phn	1.5 	&	\phn\phn	19.1 	$\pm$	\phn	0.6	&			$>$		\phs	0.39 	&	\phs	0.10 	$\pm$	0.24 	&		$-$0.63 	$\pm$	0.08 	&		$<$	27.7 	&	27.8 			&	27.7 			\\
b17	&	…	&	\phn\phn\phn	7.2 	$\pm$	\phn\phn	0.7 			&	\phn\phn	18.4 	$\pm$	\phn\phn	1.9 	&	\phn\phn	24.0 	$\pm$	\phn	1.2 	&	\phn\phn	19.3 	$\pm$	\phn	0.6	&	\phs	1.17 	$\pm$	0.18 			&	\phs	0.18 	$\pm$	0.15 	&		$-$0.28 	$\pm$	0.08 	&	27.0 			&	27.3 			&	27.2 			\\
b18	&	…	&			$<$			26.1$^{\dag}$	&			…			&	\phn\phn	11.3 	$\pm$	\phn	0.6 	&	\phn\phn\phn	5.4 	$\pm$	\phn	0.3	&	\phs		…				&			…		&		$-$0.97 	$\pm$	0.11 	&		…		&		…		&	27.5 			\\
b19	&	…	&	\phn\phn	56.5 	$\pm$	\phn\phn	5.7 			&	\phn\phn	65.2 	$\pm$	\phn\phn	6.5 	&	\phn\phn	80.2 	$\pm$	\phn	4.0 	&	\phn\phn	86.7 	$\pm$	\phn	2.6	&	\phs	0.18 	$\pm$	0.18 			&	\phs	0.14 	$\pm$	0.15 	&	\phs	0.10 	$\pm$	0.08 	&	27.9 			&	28.0 			&	28.0 			\\
b20	&	total	&	\phn\phn	93.4 	$\pm$	\phn\phn	9.4 			&	\phn\phn	64.0 	$\pm$	\phn\phn	6.6 	&	\phn\phn	18.5 	$\pm$	\phn	1.0 	&	\phn\phn\phn	8.2 	$\pm$	\phn	0.8 	&		$-$0.47 	$\pm$	0.18 			&		$-$0.85 	$\pm$	0.15 	&		$-$1.06 	$\pm$	0.15 	&	27.7 			&	27.2 			&	26.8 			\\
	&	A	&	\phn\phn	27.0 	$\pm$	\phn\phn	2.7 			&	\phn\phn	27.6 	$\pm$	\phn\phn	3.0 	&	\phn\phn	10.7 	$\pm$	\phn	0.6 	&	\phn\phn\phn	4.2 	$\pm$	\phn	0.3	&	\phs	0.03 	$\pm$	0.18 			&		$-$0.65 	$\pm$	0.16 	&		$-$1.21 	$\pm$	0.10 	&	27.3 			&	26.9 			&	26.6 			\\
	&	B	&	\phn\phn	33.8 	$\pm$	\phn\phn	3.4 			&	\phn\phn	22.4 	$\pm$	\phn\phn	2.5 	&	\phn\phn\phn	4.3 	$\pm$	\phn	0.3 	&	\phn\phn\phn	1.8 	$\pm$	\phn	0.3	&		$-$0.51 	$\pm$	0.19 			&		$-$1.12 	$\pm$	0.17 	&		$-$1.12 	$\pm$	0.23 	&	27.2 			&	26.6 			&	26.2 			\\
	&	C	&	\phn\phn	32.6 	$\pm$	\phn\phn	3.3 			&	\phn\phn	14.0 	$\pm$	\phn\phn	1.7 	&	\phn\phn\phn	3.5 	$\pm$	\phn	0.2 	&	\phn\phn\phn	2.1 	$\pm$	\phn	0.3	&		$-$1.05 	$\pm$	0.20 			&		$-$0.95 	$\pm$	0.18 	&		$-$0.66 	$\pm$	0.22 	&	27.1 			&	26.5 			&	26.2 			\\
b21	&	…	&			$<$			19.2$^{\dag}$	&	\phn\phn\phn	6.7 	$\pm$	\phn\phn	1.4 	&	\phn\phn	12.8 	$\pm$	\phn	0.7 	&	\phn\phn	13.8 	$\pm$	\phn	0.5	&			$>$			$-$1.35 	&	\phs	0.44 	$\pm$	0.28 	&	\phs	0.10 	$\pm$	0.08 	&		$<$	26.7 	&	26.8 			&	26.8 			\\
b22	&	…	&			$<$			27.9$^{\dag}$	&	\phn\phn	33.8 	$\pm$	\phn\phn	3.6 	&	\phn\phn	38.0 	$\pm$	\phn	1.9 	&	\phn\phn	17.9 	$\pm$	\phn	0.6	&			$>$		\phs	0.24 	&	\phs	0.08 	$\pm$	0.15 	&		$-$0.98 	$\pm$	0.08 	&		$<$	27.5 	&	27.6 			&	27.4 			\\
b23	&	…	&			$<$			22.9$^{\dag}$	&	\phn\phn	14.2 	$\pm$	\phn\phn	1.6 	&	\phn\phn	50.0 	$\pm$	\phn	2.5 	&	\phn\phn	38.0 	$\pm$	\phn	1.2	&			$>$			$-$0.61 	&	\phs	0.86 	$\pm$	0.16 	&		$-$0.36 	$\pm$	0.08 	&		$<$	27.6 	&	27.8 			&	27.9 			\\
b24	&	…	&	\phn\phn	30.0 	$\pm$	\phn\phn	5.0$^{\dag}$	&	\phn\phn	26.7 	$\pm$	\phn\phn	2.7 	&	\phn\phn	14.9 	$\pm$	\phn	0.8 	&	\phn\phn	12.2 	$\pm$	\phn	0.4	&		$-$0.15 	$\pm$	0.25 			&		$-$0.40 	$\pm$	0.15 	&		$-$0.26 	$\pm$	0.08 	&	28.0 			&	27.8 			&	27.6 			\\
b25	&	total	&	\phn	158.4 	$\pm$	\phn	15.9 			&	\phn\phn	99.7 	$\pm$	\phn	10.0 	&	\phn\phn	28.5 	$\pm$	\phn	1.4 	&	\phn\phn	26.5 	$\pm$	\phn	2.6 	&		$-$0.58 	$\pm$	0.18 			&		$-$0.85 	$\pm$	0.15 	&		$-$0.10 	$\pm$	0.15 	&	27.8 			&	27.3 			&	27.2 			\\
	&	A	&	\phn	101.4 	$\pm$	\phn	10.2 			&	\phn\phn	48.4 	$\pm$	\phn\phn	4.9 	&	\phn\phn	18.1 	$\pm$	\phn	0.9 	&	\phn\phn	20.3 	$\pm$	\phn	0.7	&		$-$0.92 	$\pm$	0.18 			&		$-$0.67 	$\pm$	0.15 	&	\phs	0.15 	$\pm$	0.08 	&	27.5 			&	27.1 			&	27.0 			\\
	&	B	&	\phn\phn	57.0 	$\pm$	\phn\phn	5.7 			&	\phn\phn	51.3 	$\pm$	\phn\phn	5.2 	&	\phn\phn	10.4 	$\pm$	\phn	0.5 	&	\phn\phn\phn	6.2 	$\pm$	\phn	0.3	&		$-$0.13 	$\pm$	0.18 			&		$-$1.09 	$\pm$	0.15 	&		$-$0.68 	$\pm$	0.10 	&	27.5 			&	26.9 			&	26.6 			\\
b26	&	…	&			$<$			21.4$^{\dag}$	&	\phn\phn\phn	5.8 	$\pm$	\phn\phn	1.1 	&	\phn\phn	44.8 	$\pm$	\phn	2.2 	&	\phn\phn	71.5 	$\pm$	\phn	2.2	&			$>$			$-$1.67 	&	\phs	1.39 	$\pm$	0.26 	&	\phs	0.61 	$\pm$	0.08 	&		$<$	27.7 	&	27.8 			&	28.2 			\\
b27	&	…	&			$<$			28.3$^{\dag}$	&	\phn\phn	10.4 	$\pm$	\phn\phn	1.3 	&	\phn\phn	10.7 	$\pm$	\phn	0.6 	&	\phn\phn	13.2 	$\pm$	\phn	0.5	&			$>$			$-$1.28 	&	\phs	0.02 	$\pm$	0.18 	&	\phs	0.27 	$\pm$	0.09 	&		$<$	27.0 	&	26.9 			&	26.9 			\\
b28	&	…	&	\phn\phn	68.5 	$\pm$	\phn\phn	6.9 			&	\phn\phn	51.7 	$\pm$	\phn\phn	5.2 	&	\phn\phn	17.9 	$\pm$	\phn	0.9 	&	\phn\phn	11.8 	$\pm$	\phn	0.4	&		$-$0.35 	$\pm$	0.18 			&		$-$0.72 	$\pm$	0.15 	&		$-$0.54 	$\pm$	0.08 	&	27.9 			&	27.5 			&	27.2 			\\
b29	&	…	&	\phn\phn	35.3 	$\pm$	\phn\phn	3.5 			&	\phn\phn	26.2 	$\pm$	\phn\phn	2.8 	&	\phn\phn	15.0 	$\pm$	\phn	0.8 	&	\phn\phn	14.7 	$\pm$	\phn	0.6	&		$-$0.37 	$\pm$	0.18 			&		$-$0.38 	$\pm$	0.15 	&		$-$0.03 	$\pm$	0.08 	&	27.4 			&	27.2 			&	27.1 			\\
b30	&	…	&	\phn\phn\phn	8.0 	$\pm$	\phn\phn	0.8 			&	\phn\phn	21.1 	$\pm$	\phn\phn	2.3 	&	\phn\phn	37.1 	$\pm$	\phn	1.9 	&	\phn\phn	41.6 	$\pm$	\phn	1.3	&	\phs	1.21 	$\pm$	0.19 			&	\phs	0.38 	$\pm$	0.16 	&	\phs	0.15 	$\pm$	0.08 	&	27.5 			&	27.9 			&	28.1 			\\
b31	&	…	&		1674.6 	$\pm$		167.7$^{\dag}$	&		1356.7 	$\pm$		135.7 	&		1055.9 	$\pm$		52.8 	&		1056.6 	$\pm$		32.1	&		$-$0.27 	$\pm$	0.18 			&		$-$0.17 	$\pm$	0.15 	&	\phs	0.00 	$\pm$	0.08 	&	29.8 			&	29.7 			&	29.6 			\\
b32	&	…	&	\phn\phn	27.9 	$\pm$	\phn\phn	2.8 			&	\phn\phn	28.7 	$\pm$	\phn\phn	9.5 	&	\phn\phn	26.7 	$\pm$	\phn	1.3 	&	\phn\phn	44.4 	$\pm$	\phn	1.4	&	\phs	0.03 	$\pm$	0.43 			&		$-$0.05 	$\pm$	0.44 	&	\phs	0.67 	$\pm$	0.08 	&	27.8 			&	27.7 			&	27.8 			\\
b33	&	…	&	\phn\phn\phn	8.8 	$\pm$	\phn\phn	0.9 			&	\phn\phn	17.5 	$\pm$	\phn\phn	2.0 	&	\phn\phn	18.7 	$\pm$	\phn	0.9 	&	\phn\phn	13.5 	$\pm$	\phn	0.5	&	\phs	0.85 	$\pm$	0.19 			&	\phs	0.05 	$\pm$	0.16 	&		$-$0.43 	$\pm$	0.08 	&	27.4 			&	27.7 			&	27.6 			\\
b34	&	…	&			$<$			46.5$^{\dag}$	&	\phn\phn	19.8 	$\pm$	\phn\phn	2.2 	&	\phn\phn	11.8 	$\pm$	\phn	0.6 	&	\phn\phn\phn	6.2 	$\pm$	\phn	0.4	&			$>$			$-$1.09 	&		$-$0.35 	$\pm$	0.16 	&		$-$0.84 	$\pm$	0.11 	&		$<$	27.3 	&	26.9 			&	26.7 			\\
b35	&	…	&	\phn\phn\phn	2.0 	$\pm$	\phn\phn	0.3 			&	\phn\phn\phn	4.5 	$\pm$	\phn\phn	1.3 	&	\phn\phn	25.8 	$\pm$	\phn	1.3 	&	\phn\phn	11.2 	$\pm$	\phn	0.4	&	\phs	0.99 	$\pm$	0.40 			&	\phs	1.19 	$\pm$	0.39 	&		$-$1.09 	$\pm$	0.08 	&	26.7 			&	27.4 			&	27.5 			\\
b36	&	…	&	\phn\phn	43.1 	$\pm$	\phn\phn	4.5 			&	\phn\phn	50.1 	$\pm$	\phn\phn	5.0 	&	\phn\phn	12.6 	$\pm$	\phn	0.6 	&	\phn\phn\phn	4.9 	$\pm$	\phn	0.4	&	\phs	0.19 	$\pm$	0.18 			&		$-$0.94 	$\pm$	0.15 	&		$-$1.23 	$\pm$	0.12 	&	27.5 			&	27.1 			&	26.6 			\\
b37	&	…	&	\phn\phn	59.0 	$\pm$	\phn\phn	9.8$^{\dag}$	&	\phn\phn	36.1 	$\pm$	\phn\phn	3.7 	&	\phn\phn	13.2 	$\pm$	\phn	0.7 	&	\phn\phn	12.0 	$\pm$	\phn	0.5	&		$-$0.63 	$\pm$	0.25 			&		$-$0.68 	$\pm$	0.15 	&		$-$0.12 	$\pm$	0.08 	&	27.6 			&	27.2 			&	27.1 			\\
b38	&	…	&	\phn\phn\phn	4.5 	$\pm$	\phn\phn	0.5 			&	\phn\phn\phn	8.4 	$\pm$	\phn\phn	1.5 	&	\phn\phn	13.7 	$\pm$	\phn	0.7 	&	\phn\phn	19.0 	$\pm$	\phn	0.7	&	\phs	0.77 	$\pm$	0.26 			&	\phs	0.33 	$\pm$	0.24 	&	\phs	0.43 	$\pm$	0.08 	&	27.2 			&	27.5 			&	27.7 			\\
b39	&	…	&	\phn\phn	28.8 	$\pm$	\phn\phn	2.9 			&	\phn\phn	24.0 	$\pm$	\phn\phn	2.4 	&	\phn\phn	11.2 	$\pm$	\phn	0.6 	&	\phn\phn\phn	8.9 	$\pm$	\phn	0.4	&		$-$0.23 	$\pm$	0.18 			&		$-$0.52 	$\pm$	0.15 	&		$-$0.30 	$\pm$	0.09 	&	27.4 			&	27.1 			&	27.0 			\\
b40	&	…	&	\phn	135.6 	$\pm$	\phn	13.6 			&	\phn	185.4 	$\pm$	\phn	18.5 	&	\phn	132.0 	$\pm$	\phn	6.6 	&	\phn\phn	88.5 	$\pm$	\phn	2.7	&	\phs	0.39 	$\pm$	0.18 			&		$-$0.23 	$\pm$	0.15 	&		$-$0.52 	$\pm$	0.08 	&	28.3 			&	28.2 			&	28.1 			\\
b41	&	…	&	\phn\phn	48.7 	$\pm$	\phn\phn	8.9$^{\dag}$	&	\phn\phn	55.8 	$\pm$	\phn\phn	6.5 	&	\phn\phn	25.5 	$\pm$	\phn	1.3 	&	\phn\phn	21.2 	$\pm$	\phn	0.7	&	\phs	0.17 	$\pm$	0.28 			&		$-$0.53 	$\pm$	0.17 	&		$-$0.24 	$\pm$	0.08 	&	27.6 			&	27.4 			&	27.2 			\\
b42	&	…	&	\phn\phn	12.3 	$\pm$	\phn\phn	1.2 			&	\phn\phn	10.0 	$\pm$	\phn\phn	1.5 	&	\phn\phn	14.2 	$\pm$	\phn	0.7 	&	\phn\phn	40.9 	$\pm$	\phn	1.3	&		$-$0.26 	$\pm$	0.22 			&	\phs	0.24 	$\pm$	0.20 	&	\phs	1.39 	$\pm$	0.08 	&	27.6 			&	27.6 			&	27.8 			\\
b43	&	…	&	\phn\phn	40.9 	$\pm$	\phn\phn	4.1 			&	\phn\phn	44.7 	$\pm$	\phn\phn	4.5 	&	\phn\phn	41.2 	$\pm$	\phn	2.1 	&	\phn\phn	44.9 	$\pm$	\phn	1.4	&	\phs	0.11 	$\pm$	0.18 			&		$-$0.06 	$\pm$	0.15 	&	\phs	0.11 	$\pm$	0.08 	&	28.0 			&	28.0 			&	28.0 			\\
b44	&	…	&	\phn\phn	13.8 	$\pm$	\phn\phn	1.4 			&	\phn\phn	56.5 	$\pm$	\phn\phn	5.9 	&	\phn\phn	54.2 	$\pm$	\phn	2.7 	&	\phn\phn	42.4 	$\pm$	\phn	1.3	&	\phs	1.75 	$\pm$	0.18 			&		$-$0.03 	$\pm$	0.15 	&		$-$0.32 	$\pm$	0.08 	&	27.5 			&	27.7 			&	27.6 			\\
b45	&	…	&	\phn\phn	36.7 	$\pm$	\phn\phn	3.7 			&	\phn\phn	50.8 	$\pm$	\phn\phn	5.1 	&	\phn\phn	56.4 	$\pm$	\phn	2.8 	&	\phn\phn	34.9 	$\pm$	\phn	1.1	&	\phs	0.40 	$\pm$	0.18 			&	\phs	0.07 	$\pm$	0.15 	&		$-$0.63 	$\pm$	0.08 	&	28.2 			&	28.4 			&	28.4 			\\
b46	&	…	&	\phn\phn	41.9 	$\pm$	\phn\phn	4.4 			&	\phn\phn	44.6 	$\pm$	\phn\phn	4.5 	&	\phn\phn	25.5 	$\pm$	\phn	1.3 	&	\phn\phn	14.4 	$\pm$	\phn	0.5	&	\phs	0.08 	$\pm$	0.18 			&		$-$0.38 	$\pm$	0.15 	&		$-$0.74 	$\pm$	0.08 	&	28.0 			&	27.9 			&	27.7 			\\
b47	&	…	&	\phn\phn\phn	2.0 	$\pm$	\phn\phn	0.3 			&	\phn\phn\phn	2.4 	$\pm$	\phn\phn	0.6 	&	\phn\phn	11.3 	$\pm$	\phn	0.6 	&	\phn\phn\phn	6.1 	$\pm$	\phn	0.3	&	\phs	0.21 	$\pm$	0.34 			&	\phs	1.06 	$\pm$	0.31 	&		$-$0.80 	$\pm$	0.09 	&	26.5 			&	27.0 			&	27.0 			\\
b48	&	…	&	\phn\phn	15.0 	$\pm$	\phn\phn	1.5 			&	\phn\phn	25.7 	$\pm$	\phn\phn	2.6 	&	\phn\phn	10.2 	$\pm$	\phn	0.5 	&	\phn\phn\phn	4.7 	$\pm$	\phn	0.3	&	\phs	0.67 	$\pm$	0.18 			&		$-$0.63 	$\pm$	0.15 	&		$-$1.02 	$\pm$	0.11 	&	27.1 			&	26.8 			&	26.5 			\\
\hline						
\end{tabular}

\textit{Note.} Columns are as follows:
(1) Object ID;
(2) subcomponent name of the sources;
(3)--(6) flux densities at 144\,MHz, 322\,MHz, 1.4\,GHz, and 3.0\,GHz, respectively. 
Reference for columns (3), (5), and (6) are \cite{2022A&A...659A...1S}, \cite{1995ApJ...450..559B}, and \cite{2021ApJS..255...30G}, respectively.
Data in column (3) indicated by $^{\dag}$ was obtained from the TGSS at 147.5\,MHz \citep{2017A&A...598A..78I};
(7)--(9) spectral indices between neighbouring bands;
(10)--(12) Log specific luminosity at rest 700\,MHz, 3\,GHz, and 7.0\,GHz calculated by interpolating the obtained flux densities.
\end{minipage}
\end{table*}

%% file: tab_res_nBAL.tex
\begin{table*}
\begin{minipage}{\textwidth}
\renewcommand{\tabcolsep}{0.8mm}
\caption{Flux densities, spectral indices, and radio luminosities of the non-BAL sample.}
\label{tbl:res-nBAL}
\begin{tabular}{cccccccccccccc}
\hline																																																										
ID	&	Compo-	&		$f_{0.144}$						&		$f_{0.322}$				&		$f_{1.4}$				&		$f_{3.0}$				&		$\alpha_{\rm low}$				&		$\alpha_{\rm mid}$			&		$\alpha_{\rm high}$				&		$\log L_{0.7}$		&		$\log L_{3}$		&		$\log L_{7}$		\\
	&	nent	&		(mJy)						&		(mJy)				&		(mJy)				&		(mJy)				&						&					&					&		(\WHz)		&		(\WHz)		&		(\WHz)		\\
(1)	&	(2)	&		(3)						&		(4)				&		(5)				&		(6)				&		(7)				&		(8)			&		(9)				&		(10)		&		(11)		&		(12)		\\\hline
n01	&	…	&	\phn\phn	37.8 	$\pm$	\phn\phn	3.8 			&	\phn\phn	28.9 	$\pm$	\phn\phn	3.1 	&	\phn\phn	12.2 	$\pm$	\phn	0.6 	&	\phn	10.0 	$\pm$	\phn	1.0 	&		$-$0.33 	$\pm$	0.18 		&		$-$0.59 	$\pm$	0.08 	&		$-$0.26 	$\pm$	0.15 		&	27.4 			&	27.0 			&	26.9 			\\
n02	&	…	&	\phn\phn	95.0 	$\pm$	\phn	10.6$^{\dag}$	&	\phn\phn	89.7 	$\pm$	\phn\phn	9.1 	&	\phn\phn	65.2 	$\pm$	\phn	3.3 	&	\phn	73.3 	$\pm$	\phn	7.3 	&		$-$0.07 	$\pm$	0.19 		&		$-$0.22 	$\pm$	0.08 	&	\phs	0.15 	$\pm$	0.15 		&	28.0 			&	27.9 			&	27.9 			\\
n03	&	…	&	\phn\phn	66.2 	$\pm$	\phn\phn	6.7 			&	\phn\phn	59.9 	$\pm$	\phn\phn	6.0 	&	\phn\phn	28.7 	$\pm$	\phn	1.4 	&	\phn	33.3 	$\pm$	\phn	3.3 	&		$-$0.12 	$\pm$	0.18 		&		$-$0.50 	$\pm$	0.08 	&	\phs	0.19 	$\pm$	0.15 		&	28.0 			&	27.8 			&	27.7 			\\
n04	&	…	&			$<$			24.4$^{\dag}$	&	\phn\phn	11.9 	$\pm$	\phn\phn	1.7 	&	\phn\phn	17.1 	$\pm$	\phn	0.9 	&	\phn	13.0 	$\pm$	\phn	1.3 	&			$>$		$-$0.92 	&	\phs	0.25 	$\pm$	0.10 	&		$-$0.35 	$\pm$	0.15 		&		$<$	26.9 	&	26.9 			&	26.9 			\\
n05	&	…	&			$<$			16.1$^{\dag}$	&	\phn\phn	15.8 	$\pm$	\phn\phn	1.8 	&	\phn\phn	19.8 	$\pm$	\phn	1.0 	&	\phn	10.2 	$\pm$	\phn	1.1 	&			$>$		$-$0.02 	&	\phs	0.15 	$\pm$	0.09 	&		$-$0.87 	$\pm$	0.15 		&		$<$	27.0 	&	27.1 			&	26.9 			\\
n06	&	…	&			$<$			16.2$^{\dag}$	&	\phn\phn	12.7 	$\pm$	\phn\phn	1.4 	&	\phn\phn	31.6 	$\pm$	\phn	1.6 	&	\phn	68.6 	$\pm$	\phn	6.9 	&			$>$		$-$0.31 	&	\phs	0.62 	$\pm$	0.08 	&	\phs	1.02 	$\pm$	0.15 		&		$<$	27.5 	&	27.7 			&	28.0 			\\
n07	&	…	&	\phn\phn\phn	6.8 	$\pm$	\phn\phn	0.8 			&	\phn\phn\phn	7.0 	$\pm$	\phn\phn	0.8 	&	\phn\phn	10.3 	$\pm$	\phn	0.5 	&	\phn\phn	6.8 	$\pm$	\phn	0.7 	&	\phs	0.03 	$\pm$	0.21 		&	\phs	0.26 	$\pm$	0.09 	&		$-$0.56 	$\pm$	0.16 		&	26.8 			&	26.9 			&	26.9 			\\
n08	&	…	&	\phn\phn	15.1 	$\pm$	\phn\phn	1.5 			&	\phn\phn	26.3 	$\pm$	\phn\phn	2.7 	&	\phn\phn	21.2 	$\pm$	\phn	1.1 	&	\phn	23.9 	$\pm$	\phn	2.4 	&	\phs	0.69 	$\pm$	0.18 		&		$-$0.15 	$\pm$	0.08 	&	\phs	0.16 	$\pm$	0.15 		&	27.4 			&	27.4 			&	27.4 			\\
n09	&	…	&	\phn\phn	18.5 	$\pm$	\phn\phn	1.9 			&	\phn\phn	20.5 	$\pm$	\phn\phn	2.2 	&	\phn\phn	23.9 	$\pm$	\phn	1.2 	&	\phn	31.8 	$\pm$	\phn	3.2 	&	\phs	0.13 	$\pm$	0.18 		&	\phs	0.11 	$\pm$	0.08 	&	\phs	0.37 	$\pm$	0.15 		&	27.2 			&	27.3 			&	27.4 			\\
n10	&	…	&	\phn\phn	77.7 	$\pm$	\phn\phn	9.4$^{\dag}$	&			…			&	\phn\phn	94.5 	$\pm$	\phn	4.7 	&	\phn	65.1 	$\pm$	\phn	6.5 	&	\phs		…			&			…		&		$-$0.49 	$\pm$	0.15 		&		…		&		…		&	28.0 			\\
n11	&	…	&	\phn\phn	13.4 	$\pm$	\phn\phn	1.3 			&	\phn\phn	16.9 	$\pm$	\phn\phn	1.7 	&	\phn\phn	11.1 	$\pm$	\phn	0.6 	&	\phn	13.2 	$\pm$	\phn	1.3 	&	\phs	0.29 	$\pm$	0.18 		&		$-$0.29 	$\pm$	0.08 	&	\phs	0.23 	$\pm$	0.15 		&	27.3 			&	27.2 			&	27.2 			\\
n12	&	…	&			$<$			15.8$^{\dag}$	&	\phn\phn\phn	5.3 	$\pm$	\phn\phn	1.1 	&	\phn\phn	10.2 	$\pm$	\phn	0.5 	&	\phn\phn	7.8 	$\pm$	\phn	0.8 	&			$>$		$-$1.39 	&	\phs	0.44 	$\pm$	0.14 	&		$-$0.34 	$\pm$	0.15 		&		$<$	27.2 	&	27.1 			&	27.1 			\\
n13	&	…	&			$<$			17.1$^{\dag}$	&	\phn\phn\phn	5.2 	$\pm$	\phn\phn	0.9 	&	\phn\phn	14.0 	$\pm$	\phn	0.7 	&	\phn	23.6 	$\pm$	\phn	2.4 	&			$>$		$-$1.53 	&	\phs	0.68 	$\pm$	0.13 	&	\phs	0.68 	$\pm$	0.15 		&		$<$	26.7 	&	26.9 			&	27.1 			\\
n14	&	…	&	\phn\phn	20.6 	$\pm$	\phn\phn	2.1 			&	\phn\phn	16.2 	$\pm$	\phn\phn	1.7 	&	\phn\phn	10.2 	$\pm$	\phn	0.5 	&	\phn\phn	9.0 	$\pm$	\phn	0.9 	&		$-$0.30 	$\pm$	0.18 		&		$-$0.31 	$\pm$	0.08 	&		$-$0.17 	$\pm$	0.15 		&	27.0 			&	26.8 			&	26.7 			\\
n15	&	…	&	\phn\phn	78.2 	$\pm$	\phn\phn	9.4$^{\dag}$	&	\phn\phn	74.1 	$\pm$	\phn\phn	8.3 	&	\phn\phn	26.7 	$\pm$	\phn	1.3 	&	\phn	17.8 	$\pm$	\phn	0.3 	&		$-$0.07 	$\pm$	0.21 		&		$-$0.69 	$\pm$	0.08 	&		$-$0.53 	$\pm$	0.07 		&	27.7 			&	27.3 			&	27.1 			\\
n16	&	…	&	\phn\phn	63.3 	$\pm$	\phn\phn	6.4 			&	\phn\phn	45.9 	$\pm$	\phn\phn	5.1 	&	\phn\phn	14.5 	$\pm$	\phn	0.7 	&	\phn\phn	8.5 	$\pm$	\phn	0.9 	&		$-$0.40 	$\pm$	0.19 		&		$-$0.79 	$\pm$	0.08 	&		$-$0.69 	$\pm$	0.15 		&	28.4 			&	28.0 			&	27.7 			\\
n17	&	…	&	\phn\phn	27.9 	$\pm$	\phn\phn	5.5$^{\dag}$	&	\phn\phn	62.6 	$\pm$	\phn\phn	6.3 	&	\phn\phn	59.3 	$\pm$	\phn	3.0 	&	\phn	37.2 	$\pm$	\phn	3.7 	&	\phs	1.03 	$\pm$	0.28 		&		$-$0.04 	$\pm$	0.08 	&		$-$0.61 	$\pm$	0.15 		&	27.8 			&	28.0 			&	27.9 			\\
n18	&	…	&	\phn\phn	13.6 	$\pm$	\phn\phn	1.5 			&	\phn\phn	27.5 	$\pm$	\phn\phn	2.8 	&	\phn\phn	25.8 	$\pm$	\phn	1.3 	&	\phn	22.4 	$\pm$	\phn	2.3 	&	\phs	0.88 	$\pm$	0.18 		&		$-$0.04 	$\pm$	0.08 	&		$-$0.19 	$\pm$	0.15 		&	27.6 			&	27.8 			&	27.8 			\\
n19	&	…	&	\phn\phn	18.6 	$\pm$	\phn\phn	1.9 			&	\phn\phn	26.5 	$\pm$	\phn\phn	4.0 	&	\phn\phn	13.4 	$\pm$	\phn	0.7 	&	\phn\phn	8.2 	$\pm$	\phn	0.9 	&	\phs	0.44 	$\pm$	0.23 		&		$-$0.46 	$\pm$	0.11 	&		$-$0.65 	$\pm$	0.15 		&	27.1 			&	26.9 			&	26.7 			\\
n20	&	…	&	\phn\phn	60.3 	$\pm$	\phn\phn	6.1 			&	\phn\phn	67.1 	$\pm$	\phn\phn	6.9 	&	\phn\phn	39.9 	$\pm$	\phn	2.0 	&	\phn	25.0 	$\pm$	\phn	2.5 	&	\phs	0.13 	$\pm$	0.18 		&		$-$0.35 	$\pm$	0.08 	&		$-$0.61 	$\pm$	0.15 		&	27.7 			&	27.6 			&	27.4 			\\
n21	&	…	&	\phn\phn\phn	6.5 	$\pm$	\phn\phn	0.8 			&	\phn\phn\phn	5.3 	$\pm$	\phn\phn	0.9 	&	\phn\phn	14.6 	$\pm$	\phn	0.7 	&	\phn\phn	7.4 	$\pm$	\phn	0.8 	&		$-$0.26 	$\pm$	0.26 		&	\phs	0.70 	$\pm$	0.12 	&		$-$0.89 	$\pm$	0.15 		&	26.9 			&	27.1 			&	27.1 			\\
n22	&	…	&	\phn	218.0 	$\pm$	\phn	22.9$^{\dag}$	&	\phn	147.3 	$\pm$	\phn	14.8 	&	\phn\phn	40.9 	$\pm$	\phn	2.0 	&	\phn	20.5 	$\pm$	\phn	2.1 	&		$-$0.50 	$\pm$	0.19 		&		$-$0.87 	$\pm$	0.08 	&		$-$0.91 	$\pm$	0.15 		&	28.8 			&	28.3 			&	28.0 			\\
n23	&	…	&			$<$			40.1$^{\dag}$	&	\phn\phn	10.1 	$\pm$	\phn\phn	1.2 	&	\phn\phn	11.7 	$\pm$	\phn	0.6 	&	\phn\phn	7.2 	$\pm$	\phn	0.8 	&			$>$		$-$1.76 	&	\phs	0.10 	$\pm$	0.09 	&		$-$0.65 	$\pm$	0.16 		&		$<$	27.8 	&	27.4 			&	27.3 			\\
n24	&	total	&	\phn\phn	85.6 	$\pm$	\phn\phn	8.6 			&	\phn\phn	89.9 	$\pm$	\phn\phn	9.2 	&	\phn	125.0 	$\pm$	\phn	6.3 	&	\phn	163.9 	$\pm$		16.4 	&	\phs	0.06 	$\pm$	0.18 		&	\phs	0.22 	$\pm$	0.08 	&	\phs	0.36 	$\pm$	0.15 		&	28.0 			&	28.2 			&	28.3 			\\
	&	A	&	\phn\phn	44.5 	$\pm$	\phn\phn	4.5 			&	\phn\phn	55.9 	$\pm$	\phn\phn	5.6 	&	\phn	118.4 	$\pm$	\phn	5.9 	&	\phn	161.2 	$\pm$		16.1 	&	\phs	0.28 	$\pm$	0.18 		&	\phs	0.51 	$\pm$	0.08 	&	\phs	0.41 	$\pm$	0.15 		&	27.8 			&	28.1 			&	28.2 			\\
	&	B	&	\phn\phn	29.4 	$\pm$	\phn\phn	2.9 			&	\phn\phn	22.4 	$\pm$	\phn\phn	2.4 	&	\phn\phn\phn	4.9 	$\pm$	\phn	0.3 	&	\phn\phn\phn	2.7 	$\pm$	\phn	0.7 	&		$-$0.34 	$\pm$	0.18 		&		$-$1.04 	$\pm$	0.08 	&		$-$0.77 	$\pm$	0.37 		&	27.5 			&	27.0 			&	26.6 			\\
	&	C	&	\phn\phn	11.7 	$\pm$	\phn\phn	1.2 			&	\phn\phn	11.5 	$\pm$	\phn\phn	1.9 	&	\phn\phn\phn	1.8 	$\pm$	\phn	0.2 	&			$<$		0.8 	&		$-$0.02 	$\pm$	0.24 		&		$-$1.27 	$\pm$	0.13 	&			$<$		$-$1.02 	&	27.2 			&	26.6 			&		$<$	26.2 	\\
n25	&	…	&	\phn\phn\phn	0.9 	$\pm$	\phn\phn	0.2 			&	\phn\phn\phn	2.4 	$\pm$	\phn\phn	0.5 	&	\phn\phn	12.8 	$\pm$	\phn	0.7 	&	\phn	13.6 	$\pm$	\phn	1.4 	&	\phs	1.17 	$\pm$	0.34 		&	\phs	1.14 	$\pm$	0.14 	&	\phs	0.07 	$\pm$	0.15 		&	26.2 			&	27.0 			&	27.2 			\\
n26	&	…	&	\phn	114.3 	$\pm$	\phn	11.5 			&	\phn\phn	75.1 	$\pm$	\phn\phn	8.0 	&	\phn\phn	38.4 	$\pm$	\phn	1.9 	&	\phn	23.8 	$\pm$	\phn	2.4 	&		$-$0.52 	$\pm$	0.18 		&		$-$0.46 	$\pm$	0.08 	&		$-$0.63 	$\pm$	0.15 		&	28.1 			&	27.8 			&	27.6 			\\
n27	&	total	&	\phn	485.8 	$\pm$	\phn	49.0 			&	\phn	285.2 	$\pm$	\phn	28.7 	&	\phn\phn	77.6 	$\pm$	\phn	3.9 	&	\phn	35.4 	$\pm$	\phn	3.5 	&		$-$0.68 	$\pm$	0.18 		&		$-$0.89 	$\pm$	0.08 	&		$-$1.03 	$\pm$	0.15 		&	28.9 			&	28.4 			&	28.1 			\\
	&	A	&			…					&	\phn\phn	72.0 	$\pm$	\phn\phn	7.7 	&	\phn\phn	16.4 	$\pm$	\phn	0.8 	&	\phn	15.5 	$\pm$	\phn	1.6 	&			…			&		$-$1.01 	$\pm$	0.08 	&		$-$0.07 	$\pm$	0.15 		&		…		&	27.8 			&	27.5 			\\
	&	B	&			…					&	\phn	138.8 	$\pm$	\phn	14.0 	&	\phn\phn	38.5 	$\pm$	\phn	1.9 	&	\phn	10.1 	$\pm$	\phn	1.1 	&			…			&		$-$0.87 	$\pm$	0.08 	&		$-$1.75 	$\pm$	0.15 		&		…		&	28.1 			&	27.7 			\\
	&	C	&			…					&	\phn\phn	74.3 	$\pm$	\phn\phn	7.6 	&	\phn\phn	22.7 	$\pm$	\phn	1.1 	&	\phn	9.8 	$\pm$	\phn	1.0 	&			…			&		$-$0.81 	$\pm$	0.08 	&		$-$1.11 	$\pm$	0.15 		&		…		&	27.9 			&	27.5 			\\
n28	&	…	&	\phn\phn\phn	5.6 	$\pm$	\phn\phn	0.6 			&	\phn\phn\phn	5.7 	$\pm$	\phn\phn	0.8 	&	\phn\phn\phn	8.2 	$\pm$	\phn	0.4 	&	\phn	18.5 	$\pm$	\phn	1.9 	&	\phs	0.02 	$\pm$	0.22 		&	\phs	0.24 	$\pm$	0.10 	&	\phs	1.07 	$\pm$	0.15 		&	26.6 			&	26.8 			&	27.0 			\\
n29	&	…	&	\phn\phn	52.4 	$\pm$	\phn\phn	8.7$^{\dag}$	&	\phn\phn	66.6 	$\pm$	\phn\phn	6.7 	&	\phn\phn	55.8 	$\pm$	\phn	2.8 	&	\phn	67.6 	$\pm$	\phn	6.8 	&	\phs	0.31 	$\pm$	0.25 		&		$-$0.12 	$\pm$	0.08 	&	\phs	0.25 	$\pm$	0.15 		&	27.6 			&	27.6 			&	27.7 			\\
n30	&	…	&		4122.7 	$\pm$		412.4 			&		2750.9 	$\pm$		275.5 	&		1165.4 	$\pm$		58.3 	&		691.8 	$\pm$		69.2 	&		$-$0.50 	$\pm$	0.18 		&		$-$0.58 	$\pm$	0.08 	&		$-$0.68 	$\pm$	0.15 		&	30.1 			&	29.7 			&	29.5 			\\
n31	&	…	&	\phn	281.5 	$\pm$	\phn	28.2 			&	\phn	447.6 	$\pm$	\phn	44.8 	&	\phn	261.7 	$\pm$		13.1 	&		143.4 	$\pm$		14.4 	&	\phs	0.58 	$\pm$	0.18 		&		$-$0.37 	$\pm$	0.08 	&		$-$0.79 	$\pm$	0.15 		&	28.5 			&	28.4 			&	28.2 			\\
n32	&	…	&			$<$			39.7$^{\dag}$	&			$<$		4.0 	&	\phn\phn	16.5 	$\pm$	\phn	0.8 	&	\phn	92.6 	$\pm$	\phn	9.3 	&	\phs		…			&			$>$	0.97 	&	\phs	2.26 	$\pm$	0.15 		&		$<$	27.6		&		$<$	27.0 	&	27.7 			\\
n33	&	…	&	\phn	315.5 	$\pm$	\phn	32.4$^{\dag}$	&	\phn	357.5 	$\pm$	\phn	35.8 	&	\phn	181.2 	$\pm$	\phn	9.1 	&		151.9 	$\pm$		15.4 	&	\phs	0.16 	$\pm$	0.18 		&		$-$0.46 	$\pm$	0.08 	&		$-$0.23 	$\pm$	0.15 		&	28.6 			&	28.4 			&	28.3 			\\
n34	&	…	&	\phn\phn\phn	4.0 	$\pm$	\phn\phn	0.4 			&	\phn\phn	10.5 	$\pm$	\phn\phn	1.1 	&	\phn\phn	33.2 	$\pm$	\phn	1.7 	&	\phn	35.9 	$\pm$	\phn	3.6 	&	\phs	1.21 	$\pm$	0.18 		&	\phs	0.79 	$\pm$	0.08 	&	\phs	0.10 	$\pm$	0.15 		&	27.2 			&	27.8 			&	28.0 			\\
n35	&	…	&	\phn\phn	12.3 	$\pm$	\phn\phn	1.2 			&	\phn\phn	10.7 	$\pm$	\phn\phn	1.5 	&	\phn\phn	13.2 	$\pm$	\phn	0.7 	&	\phn\phn	9.2 	$\pm$	\phn	0.9 	&		$-$0.17 	$\pm$	0.21 		&	\phs	0.14 	$\pm$	0.10 	&		$-$0.48 	$\pm$	0.15 		&	27.4 			&	27.4 			&	27.4 			\\
n36	&	…	&	\phn\phn	37.2 	$\pm$	\phn\phn	3.8 			&	\phn\phn	25.1 	$\pm$	\phn\phn	2.6 	&	\phn\phn	21.7 	$\pm$	\phn	1.1 	&	\phn	38.3 	$\pm$	\phn	3.9 	&		$-$0.49 	$\pm$	0.18 		&		$-$0.10 	$\pm$	0.08 	&	\phs	0.75 	$\pm$	0.15 		&	27.9 			&	27.8 			&	27.8 			\\
n37	&	…	&	\phn\phn	45.9 	$\pm$	\phn\phn	4.6 			&	\phn\phn	48.9 	$\pm$	\phn\phn	5.0 	&	\phn\phn	47.5 	$\pm$	\phn	2.4 	&	\phn	45.7 	$\pm$	\phn	4.6 	&	\phs	0.08 	$\pm$	0.18 		&		$-$0.02 	$\pm$	0.08 	&		$-$0.05 	$\pm$	0.15 		&	27.8 			&	27.8 			&	27.8 			\\
n38	&	…	&	\phn\phn	42.6 	$\pm$	\phn\phn	4.3 			&	\phn\phn	38.8 	$\pm$	\phn\phn	3.9 	&	\phn\phn	47.7 	$\pm$	\phn	2.4 	&	\phn	50.2 	$\pm$	\phn	5.0 	&		$-$0.12 	$\pm$	0.18 		&	\phs	0.14 	$\pm$	0.08 	&	\phs	0.07 	$\pm$	0.15 		&	28.2 			&	28.2 			&	28.2 			\\
n39	&	…	&	\phn\phn	10.8 	$\pm$	\phn\phn	1.1 			&	\phn\phn	13.5 	$\pm$	\phn\phn	1.7 	&	\phn\phn	12.7 	$\pm$	\phn	0.6 	&	\phn	14.2 	$\pm$	\phn	1.4 	&	\phs	0.28 	$\pm$	0.20 		&		$-$0.05 	$\pm$	0.09 	&	\phs	0.15 	$\pm$	0.15 		&	27.1 			&	27.1 			&	27.2 			\\
n40	&	…	&	\phn\phn\phn	2.8 	$\pm$	\phn\phn	0.3 			&	\phn\phn\phn	3.2 	$\pm$	\phn\phn	0.6 	&	\phn\phn	11.0 	$\pm$	\phn	0.6 	&	\phn\phn	9.5 	$\pm$	\phn	1.0 	&	\phs	0.15 	$\pm$	0.28 		&	\phs	0.84 	$\pm$	0.13 	&		$-$0.20 	$\pm$	0.15 		&	26.3 			&	26.8 			&	26.8 			\\
n41	&	…	&	\phn\phn	10.0 	$\pm$	\phn\phn	1.0 			&	\phn\phn	14.3 	$\pm$	\phn\phn	1.4 	&	\phn\phn	16.7 	$\pm$	\phn	0.8 	&	\phn\phn	7.5 	$\pm$	\phn	0.8 	&	\phs	0.44 	$\pm$	0.18 		&	\phs	0.11 	$\pm$	0.08 	&		$-$1.05 	$\pm$	0.15 		&	27.0 			&	27.1 			&	26.8 			\\
n42	&	…	&	\phn	136.0 	$\pm$	\phn	13.6 			&	\phn\phn	80.2 	$\pm$	\phn\phn	8.1 	&	\phn\phn	22.7 	$\pm$	\phn	1.1 	&	\phn	16.1 	$\pm$	\phn	1.6 	&		$-$0.66 	$\pm$	0.18 		&		$-$0.86 	$\pm$	0.08 	&		$-$0.45 	$\pm$	0.15 		&	28.2 			&	27.7 			&	27.5 			\\
n43	&	…	&	\phn	113.5 	$\pm$	\phn	11.4 			&	\phn\phn	70.2 	$\pm$	\phn\phn	7.1 	&	\phn\phn	18.9 	$\pm$	\phn	1.0 	&	\phn	11.9 	$\pm$	\phn	1.2 	&		$-$0.60 	$\pm$	0.18 		&		$-$0.89 	$\pm$	0.08 	&		$-$0.61 	$\pm$	0.15 		&	28.5 			&	28.0 			&	27.7 			\\
n44	&	…	&	\phn	196.3 	$\pm$	\phn	19.8 			&	\phn	113.7 	$\pm$	\phn	11.6 	&	\phn\phn	50.0 	$\pm$	\phn	2.5 	&	\phn	74.2 	$\pm$	\phn	7.5 	&		$-$0.68 	$\pm$	0.18 		&		$-$0.56 	$\pm$	0.08 	&	\phs	0.52 	$\pm$	0.15 		&	28.7 			&	28.3 			&	28.2 			\\
n45	&	…	&	\phn\phn\phn	2.8 	$\pm$	\phn\phn	0.4 			&	\phn\phn\phn	7.2 	$\pm$	\phn\phn	1.2 	&	\phn\phn	43.9 	$\pm$	\phn	2.2 	&	\phn	50.6 	$\pm$	\phn	5.1 	&	\phs	1.18 	$\pm$	0.28 		&	\phs	1.23 	$\pm$	0.12 	&	\phs	0.19 	$\pm$	0.15 		&	26.8 			&	27.6 			&	27.8 			\\
n46	&	…	&	\phn\phn	30.3 	$\pm$	\phn\phn	3.1 			&	\phn\phn	31.6 	$\pm$	\phn\phn	3.4 	&	\phn\phn\phn	9.5 	$\pm$	\phn	0.5 	&	\phn\phn	7.9 	$\pm$	\phn	0.8 	&	\phs	0.06 	$\pm$	0.18 		&		$-$0.82 	$\pm$	0.08 	&		$-$0.23 	$\pm$	0.15 		&	27.5 			&	27.1 			&	26.9 			\\
n47	&	…	&			$<$			25.8$^{\dag}$	&	\phn\phn\phn	7.5 	$\pm$	\phn\phn	1.0 	&	\phn\phn	15.2 	$\pm$	\phn	0.8 	&	\phn	13.3 	$\pm$	\phn	1.4 	&			$>$		$-$1.59 	&	\phs	0.48 	$\pm$	0.09 	&		$-$0.17 	$\pm$	0.15 		&		$<$	27.9 	&	27.6 			&	27.7 			\\
n48	&	…	&	\phn\phn	15.6 	$\pm$	\phn\phn	1.6 			&	\phn\phn	11.8 	$\pm$	\phn\phn	1.3 	&	\phn\phn	36.6 	$\pm$	\phn	1.8 	&	\phn	61.4 	$\pm$	\phn	6.2 	&		$-$0.34 	$\pm$	0.19 		&	\phs	0.77 	$\pm$	0.08 	&	\phs	0.68 	$\pm$	0.15 		&	27.4 			&	27.6 			&	27.9 			\\
\hline				
\end{tabular}

\textit{Note.} Columns are as follows:
(1) Object ID;
(2) subcomponent name of the sources;
(3)--(6) flux densities at 144\,MHz, 322\,MHz, 1.4\,GHz, and 3.0\,GHz, respectively.
Reference for columns (3), (5), and (6) are \cite{2022A&A...659A...1S}, \cite{1995ApJ...450..559B}, and \cite{2021ApJS..255...30G}, respectively.
Data in column (3) indicated by $^{\dag}$ was obtained from the TGSS at 147.5\,MHz \citep{2017A&A...598A..78I};
(7)--(9) spectral indices between neighbouring bands;
(10)--(12) Log specific luminosity at rest 700\,MHz, 3\,GHz, and 7.0\,GHz calculated by interpolating the obtained flux densities.
\end{minipage}
\end{table*}

%% file: tab_fit_BAL.tex
\begin{table*}
\begin{minipage}{0.92\textwidth}
\caption{Spectral classification of the BAL and non-BAL samples.}
\label{tbl:res-fit}
\begin{tabular}{cccccccccccccc}
\hline																																					
ID	&	method	&	$\alpha$	&	spectral 	&			$S_\mathrm{p}$	&			$\nu_\mathrm{p}$	&		$L_\mathrm{bol}$	&	ID	&	method	&	$\alpha$	&	spectral	&			$S_\mathrm{p}$	&			$\nu_\mathrm{p}$	&		$L_\mathrm{bol}$	\\
	&		&		&	type	&			(mJy)	&			(GHz)	&		(W)	&		&		&		&	 type	&			(mJy)	&			(GHz)	&		(W)	\\
(1)	&	(2)	&	(3)	&	(4)	&			(5)	&			(6)	&		(7)	&	(8)	&	(9)	&	(10)	&	(11)	&			(12)	&			(13)	&		(14)	\\\hline
b01	&	P	&	$-$0.57	&	s	&	\phn	$>$	112	&		$<$	0.48	&	$<$	36.3 	&	n01	&	P	&	$-$0.46	&	f	&			…	&			…	&		…	\\
b02	&	P	&	$-$0.26	&	f	&			…	&			…	&		…	&	n02	&	P	&	$-$0.15	&	f	&			…	&			…	&		…	\\
b03	&	H	&	$-$1.16	&	s	&	\phn	$>$	174	&		$<$	0.51	&	$<$	36.6 	&	n03	&	P	&	$-$0.26	&	f	&			…	&			…	&		…	\\
b04	&	H	&	$-$0.83	&	p	&	\phn\phn	\phs	13	&		\phs	1.54	&	\phs	35.7 	&	n04	&	H	&	$-$0.36	&	f	&			…	&			…	&		…	\\
b05	&	H	&	$-$0.82	&	s	&	\phn\phn	$>$	82	&		$<$	0.48	&	$<$	36.2 	&	n05	&	H	&	$-$0.64	&	p	&	\phn\phn	\phs	22	&	\phn	\phs	2.23	&	\phs	36.1 	\\
b06	&	H	&	$-$1.11	&	p	&	\phn\phn	\phs	82	&		\phs	6.78	&	\phs	37.4 	&	n06	&	H	&	$-$0.62	&	p	&	\phn\phn	\phs	70	&		\phs	11.81	&	\phs	37.7 	\\
b07	&	H	&	$-$0.93	&	p	&	\phn\phn	\phs	59	&		\phs	0.53	&	\phs	36.0 	&	n07	&	P	&	\phs0.03	&	f	&			…	&			…	&		…	\\
b08	&	H	&	$-$0.33	&	f	&			…	&			…	&		…	&	n08	&	P	&	\phs0.13	&	f	&			…	&			…	&		…	\\
b09	&	H	&	$-$2.06	&	p	&	\phn\phn	\phs	62	&		\phs	4.80	&	\phs	37.0 	&	n09	&	P	&	\phs0.17	&	f	&			…	&			…	&		…	\\
b10	&	H	&	$-$0.76	&	s	&	\phn\phn	$>$	57	&		$<$	0.18	&	$<$	35.6 	&	n10	&	H	&	$-$0.51	&	p	&	\phn	\phs	183	&	\phn	\phs	1.60	&	\phs	37.1 	\\
b11	&	P	&	$-$0.60	&	p	&	\phn\phn	$>$	28	&		$<$	4.46	&	$<$	36.6 	&	n11	&	P	&	$-$0.05	&	f	&			…	&			…	&		…	\\
b12	&	H	&	$-$1.78	&	s	&	\phn	$>$	797	&		$<$	0.24	&	$<$	36.8 	&	n12	&	H	&	$-$0.34	&	f	&			…	&			…	&		…	\\
b13	&	H	&	$-$1.16	&	dp	&		\phs	1139	&		\phs	1.19	&	\phs	37.6 	&	n13	&	P	&	\phs0.68	&	f	&			…	&			…	&		…	\\
b14	&	H	&	$-$1.30	&	p	&	\phn\phn	\phs	25	&		\phs	5.20	&	\phs	36.9 	&	n14	&	P	&	$-$0.29	&	f	&			…	&			…	&		…	\\
b15	&	H	&	$-$0.43	&	f	&			…	&			…	&		…	&	n15	&	H	&	$-$0.65	&	p	&	\phn	\phs	107	&	\phn	\phs	0.55	&	\phs	36.1 	\\
b16	&	H	&	$-$1.39	&	p	&	\phn\phn	\phs	29	&		\phs	3.71	&	\phs	36.8 	&	n16	&	H	&	$-$0.69	&	p	&	\phn\phn	\phs	78	&	\phn	\phs	0.25	&	\phs	36.3 	\\
b17	&	H	&	$-$0.68	&	p	&	\phn\phn	\phs	27	&		\phs	3.50	&	\phs	36.4 	&	n17	&	H	&	$-$0.54	&	p	&	\phn\phn	\phs	75	&	\phn	\phs	2.62	&	\phs	37.0 	\\
b18	&	H	&	$-$1.40	&	s	&	\phn\phn	$>$	11	&		$<$	3.59	&	$<$	36.5 	&	n18	&	H	&	$-$0.19	&	f	&			…	&			…	&		…	\\
b19	&	P	&	$-$0.06	&	f	&			…	&			…	&		…	&	n19	&	H	&	$-$0.65	&	p	&	\phn\phn	\phs	27	&	\phn	\phs	0.94	&	\phs	35.7 	\\
b20	&	P	&	$-$0.78	&	s	&	\phn\phn	$>$	93	&		$<$	0.41	&	$<$	35.9 	&	n20	&	H	&	$-$0.61	&	p	&	\phn\phn	\phs	69	&	\phn	\phs	1.12	&	\phs	36.3 	\\
b21	&	H	&	$-$1.17	&	p	&	\phn\phn	\phs	11	&		\phs	4.03	&	\phs	35.9 	&	n21	&	P	&	\phs0.13	&	f	&			…	&			…	&		…	\\
b22	&	H	&	$-$0.79	&	p	&	\phn\phn	\phs	49	&		\phs	2.24	&	\phs	36.6 	&	n22	&	P	&	$-$0.85	&	s	&	\phn	$>$	218	&	\phn	$<$	0.58	&	$<$	37.0	\\
b23	&	H	&	$-$1.20	&	p	&	\phn\phn	\phs	46	&		\phs	6.28	&	\phs	37.2 	&	n23	&	H	&	$-$0.36	&	f	&			…	&			…	&		…	\\
b24	&	H	&	$-$0.43	&	f	&			…	&			…	&		…	&	n24	&	H	&	$-$0.27	&	f	&			…	&			…	&		…	\\
b25	&	P	&	$-$0.63	&	s	&	\phn	$>$	158	&		$<$	0.15	&	$<$	35.7 	&	n25	&	H	&	…	&	f	&			…	&			…	&		…	\\
b26	&	H	&	$-$1.72	&	p	&	\phn\phn	\phs	86	&		\phs	9.44	&	\phs	37.8 	&	n26	&	P	&	$-$0.48	&	f	&			…	&			…	&		…	\\
b27	&	P	&	 \phs0.14	&	f	&			…	&			…	&		…	&	n27	&	P	&	$-$0.88	&	s	&	\phn	$>$	486	&	\phn	$<$	0.27	&	$<$	36.9	\\
b28	&	P	&	$-$0.55	&	s	&	\phn\phn	$>$	68	&		$<$	0.18	&	$<$	35.7 	&	n28	&	DP	&	$-$0.22	&	ds	&	\phn\phn\phn	$>$	6	&	\phn	$<$	0.42	&	$<$	34.8	\\
b29	&	P	&	$-$0.24	&	f	&			…	&			…	&		…	&	n29	&	P	&	\phs0.07	&	f	&			…	&			…	&		…	\\
b30	&	H	&	$-$0.64	&	p	&	\phn\phn	\phs	42	&		\phs	6.83	&	\phs	37.3 	&	n30	&	P	&	$-$0.53	&	s	&		$>$	4123	&	\phn	$<$	0.22	&	$<$	37.8	\\
b31	&	P	&	$-$0.22	&	f	&			…	&			…	&		…	&	n31	&	H	&	$-$0.79	&	p	&	\phn	\phs	439	&	\phn	\phs	1.26	&	\phs	37.2 	\\
b32	&	P	&	 \phs0.01	&	f	&			…	&			…	&		…	&	n32	&	P	&	\phs1.03	&	f	&			…	&			…	&		…	\\
b33	&	H	&	$-$0.59	&	dp	&	\phn\phn	\phs	22	&		\phs	3.15	&	\phs	36.7 	&	n33	&	H	&	$-$0.37	&	f	&			…	&			…	&		…	\\
b34	&	H	&	$-$1.07	&	p	&	\phn\phn	\phs	36	&		\phs	1.60	&	\phs	36.1 	&	n34	&	H	&	…	&	f	&			…	&			…	&		…	\\
b35	&	H	&	$-$1.09	&	p	&	\phn\phn	\phs	25	&		\phs	5.18	&	\phs	36.8 	&	n35	&	P	&	$-$0.06	&	f	&			…	&			…	&		…	\\
b36	&	H	&	$-$1.47	&	p	&	\phn\phn	\phs	50	&		\phs	0.53	&	\phs	35.8 	&	n36	&	DP	&	$-$0.46	&	ds	&	\phn\phn	$>$	37	&	\phn	$<$	0.55	&	$<$	36.1	\\
b37	&	P	&	$-$0.56	&	s	&	\phn\phn	$>$	59	&		$<$	0.45	&	$<$	35.9 	&	n37	&	P	&	$-$0.14	&	f	&			…	&			…	&		…	\\
b38	&	H	&	$-$0.35	&	f	&			…	&			…	&		…	&	n38	&	P	&	$-$0.06	&	f	&			…	&			…	&		…	\\
b39	&	P	&	$-$0.37	&	f	&			…	&			…	&		…	&	n39	&	H	&	$-$0.17	&	f	&			…	&			…	&		…	\\
b40	&	H	&	$-$0.23	&	f	&			…	&			…	&		…	&	n40	&	P	&	\phs0.47	&	f	&			…	&			…	&		…	\\
b41	&	H	&	$-$0.52	&	p	&	\phn\phn	\phs	59	&		\phs	0.74	&	\phs	36.1 	&	n41	&	H	&	$-$1.05	&	p	&	\phn\phn	\phs	17	&	\phn	\phs	1.71	&	\phs	35.9 	\\
b42	&	DP	&	$-$0.44	&	ds	&	\phn\phn	$>$	12	&		$<$	0.58	&	$<$	35.8 	&	n42	&	P	&	$-$0.72	&	s	&	\phn	$>$	136	&	\phn	$<$	0.49	&	$<$	36.5	\\
b43	&	P	&	$-$0.19	&	f	&			…	&			…	&		…	&	n43	&	P	&	$-$0.76	&	s	&	\phn	$>$	113	&	\phn	$<$	0.57	&	$<$	36.7	\\
b44	&	H	&	$-$1.90	&	p	&	\phn\phn	\phs	63	&		\phs	3.86	&	\phs	36.9 	&	n44	&	DP	&	$-$0.68	&	ds	&	\phn	$>$	196	&	\phn	$<$	0.57	&	$<$	36.9	\\
b45	&	H	&	$-$1.65	&	p	&	\phn\phn	\phs	56	&		\phs	3.03	&	\phs	37.2 	&	n45	&	H	&	$-$0.79	&	p	&	\phn\phn	\phs	55	&			\phs10.63	&	\phs	37.4 	\\
b46	&	H	&	$-$1.12	&	p	&	\phn\phn	\phs	43	&		\phs	0.76	&	\phs	36.3 	&	n46	&	H	&	$-$0.64	&	p	&	\phn\phn	\phs	39	&	\phn	\phs	0.67	&	\phs	35.9 	\\
b47	&	H	&	$-$1.20	&	p	&	\phn\phn	\phs	10	&		\phs	5.10	&	\phs	36.3 	&	n47	&	P	&	\phs0.28	&	f	&			…	&			…	&		…	\\
b48	&	H	&	$-$1.02	&	p	&	\phn\phn	\phs	26	&		\phs	1.01	&	\phs	35.7 	&	n48	&	DP	&	…	&	ds	&	\phn\phn	$>$	16	&	\phn	$<$	0.50	&	$<$	35.6	\\
\hline				
\end{tabular}

\textit{Note.} Columns are as follows:
(1) and (8) Object ID;
(2) and (9) function used for fitting to the radio spectrum where P, DP, and H represent pawer law, double power law, and hyperbola, respectively;
(3) and (10) estimated spectral index of the optically thin regime employed for classifying the spectral type;
(4) and (11) result of the spectral classification;
(5)--(7) and (12)--(14) estimated peak flux density, $S_\mathrm{p}$, peak frequency, $\nu_\mathrm{p}$, and bolometric radio luminosity, $L_\mathrm{bol}$, of steep or peaked spectral sources.
\end{minipage}
\end{table*}